\documentclass[apj,tighten,numberedappendix,appendixfloats,iop]{emulateapj}
\usepackage%
[
pagebackref, 
colorlinks=true,
linkcolor=red,filecolor=red,
citecolor=blue,
urlcolor=cyan,
pdftitle={Outward Migration of Jupiter and Saturn},
pdfauthor={D'Angelo, Marzari},
pdfsubject={Outward Migration of Jupiter and Saturn},
pdfcreator={KAZEditor},
pdfproducer={Kazzimiei},
pdfkeywords={Accretion, accretion disks; Hydrodynamics; Methods: numerical; Planet-disk interactions; Planets and satellites: formation; Protoplanetary disks},
bookmarks=true,
breaklinks=true,
hypertexnames=false,
pdfstartview=FitV,
hyperfootnotes=true,
bookmarksopen=false,
pdfpagemode=None]{hyperref}
\usepackage{breakurl}
\newcommand{\Mp}{\mbox{$M_{p}$}}                     
\newcommand{\Ms}{\mbox{$M_{s}$}}                     
\newcommand{\Msun}{\mbox{$M_{\odot}$}}                     
\newcommand{\Rhill}{\mbox{$R_\mathrm{H}$}}           
\newcommand{\AU}{\mbox{AU}}                          
\newcommand{\figlen}{0.9\linewidth}

\slugcomment{The Astrophysical Journal, 757:50, 2012 September 20}

\shorttitle{Outward Migration of Jupiter and Saturn}
\shortauthors{D'Angelo \& Marzari}

\begin{document}

\title{Outward Migration of Jupiter and Saturn in Evolved Gaseous Disks}
\author{Gennaro D'Angelo\altaffilmark{1,2,4} and Francesco Marzari\altaffilmark{3}}
\altaffiltext{1}{NASA Ames Research Center, MS 245-3, Moffett Field, CA 94035, USA 
(\href{mailto:gennaro.dangelo@nasa.gov}{gennaro.dangelo@nasa.gov})}
\altaffiltext{2}{SETI Institute, 189 Bernardo Avenue, Mountain View, CA 94043, USA}
\altaffiltext{3}{Department of Physics, University of Padova, Via Marzolo 8,  Padova  I-35131, Italy
(\href{mailto:francesco.marzari@pd.infn.it}{francesco.marzari@pd.infn.it})}
\altaffiltext{4}{Visiting Research Scientist, Los Alamos National Laboratory, Los Alamos, NM 87545, USA}

\begin{abstract}
The outward migration of a pair of resonant-orbit planets, 
driven by tidal interactions with a gas-dominated disk, is studied 
in the context of evolved solar nebula models.
The planets' masses, $M_{1}$ and $M_{2}$, correspond to those 
of Jupiter and Saturn.
Hydrodynamical calculations in two and three dimensions 
are used to quantify the migration
rates and analyze the conditions under which the outward 
migration mechanism may operate.
The planets are taken to be fully formed 
after $10^{6}$ and before $3\times 10^{6}$ years. 
The orbital evolution of the
planets in an evolving disk is then  calculated until the disk's gas is 
completely dissipated. 
Orbital locking in the 3:2 mean motion resonance
may lead to outward migration under appropriate
conditions of disk viscosity and temperature. However, resonance locking 
does not necessarily result in outward migration. 
This is the case, for example, if
convergent migration leads to locking in the 2:1 mean motion resonance,
as post-formation disk conditions seem to suggest.
Accretion of gas on the planets may deactivate the outward migration
mechanism by raising the mass ratio $M_{2}/M_{1}$ and/or by reducing 
the accretion rate toward the star, hence depleting the inner disk.
For migrating planets locked in the 3:2 mean motion resonance, there
are stalling radii that depend on disk viscosity and on stellar irradiation,
when it determines the disk's thermal balance.
Planets locked in the 3:2 orbital resonance that start moving outward 
from within $1$--$2\,\AU$ may reach beyond $\approx 5\,\AU$ only under 
favorable conditions. 
However, within the explored space of disk parameters, only a small fraction 
-- less than a few percent -- of the models predict that the interior planet reaches 
beyond $\approx 4\,\AU$.
\end{abstract}

\keywords{accretion, accretion disks --- hydrodynamics --- methods: numerical --- planet-disk interactions ---planets and satellites: formation --- protoplanetary disks}

\section{Introduction}
\label{sec:Intro}

\defcitealias{gennaro2008}{DL08}
\defcitealias{lubow2006}{LD06}

The architecture of the solar system bears some evidence that Jupiter 
and Saturn may have been closer to each other in the past
\citep[e.g.,][]{malhotra1993,malhotra1995,tsiganis2005,morbidelli2005,gomes2005}. 
They later moved away from each other because 
of gravitational interactions with the remnants of the disk of planetesimals
from which these planets had formed \citep[e.g.,][]{fernandez1984,hahn1999}.
The planetesimal-driven migration of Jupiter and Saturn occurred relatively late, 
after the gaseous component of the solar nebula had dispersed, 
and the extent of their radial displacements was probably less than 
$\sim 1\,\AU$ \citep[e.g.,][]{franklin2004,minton2009}. 

Recently, \citet{walsh2011} proposed a scenario in which orbital migration 
of Jupiter and Saturn occurred much earlier in the solar system history 
and was driven by tidal torques in a gas-dominated nebula. 
The progenitors of Jupiter and Saturn underwent rapid convergent migration 
toward the Sun, until Saturn became trapped in the 2:3 mean motion resonance 
with Jupiter. By that time and under the applied conditions, Jupiter had reached 
$\approx 1.5\,\AU$ and Saturn $\approx 2.0\,\AU$. 
Once the resonant configuration was established, the planets reversed 
the direction of motion and began migrating outward, 
preserving the 2:3 commensurability.
This scenario may help explain some features of the inner solar system,
including the Mars-to-Earth mass ratio and the radial variation of composition
in the asteroid belt \citep[see][for details]{walsh2011}.

The outward migration is a direct result of the ``compact'' orbital configuration.
Qualitatively, the negative torque balance that would result for a 
single-planet is tipped in favor of the positive torque (from the 
inner disk) because the negative torque (from the outer disk) is abated 
by a local reduction of the surface density.
This situation requires that the planets be massive enough to significantly perturb, 
via tidal interaction, the
disk's surface density and that their density gaps overlap. These requirements
are typically realized if the orbital separation is at most several times the sum 
of the planets' Hill radii. Therefore, depending on the masses, a (near) 3:2 
commensurability is favorable to sustain outward migration of 
a Jupiter--Saturn pair, whereas for more massive planets, by a factor 
of about three, a (near) 2:1 commensurability may promote outward migration.

A study by \citet{pierens2011} lends support, under appropriate conditions, 
to the inward-outward migration scenario of the Jupiter--Saturn system proposed 
by \citet{walsh2011}. 
One scope of this paper is to revisit this idea in the context of evolved 
models of a gas-dominated solar nebula. In particular, we concentrate on the
outward migration of a pair of giant planets, whose masses correspond to those of
Jupiter and Saturn, after their orbits become locked in the 3:2 mean motion resonance, 
compatibly with the formation timescales of both Jupiter and Saturn, estimated from 
core-nucleated accretion models.

We also wish to provide some constraints on the range of radial migration
of Jupiter (and Saturn), as a function of the solar nebula properties, under 
the assumption that the 3:2 orbital resonance is maintained throughout
the disk's evolution.
Conditions that may break the resonance locking between the two planets 
or that may inhibit or prevent outward migration are analyzed as well. 
In particular, we focus on the process of gas
accretion that, on one hand, may alter the planets' mass ratio and, on the
other, may reduce the disk density inside the orbit of the interior planet.
Both effects act to change the balance of the torques exerted on the planets.
In addition, we examine the disk conditions under which convergent 
migration leads to capture of the exterior planet in the 1:2 orbital 
resonance with the interior planet, a configuration that does not 
promote outward migration of a Jupiter--Saturn pair, and which may
leave the planets stranded in the inner disk region.
The possibility that Saturn forms within the 1:2 commensurability 
with Jupiter is also analyzed.

The layout of the paper is as follows. In Section~\ref{sec:DEM}, we describe
dynamics and thermodynamics of disk models and report on their evolution. 
In Section~\ref{sec:TI}, the tidal interaction calculations in two and three
dimensions are presented, along with the calculations of the migration
rates of a 3:2 resonant-orbit pair. Section~\ref{sec:LTM} is dedicated to
the long-term orbital evolution of two planets locked in the 3:2 orbital resonance.
Two possible effects of gas accretion are analyzed in Sections~\ref{sec:PPoA}
and \ref{sec:SPoA}, while conditions for capture in the 2:1 
mean motion resonance and some related issues are examined in Section~\ref{sec:21}.
Section~\ref{sec:SaD} contains the discussion and the summary of the results.

\section{Long-Term Disk Evolution Models}
\label{sec:DEM}

In this section we describe the dynamics and thermodynamics of solar nebula
models. For tested parameters, we report on the disk evolution until the gas is 
almost entirely dispersed, that is until the disk mass, 
$M_{\mathrm{D}}$, is less than $10^{-5}$ times the mass of the star. 
By assumption, successful sets of parameters representing a solar nebula 
model are those that provide a disk lifetime, $\tau_{\mathrm{D}}$, 
no greater than $\sim 2\times 10^{7}$ years. 
Although the gas mass of disks is notoriously difficult to ascertain,
according to observations \citep[see, e.g., reviews by][]{roberge2010,williams2011},
the presence of gas in the inner regions of protoplanetary disks
appears to last $\lesssim 10^{7}$ years \citep[see also][]{haisch2001}.

\subsection{Disk Dynamics}
\label{sec:DD}

Consider a gaseous disk orbiting a central star of mass $\Ms$. 
In the framework
of one-dimensional (1D) modeling, we assume azimuthal symmetry around
the star and use vertically averaged quantities as a function of  the radial 
distance $r$.
For the current purposes,
we assume that the evolution of the disk is driven by viscous torques, 
$\mathcal{T}_{\nu}$, and wind dispersal, $\dot{M}_{\mathrm{w}}$
at the disk's surface. 
The torque exerted on a disk ring of radius $r$, by material orbiting
inside the ring, is 
$\mathcal{T}_{\nu}=-2\pi r^{3}\nu\Sigma \partial\Omega/\partial r$ 
\citep{lynden-bell1974},
where $\nu$ is the kinematic viscosity of the gas,  $\Sigma$ the surface 
density,  and $\Omega$ the angular velocity. If $\Omega$ is identified
as the Keplerian velocity 
(i.e., if effects of gas and magnetic pressure gradients are neglected), 
then $\mathcal{T}_{\nu}=3\pi \nu\Sigma \mathcal{H}$, where 
$\mathcal{H}=r^{2}\Omega$ is the specific angular momentum of
the gas. 
Along with viscous diffusion, the disk is dispersed by a wind, whose origin 
is gas photo-evaporation from the disk surface produced by photons emitted 
by the central star. Hence, we write 
$\dot{M}_{\mathrm{w}}=2\pi\int \dot{\Sigma}_{\mathrm{pe}} r dr$,
where $\dot{\Sigma}_{\mathrm{pe}}$ is the mass per unit surface area and
unit time removed from the disk.

The continuity equation for the disk requires that
\begin{equation}
\frac{\partial}{\partial t}\Sigma + \frac{1}{r}\frac{\partial}{\partial r}(r\Sigma u_{r}) =%
-\dot{\Sigma}_{\mathrm{pe}},
\label{eq:cont}
\end{equation}
where $u_{r}$ is radial velocity of the gas.  
On the left-hand side, one can recognize the mass per unit time 
flowing through a circumference of radius $r$,
$\mathcal{F}=2\pi r \Sigma u_{r}$. By using the relation
$\mathcal{F}=-\partial \mathcal{T}_{\nu}/\partial \mathcal{H}$ 
\citep[see][]{lynden-bell1974} and since we assume Keplerian rotation
($\partial \mathcal{H}/\partial r = r\Omega/2$),
Equation~(\ref{eq:cont}) becomes
\begin{equation}
\pi r \frac{\partial}{\partial t}(\Sigma+\Sigma_{\mathrm{pe}}) - %
\frac{\partial}{\partial r}\left(\frac{1}{r\Omega}\frac{\partial \mathcal{T}_{\nu}}{\partial r}\right) = 0.
\label{eq:diskeq}
\end{equation}
To seek for numerical solutions of Equation~(\ref{eq:diskeq}), it is 
convenient to use $\mathcal{H}$ as independent variable 
and $\mathcal{S}=\mathcal{H}^{3}\Sigma$ as dependent variable and 
then solve
\begin{equation}
\frac{\partial}{\partial t}(\mathcal{S}+\mathcal{S}_{\mathrm{pe}}) - %
\frac{3}{4}(G\Ms)^2%
\frac{\partial^{2}}{\partial \mathcal{H}^{2}}\left(%
\frac{\mathcal{\nu\mathcal{S}}}{\mathcal{H}^2}\right) = 0.
\label{eq:numeq}
\end{equation}
In the above equation, $G$ is the gravitational constant and
$\dot{\mathcal{S}}_{\mathrm{pe}}=\mathcal{H}^{3}\dot{\Sigma}_{\mathrm{pe}}$.
Note that quantity $\mathcal{S}/\mathcal{H}^{2}$ is the angular momentum
per unit surface area.
In writing Equation~(\ref{eq:cont}), we neglected the effects of the star's growth,
which would introduce a term on the right-hand side of order 
$\Sigma\dot{M}_{s}/\Ms$ \citep[see][]{ruden1991}. Given the initial values of 
$\dot{M}_{s}/\Ms$ considered here (see Section~\ref{sec:NPP}) and the decline 
of $\dot{M}_{s}$ with time, this term would affect $\Sigma$ only over a time
scale of order $10^{7}$ years, or longer.

Photo-evaporation involves contributions from  far-ultraviolet (FUV), 
extreme-ultraviolet (EUV), and X-ray radiation 
emitted by the star \citep[see][and references therein]{dullemond2007,clarke2011}. 
FUV radiation may be especially
important in removing gas at large distances from the star, reducing the gas supply
to the inner parts of the disk. However, a self-consistent calculation of 
FUV photo-evaporation rates requires solving for the detailed vertical structure 
of the disk \citep[e.g.,][]{gorti2009a,gorti2009b}. 
Photo-evaporation by EUV photons is more tractable since they ionize hydrogen 
at the very upper layers of the disk. 
Here we follow a simple approach and adopt the formulation of the EUV 
photo-evaporation rate proposed by \citet{dullemond2007}:
\begin{equation}
\frac{\dot{\Sigma}_{\mathrm{pe}}}{\dot{\Sigma}^{\mathrm{g}}_{\mathrm{pe}}}%
=\left\{%
\begin{array}{ll}
          \exp{\left[\frac{1}{2}\left(1-\frac{r_{\mathrm{g}}}{r}\right)\right]}%
          \left(\frac{r_{\mathrm{g}}}{r}\right)^{2}
                                                                    & \mathrm{for}\ r\le r_{\mathrm{g}},\\
          \left(\frac{r_{\mathrm{g}}}{r}\right)^{5/2}
                                                                    & \mathrm{for}\ r> r_{\mathrm{g}}.
\end{array}
\right.
\label{eq:Spe}
\end{equation}
The radius $r_{\mathrm{g}}\approx 10 \left(\Ms/\Msun\right)\,\AU$ is the gravitational 
radius, beyond which gas at the disk surface is unbound 
\citep[see, e.g.,][and references therein]{armitage2011}. 
The photo-evaporation rate at $r_{\mathrm{g}}$ is
\begin{equation}
\dot{\Sigma}^{\mathrm{g}}_{\mathrm{pe}}=%
1.16\times 10^{-11}\sqrt{f_{41}}\left(\frac{1\,\AU}{r_{\mathrm{g}}}\right)^{3/2}\,%
\left(\frac{\Msun}{\AU^{2}\,\mathrm{yr}}\right),
\label{eq:Speg}
\end{equation}
where $f_{41}$ is the rate of EUV ionizing photons emitted by the star in units of
$10^{41}\,\mathrm{s}^{-1}$. The total mass loss rate due to photo-evaporation
is found by integrating Equation~(\ref{eq:Spe}) over the entire disk according to
the definition given above, hence
$\dot{M}_{\mathrm{w}}=\left(0.55977\sqrt{e}+4\pi\right)\,r^{2}_{\mathrm{g}}\,%
\dot{\Sigma}^{\mathrm{g}}_{\mathrm{pe}}$ or
\begin{equation}
\dot{M}_{\mathrm{w}}=%
1.56\times 10^{-10}\sqrt{f_{41}\left(\frac{r_{\mathrm{g}}}{1\,\AU}\right)}\,%
\Msun\,\mathrm{yr}^{-1}.
\label{eq:dotMw}
\end{equation}

The maximum of $\dot{\Sigma}_{\mathrm{pe}}$ occurs at
$r=r_{\mathrm{g}}/4$.
Locally, gas is removed via photo-evaporation and supplied by viscous
diffusion, i.e., accretion through the disk, $\dot{M}=-\mathcal{F}$ 
(note that $\mathcal{F}$ is positive for an outward transfer of mass). 
Recalling the relations reported above, we can write
$\partial\mathcal{T}_{\nu}/\partial\mathcal{H}=%
3\pi\partial(\nu\Sigma\mathcal{H})/\partial\mathcal{H}$,  and thus
\begin{equation}
\dot{M}=3\pi\left[\nu\Sigma+2 r \frac{\partial}{\partial r}(\nu\Sigma)\right].
\label{eq:dotM}
\end{equation}
For $\nu\Sigma$ nearly independent of $r$, i.e., in a stationary disk
\citep[][see also Equation~\ref{eq:cont} with the right-hand side set to zero]{pringle1981}, 
$\dot{M}=3\pi\nu\Sigma$ is nearly constant throughout the disk. 
Therefore, if $\Ms=1\,\Msun$, we expect gas depletion 
induced by photo-evaporation to occur first around $\sim 3\,\AU$.

\subsection{Disk Thermodynamics}
\label{sec:DT}

In order to determine the thermal energy budget of the disk during 
its evolution, we assume that there is a balance among three terms: 
viscous heating, irradiation heating by the central star, and radiative 
cooling from the disk's surface. 
Viscous dissipation produces an energy flux equal to
$Q_{\nu}=\nu\Sigma \left(r\partial\Omega/\partial r\right)^{2}$
\citep[see, e.g.,][]{m&m}, which
in case of the Keplerian rotation becomes
\begin{equation}
Q_{\nu}=\frac{9}{4}\nu\Sigma\Omega^{2}.
\label{eq:Qnu}
\end{equation}
Since $Q_{\nu}\propto 1/r^{3}$, for a disk with 
$\partial (\nu\Sigma)/\partial r\approx 0$, viscous dissipation becomes 
an ever less important source of heating as the distance from the star increases.

We follow the formulation of \citet{hubeny1990} for an irradiated
disk and write the energy flux escaping from both sides of the disk 
surface as
\begin{equation}
Q_{\mathrm{cool}}=2\sigma_{\mathrm{SB}}\,T^{4}\left(%
\frac{3}{8}\tau_{\mathrm{R}}+\frac{1}{2}+\frac{1}{4\tau_{\mathrm{P}}}%
\right)^{-1},
\label{eq:Qcool}
\end{equation}
whereas the heating flux arising from stellar irradiation can be written as
\begin{equation}
Q_{\mathrm{irr}}=2\sigma_{\mathrm{SB}}\,T^{4}_{\mathrm{irr}}\left(%
\frac{3}{8}\tau_{\mathrm{R}}+\frac{1}{2}+\frac{1}{4\tau_{\mathrm{P}}}%
\right)^{-1}.
\label{eq:Qirr}
\end{equation}
In the above equations, $\sigma_{\mathrm{SB}}$ is the Stefan-Boltzmann constant, 
$T$ the mid-plane temperature, and $T_{\mathrm{irr}}$ the irradiation 
temperature.  Note that, for an irradiated disk, the constant in parenthesis 
on the right-hand side of Equation~(\ref{eq:Qcool}) is generally slightly different from 
that of a non-irradiated disk \citep[compare with Equation~14 of][]{gennaro2003b}.
As in \citet{menou2004}, we set
\begin{equation}
T^{4}_{\mathrm{irr}}=(1-\epsilon)\,T^{4}_{s}%
                                  \left(\frac{R_{s}}{r}\right)^{2}W_{\mathrm{G}},
\label{eq:Tirr}
\end{equation}
where $\epsilon$ is a measure of the disk's albedo, for which we adopt the
value $1/2$, and $T_{s}$ and $R_{s}$ are the effective temperature and 
radius of the star, respectively.
This interpretation of the irradiation temperature, however, neglects 
the contribution of luminosity released by stellar accretion \citep[e.g.,][]{hartmann2011}.
In an actively accreting disk, quantity $T^{4}_{s}$ should be replaced with
$T^{4}_{*}=T^{4}_{s}+T^{4}_{\mathrm{acc}}$, where $T^{4}_{\mathrm{acc}}$ 
quantifies the luminosity due to accretion 
$L_{\mathrm{acc}}=G\Ms\dot{M}_{s}/(2R_{s})$ \citep{pringle1981}
and thus
\begin{equation}
T^{4}_{\mathrm{acc}}=\frac{1}{8\pi}%
\left(\frac{G\Ms\dot{M}_{s}}{\sigma_{\mathrm{SB}}R^{3}_{s}}\right),
\label{eq:Tacc}
\end{equation}
where the accretion rate $\dot{M}_{s}$, computed as 
$-\partial \mathcal{T}_{\nu}/\partial \mathcal{H}$ (see Section~\ref{sec:DD})
at the disk's inner radius, varies with time.

The quantity $W_{\mathrm{G}}$ in Equation~(\ref{eq:Tirr}) is a geometrical 
factor that accounts for illumination of disk portions close to (first term) and 
far from (second term) the star \citep[see][]{chang1997}
\begin{equation}
W_{\mathrm{G}}=0.4 \left(\frac{R_{s}}{r}\right)+%
\frac{H}{r}\left(\frac{d\ln{H}}{d\ln{r}}-1\right).
\label{eq:WG}
\end{equation}
The adiabatic scale-height of the disk, 
$H=\sqrt{\gamma\,k_{\mathrm{B}}T/(\mu m_{\mathrm{H}})}/\Omega$, is
derived from the requirement of vertical hydrostatic equilibrium. 
The adiabatic index, $\gamma$, is $1.4$,
the mean molecular wight, $\mu$, is $2.39$, $k_{\mathrm{B}}$ 
is the Boltzmann constant, and $m_{\mathrm{H}}$ the hydrogen mass.

If the second term on the right-had side of Equation~(\ref{eq:WG}) is negative, 
the disk is self-shadowed and that term should be dropped. 
A self-consistent calculation of this term from 1D,
vertically averaged models may lead to numerical instabilities 
\citep[see, e.g.,][]{hueso2005} . In fact, meaningful determinations of this term 
involve solving for the vertical thermal structure of the disk. 
Therefore, the last term in parenthesis on the right-hand side of 
Equation~(\ref{eq:WG}) is written as $\eta$ and approximated to $2/7$ 
\citep[see, e.g.,][]{dalessio1998,menou2004,hueso2005,rafikov2006}.

The optical depths
$\tau_{\mathrm{R}}=\kappa_{\mathrm{R}}\Sigma/2$ and
$\tau_{\mathrm{P}}=\kappa_{\mathrm{P}}\Sigma/2$ in 
Equations~(\ref{eq:Qcool}) and (\ref{eq:Qirr}) are based,
respectively, on Rosseland ($\kappa_{\mathrm{R}}$) and Planck 
($\kappa_{\mathrm{P}}$) mean opacities.
Both  $\kappa_{\mathrm{R}}$ and $\kappa_{\mathrm{P}}$ depend
on $T$ and the mass density $\rho=\Sigma/(2H)$.
We adopt grain opacities from \citet{pollack1994}, at temperatures 
below the vaporization temperatures of silicates, and gas opacities
from \citet{ferguson2005} for solar abundances, when all grain species 
have evaporated.

The thermal energy budget is given by
\begin{equation}
Q_{\nu}+Q_{\mathrm{irr}}-Q_{\mathrm{cool}}=0.
\label{eq:EEq}
\end{equation}
Note that if $Q_{\nu}\ll Q_{\mathrm{irr}}$, a situation that may occur
in an evolved disk, Equation~(\ref{eq:EEq}) results in a gas temperature 
$T=T_{\mathrm{irr}}$, that is
\begin{equation}
T=T_{*}\sqrt{\frac{R_{s}}{r}}%
    \left[(1-\epsilon)\,W_{\mathrm{G}}\right]^{1/4}.
\label{eq:Tsemp}
\end{equation}
The factor $W_{\mathrm{G}}$ is typically a weakly dependent function
of $T$. 
If $W_{\mathrm{G}}$ is a constant, then $T\propto r^{-1/2}$.
If  $W_{\mathrm{G}}\propto R_{s}/r$ (e.g., at radii $r\sim R_{s}$), 
then $T\propto r^{-3/4}$.
If $W_{\mathrm{G}}\propto H/r$ (as we assume for $r\gg R_{s}$), 
then $W^{1/4}_{\mathrm{G}}\propto T^{1/8}$,
the temperature is $T\propto r^{-3/7}$ \citep[see also][]{chambers2009}, 
and the disk's aspect ratio is
\begin{equation}
\left(\frac{H}{r}\right)^{7}=\eta\left(1-\epsilon\right)%
\left(\frac{\gamma\,k_{\mathrm{B}}T_{*}}{\mu m_{\mathrm{H}}}\right)^{4}\!%
\left(\frac{R_{s}}{G\Ms}\right)^{4}\!%
\left({\frac{r}{R_{s}}}\right)^{2}.
\label{eq:HRsemp}
\end{equation}
The choice of the parameter $\eta$ may have some impact on the disk's 
thermal budget, yet Equation~(\ref{eq:HRsemp}) suggests that
this impact is low.

\subsection{Numerical Procedures and Parameters}
\label{sec:NPP}

\begin{figure*}[t!]
\centering%
\resizebox{\figlen}{!}{%
\includegraphics[clip,bb=20 10 481 337]{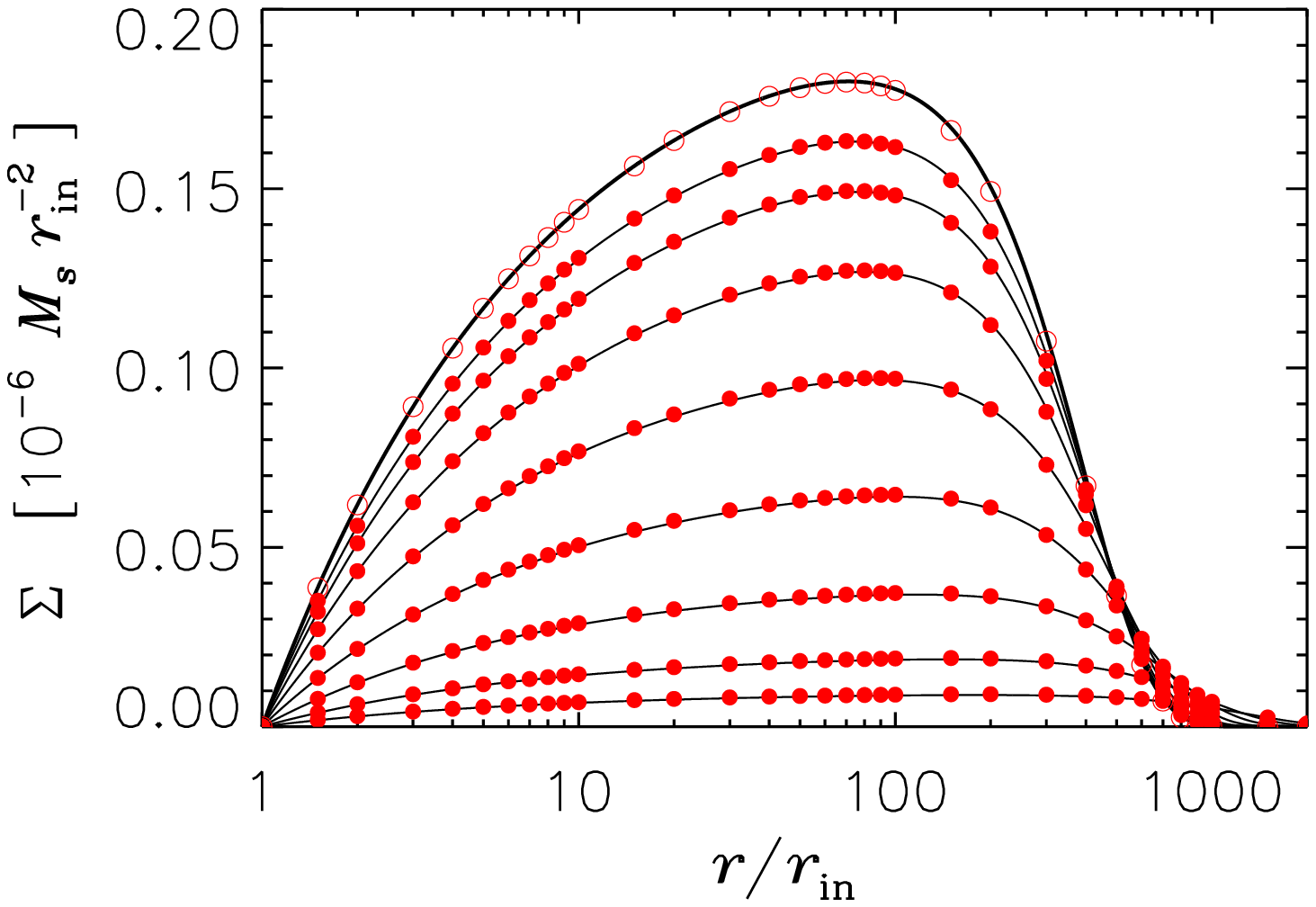}%
\includegraphics[clip,bb=20 10 481 337]{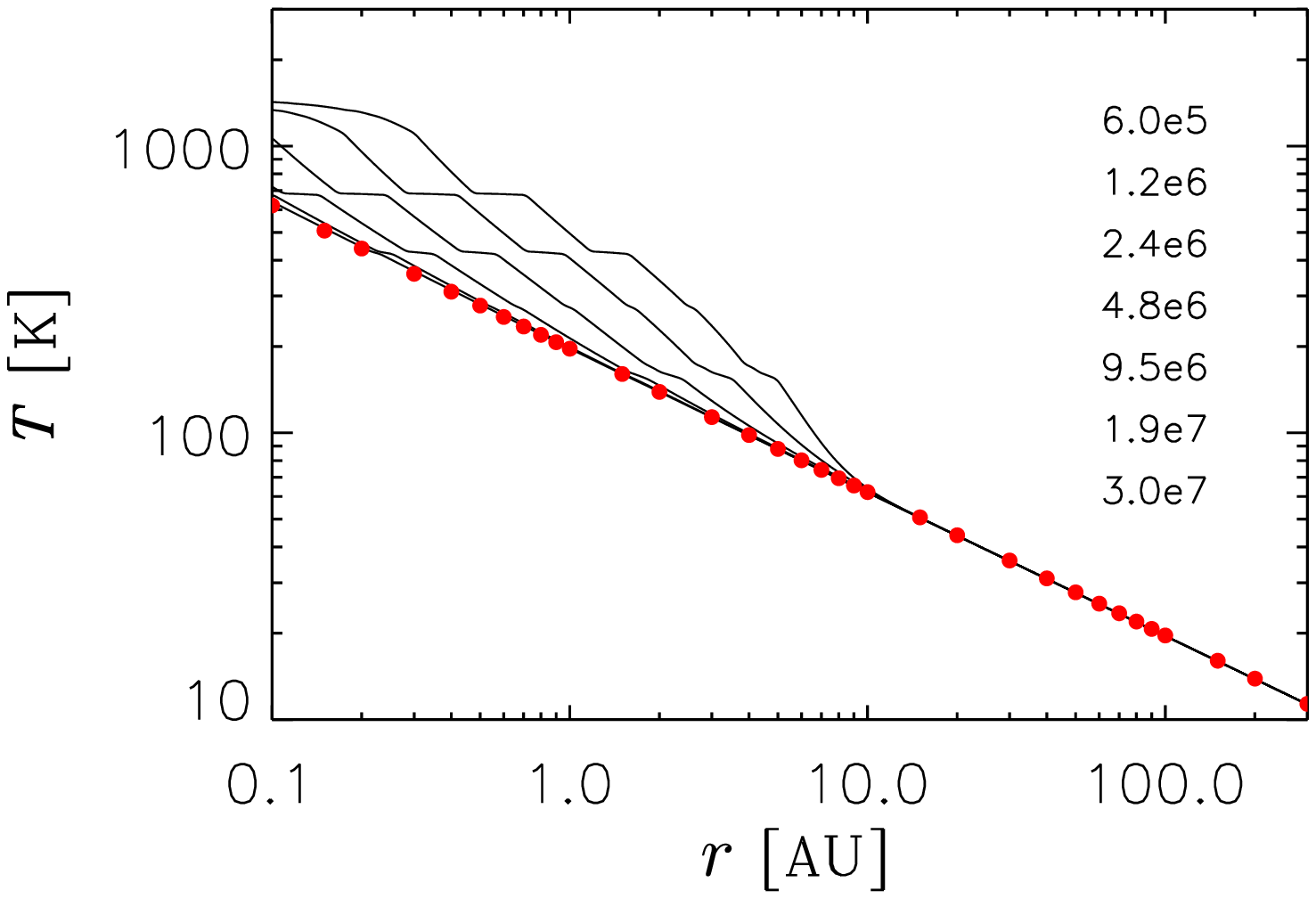}}
\caption{%
              \textit{Left:}
              evolution of a viscous disk obtained by solving Equation~(\ref{eq:numeq}),
              with $\dot{\mathcal{S}}_{\mathrm{pe}}=0$, by means of the Dormand--Prince 
              method. The initial condition (open circles) is the analytic solution of 
              \citet{lynden-bell1974} for a disk with no central couple, 
              $M_{\mathrm{D}}=0.1\,\Msun$, $\dot{M}_{s}=10^{-7}\,\mathrm{\Msun\,yr^{-1}}$, 
              and $\nu=8\times 10^{-6}\,r^{2}_{\mathrm{in}}\,\Omega_{\mathrm{in}}$.
              The solid lines represent the numerical solution at different times and the
              filled circles are computed using the analytic solution.
              \textit{Right:}
              evolution of temperature obtained by solving Equation~(\ref{eq:EEq}) for
              the disk in the left panel. Times in the legend are in years.
              For testing purposes, $W_{\mathrm{G}}$ is set equal to $0.05$.
              The temperature predicted by  Equation~(\ref{eq:Tsemp}) is indicated
              by filled circles.
             }
\label{fig:t1}
\end{figure*}
Equation~(\ref{eq:numeq}) is evolved in time using an implicit numerical scheme,
which avoids the sometimes prohibitively short time steps required by an explicit 
approach, especially when the inner disk radius extends very close to the star
\citep[see][]{bath1981}.
We either use a second-order Crank--Nicolson method \citep[e.g.,][]{pressF1992}
or a fourth/fifth-order Dormand--Prince method with an adaptive step-size control 
based on the global accuracy of the solution \citep{hairer1993}. 
In the latter case, the evaluation of derivatives (in the Runge--Kutta sequence) 
is performed by means of a backward Euler (implicit) method.
A zero-torque boundary condition, $\mathcal{S}=0$, is applied at the disk's
inner edge. At the outer edge, the applied boundary condition is such that
$\partial\dot{M}/\partial\mathcal{H}=%
\partial^{2}{\mathcal{T}_{\nu}}/\partial\mathcal{H}^{2}$ is constant.
Figure~\ref{fig:t1} (\textit{left}) shows a comparison between numerical
(lines) and analytic (circles) solutions of Equation~(\ref{eq:numeq}) 
(see figure's caption for details). 

Equation~(\ref{eq:EEq}) is solved for the mid-plane temperature, $T$,
at each radius, using a root-finding algorithm based on the 
Brent's method \citep{brent1973}. Convergence of the root-finding
process is achieved within a tolerance of $10^{-3}\,\mathrm{K}$.
An iterative procedure is implemented for each determination of $T$, so
that the applied value of $H$ and that corresponding to the converged 
temperature do not differ by more than $1$\%.
In Figure~\ref{fig:t1} (\textit{right}), the evolution of temperature is shown 
for the disk considered in the left panel. In this test, we set 
$W_{\mathrm{G}}=0.05$, so that temperature evolves toward
that in Equation~(\ref{eq:Tsemp}), indicated as filled circles.
The temperature profiles show major opacity transitions at 
$T\approx 160$, $420$, $680$, and $1400\,\mathrm{K}$, caused 
by vaporization of, respectively, water ice, refractory organics, troilite, 
and silicate grains \citep[see][]{pollack1994}.
Note that heating via viscous dissipation is basically confined within
$\sim 10\,\AU$ (see Section~\ref{sec:DT}).

The solar nebula extends from $r_{\mathrm{in}}=0.01\,\AU$ to 
$r_{\mathrm{out}}=1850\,\AU$ and is discretized over $10000$ 
grid points. 
The large outer radius is chosen to not interfere with viscous 
spreading of the disk.
The numerical resolution is variable and such that 
$\Delta r/r\simeq 1.2\times 10^{-3}$. In this study, we assume that
$\Ms=1\,\Msun$, $T_{s}=4280\,\mathrm{K}$, and $R_{s}=2\,R_{\odot}$
\citep{siess2000}.
The initial surface density distribution of the gas obeys the relation
$\Sigma=\Sigma^{0}_{1}(r_{1}/r)^{\beta}$, where $\beta=1/2$, $1$, or 
$3/2$, within at least $\sim 10\,\AU$. Farther away from the star, 
$\Sigma$ is exponentially tapered.  The extremes of the ``slope'' $\beta$ 
bracket values derived for the solar nebula by \citet{davis2005} and by
\citet{weidenschilling1977a} and \citet{hayashi1981}.
The quantity $\Sigma^{0}_{1}$, the surface density at $r_{1}=1\,\AU$, 
is such that the initial disk mass is 
$M^{0}_{\mathrm{D}}\simeq 0.022$, $0.044$, or $0.088\,\Ms$.
(These will be regarded as nominal values. The total initial disk mass 
differs somewhat for the different values of $\beta$ because of the
tapering procedure).
The photo-evaporation rate (Equation~\ref{eq:Spe}) is specified by 
imposing $f_{41}$ in Equation~(\ref{eq:Speg}). 
Here we use $f_{41}=0.1$, $1$, $10$, $100$, and $1000$ 
\citep{alexander2005}.
The kinematic viscosity is $\nu=\nu_{1}(r/r_{1})^{\beta}$ and
$\nu_{1}=4\times 10^{-6}$, $8\times 10^{-6}$, and
$1.6\times 10^{-5}\,r^{2}_{1}\,\Omega_{1}$, where $\Omega_{1}$ is
the rotation rate at $r=r_{1}$. As a reference, in a disk with constant
aspect ratio $H/r=0.04$, $\nu_{1}=8\times 10^{-6}\,r^{2}_{1}\,\Omega_{1}$
corresponds to a turbulence parameter \citep{S&S1973} 
$\alpha_{\mathrm{t}}=0.005$.
The initial accretion rate onto the star ranges from a few times 
$10^{-8}$ to a few times $10^{-7}\,\Msun\,\mathrm{yr}^{-1}$.
For comparison, the mass loss rate in Equation~(\ref{eq:dotMw})
is between $\sim 10^{-10}$ and $\sim 10^{-8}\,\Msun\,\mathrm{yr}^{-1}$

\subsection{Model Results}
\label{sec:MR}

The majority of disk models have an initial gas inventory of at least 
$\sim 0.02\,\Msun$ within a distance of $40\,\AU$ from the Sun,
as required by a canonical minimum mass solar nebula 
\citep[MMSN; e.g.,][]{weidenschilling1977a,hayashi1981}. This value is
also consistent with the more recent MMSN model adopted by \citet{chiang2010}.
Due to the steepness of the surface density,
disk models with the lowest initial mass and parameter $\beta=3/2$ have
only $0.01\,\Msun$ worth of gas within $40\,\AU$ of the Sun.

Gas is removed via the combined action of accretion onto the star, 
$\dot{M}$ (Equation~\ref{eq:dotM}), and photo-evaporation, 
$\dot{M}_{\mathrm{w}}$ (Equation~\ref{eq:dotMw}).
In particular, Equation~(\ref{eq:dotMw}) sets an upper limit to dispersal 
timescale, $\tau_{\mathrm{D}}$, 
ranging from $\sim 1.4\,\mathrm{Myr}$ for 
$M^{0}_{\mathrm{D}}\simeq 0.022\,\Ms$ (when $f_{41}=1000$)
to $\sim 560\,\mathrm{Myr}$ for $M^{0}_{\mathrm{D}}\simeq 0.088\,\Ms$
(when $f_{41}=0.1$). 
For computational purposes, $\tau_{\mathrm{D}}$
is defined as the time past which $M_{\mathrm{D}}\lesssim10^{-5}\,\Ms$.

\begin{figure*}[t!]
\centering%
\resizebox{\figlen}{!}{%
\includegraphics[clip,bb=50 70 400 302]{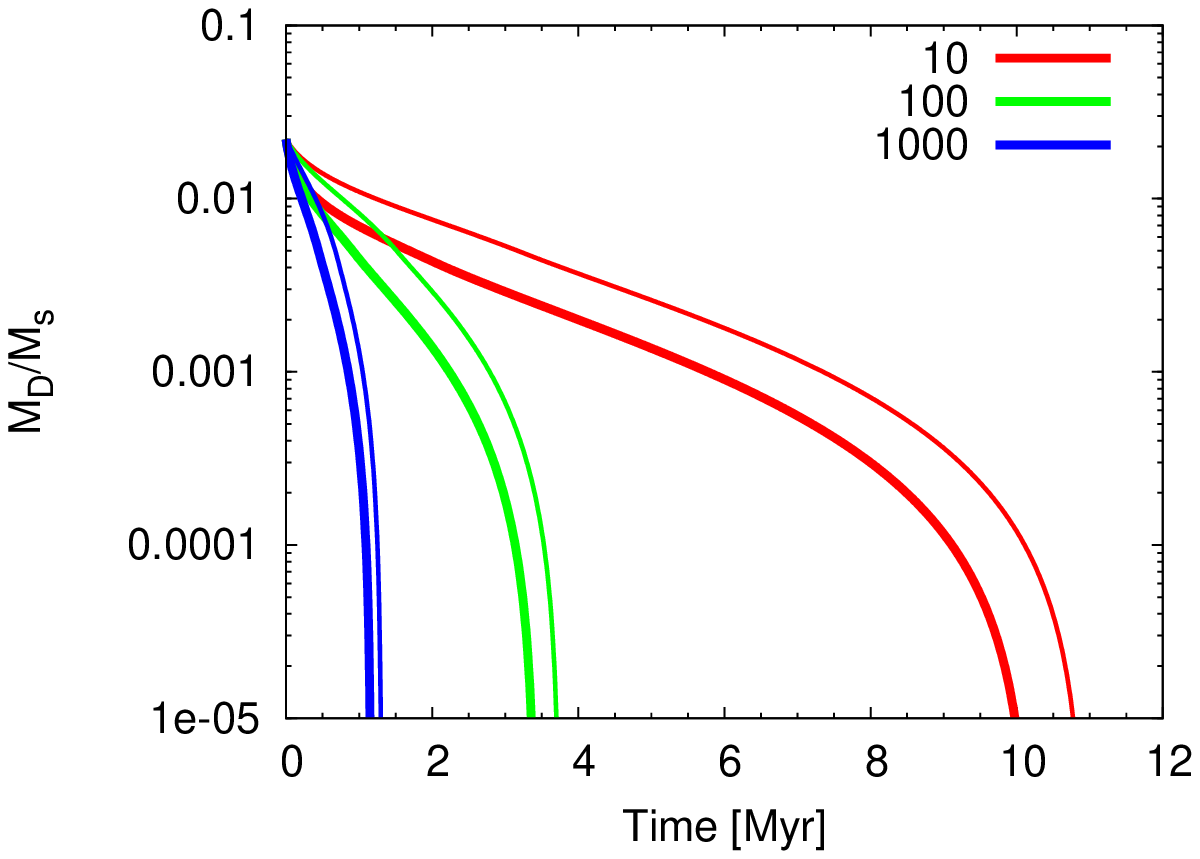}%
\includegraphics[clip,bb=50 70 400 302]{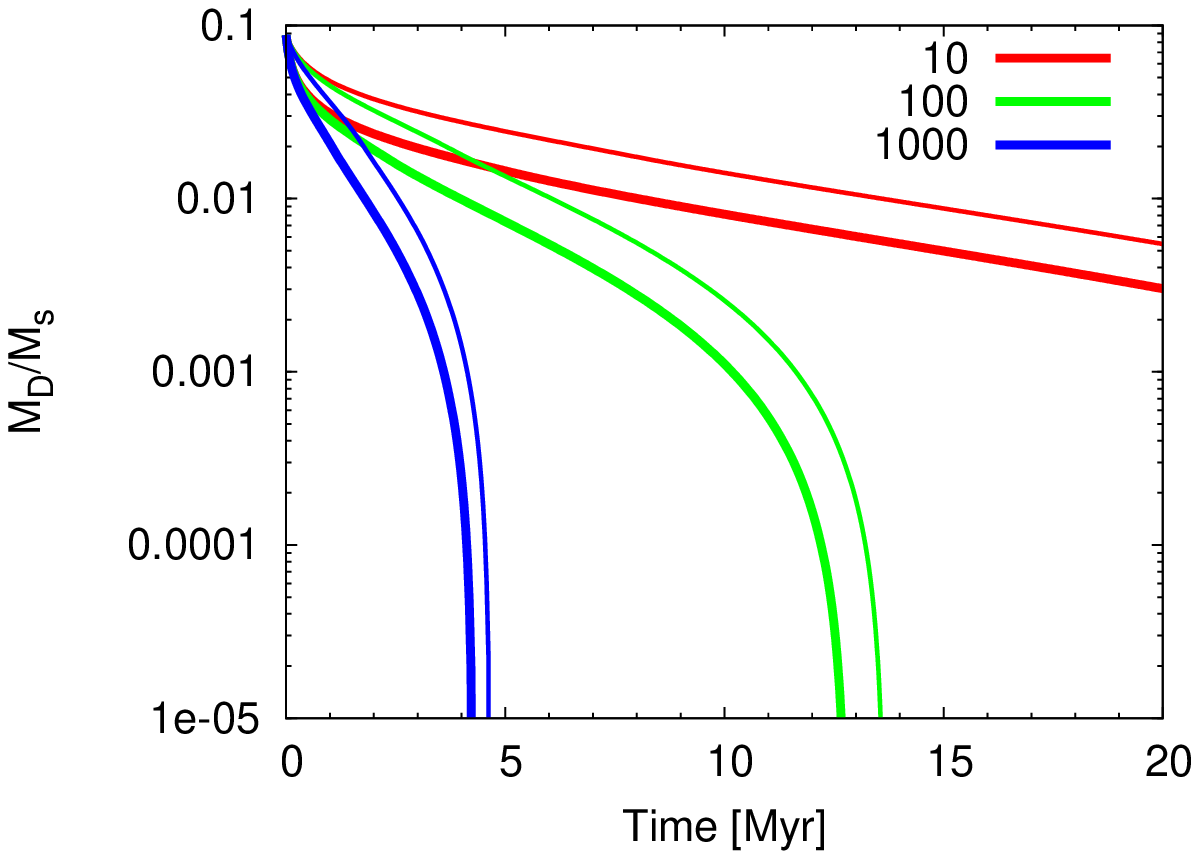}}
\resizebox{\figlen}{!}{%
\includegraphics[clip,bb=50 70 400 302]{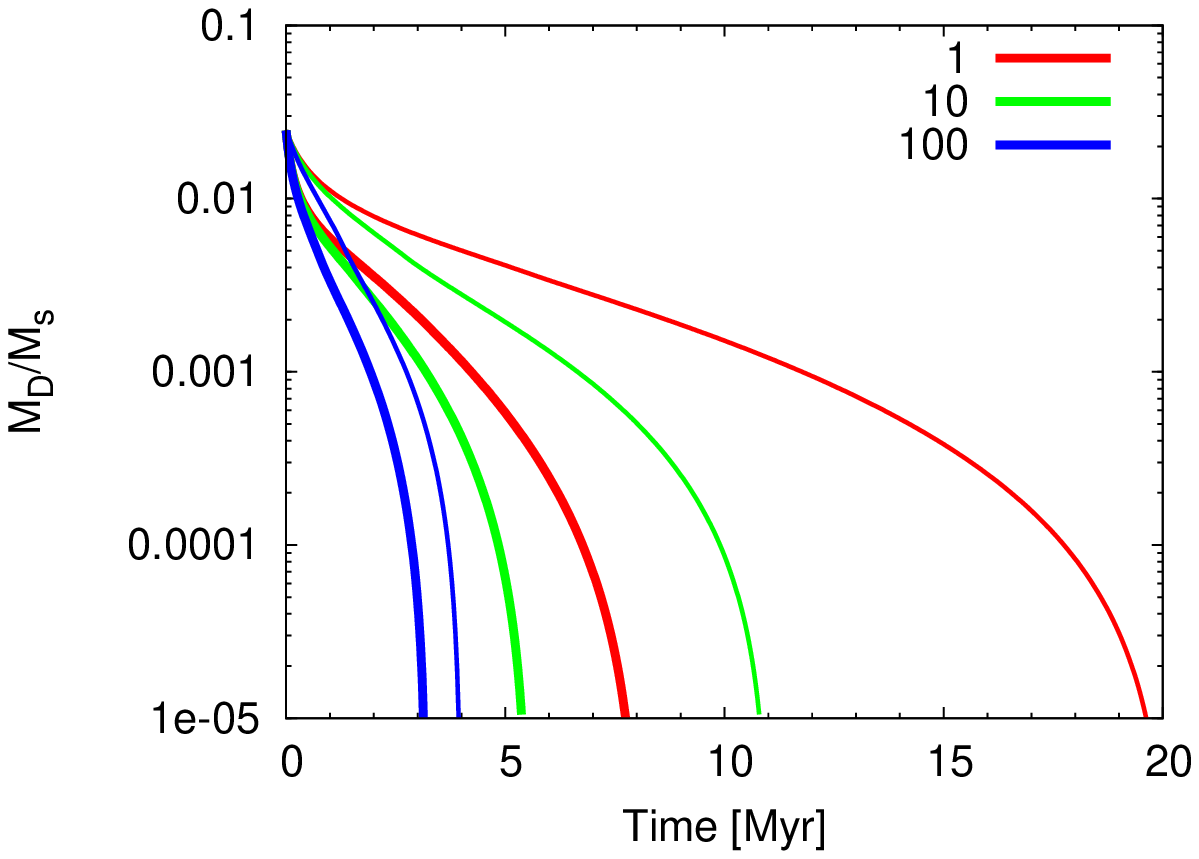}%
\includegraphics[clip,bb=50 70 400 302]{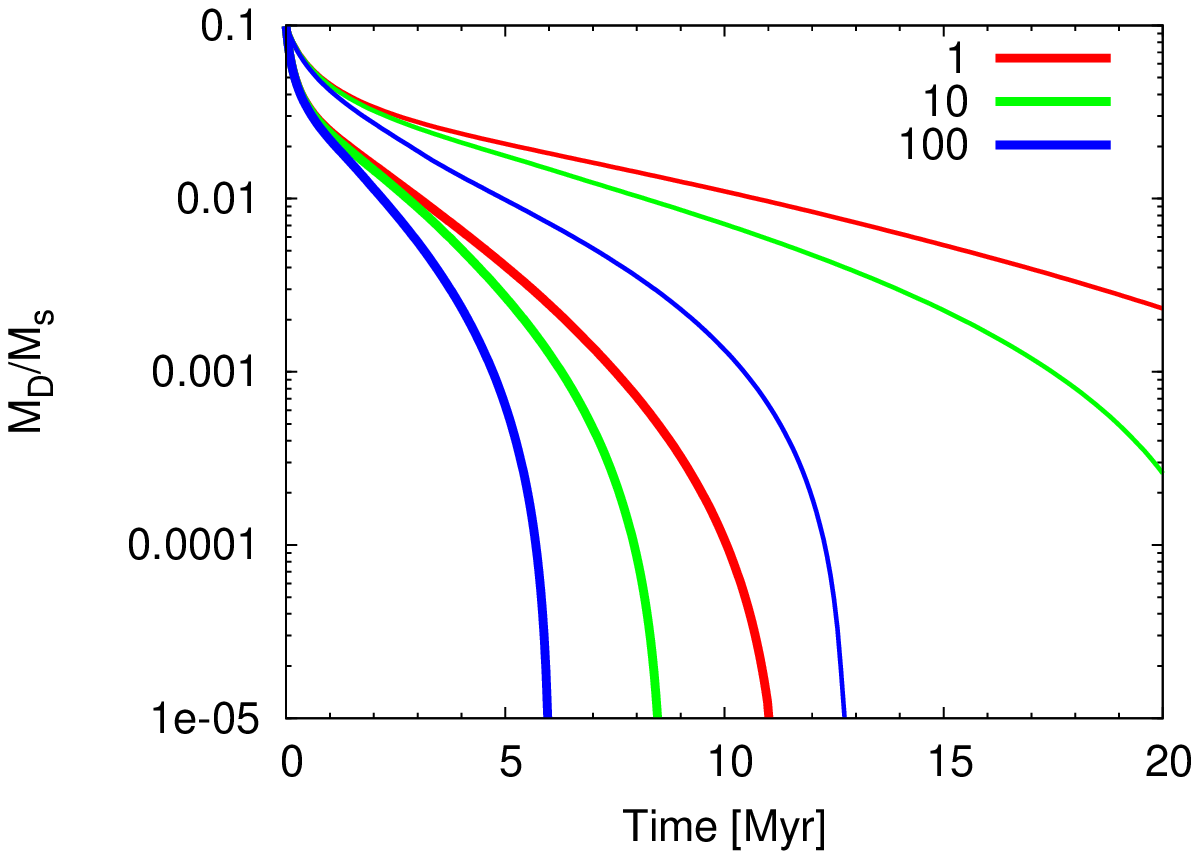}}
\resizebox{\figlen}{!}{%
\includegraphics[clip,bb=50 49 400 302]{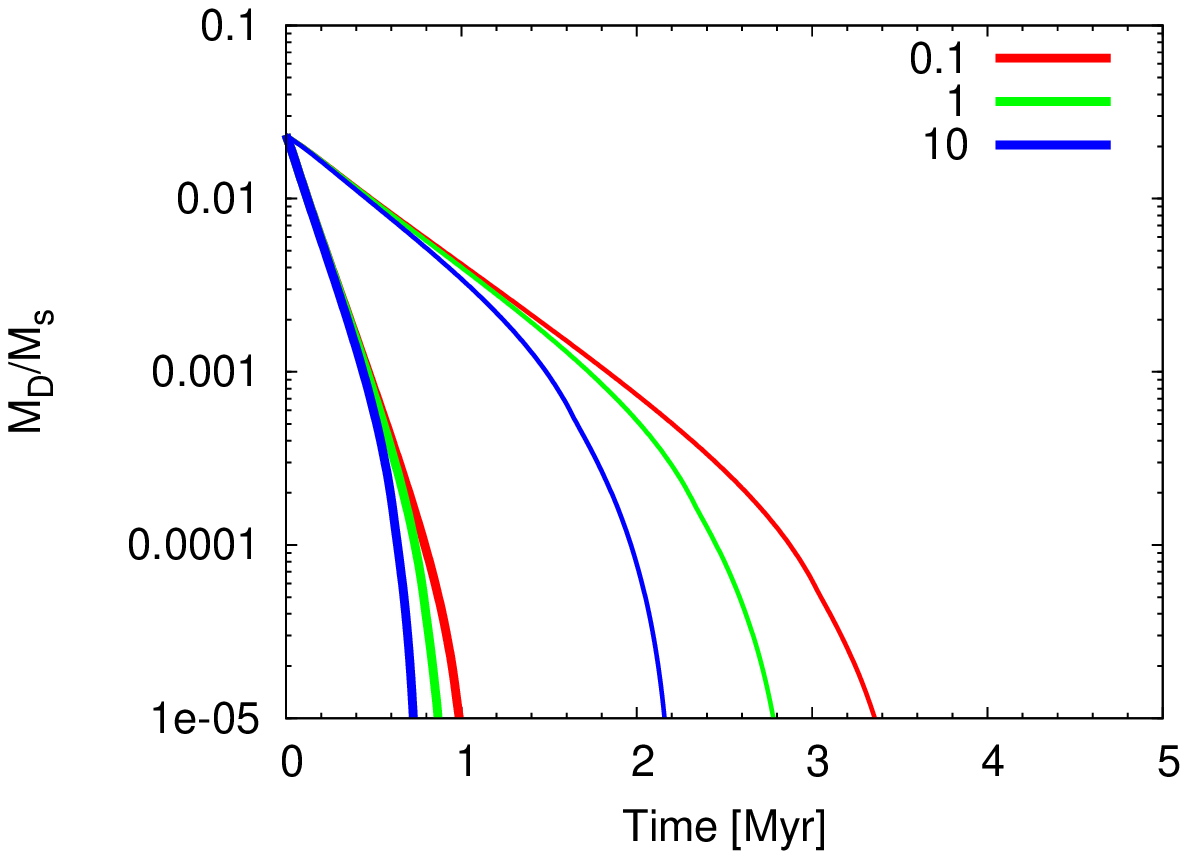}%
\includegraphics[clip,bb=50 49 400 302]{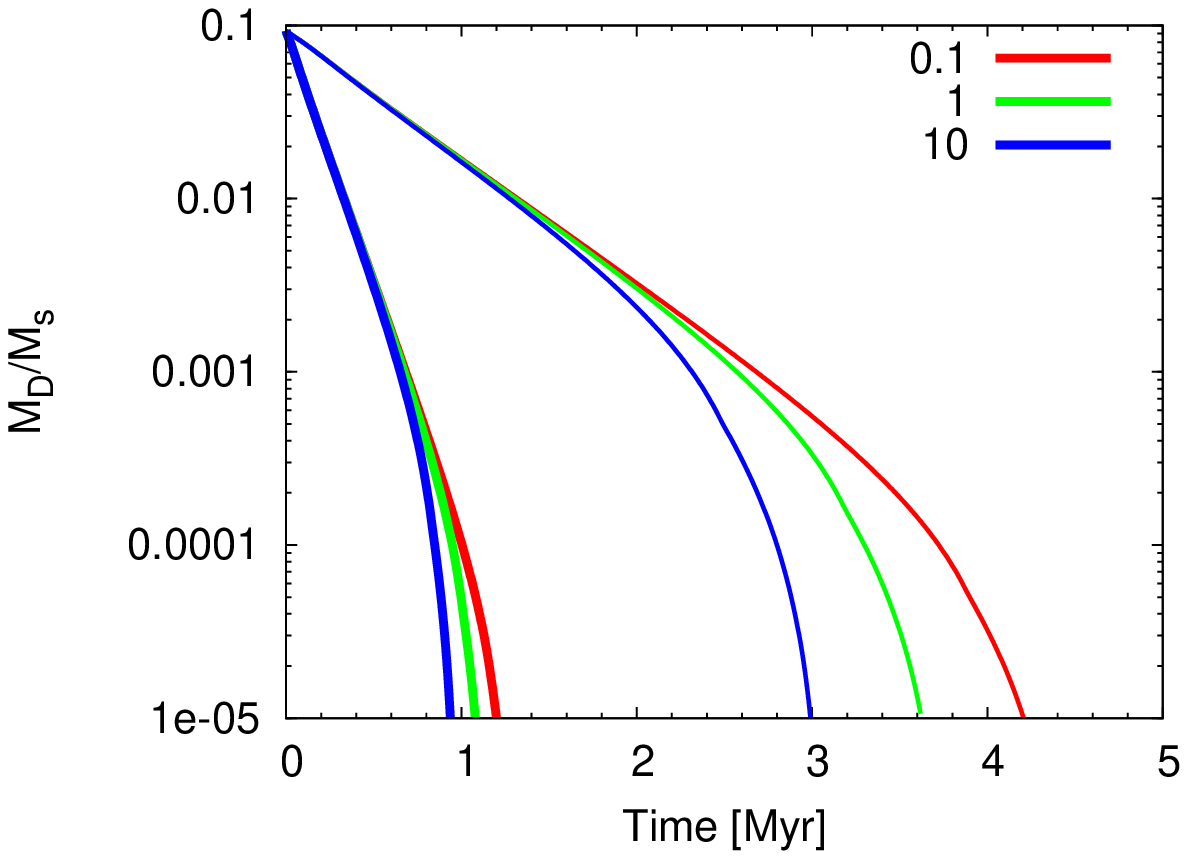}}
\caption{%
              Mass evolution for disks with initial (nominal) masses 
              $M^{0}_{\mathrm{D}}\simeq 0.022\,\Ms$ (\textit{left}) and $0.088\,\Ms$
              (\textit{right}). The initial $\Sigma$ has $\beta=1/2$ (\textit{top}), $1$ 
              (\textit{center}), and $3/2$ (\textit{bottom}).   
              Thin and thick lines represent models with viscosity 
              $\nu_{1}=4\times 10^{-6}$ and 
              $1.6\times 10^{-5}\,r^{2}_{1}\,\Omega_{1}$, respectively.
              Different line colors 
              correspond to different rates of EUV ionizing photons
              emitted by the star in units of $10^{41}\,\mathrm{s^{-1}}$, as reported in 
              the legends.
             }
\label{fig:MDvst}
\end{figure*}
The evolution of the disk mass for some selected cases is illustrated in 
Figure~\ref{fig:MDvst} for each reference viscosity (see figure's caption
for details). A complete list of the disk lifetimes is reported in Table~\ref{tbl:tD}.
The behavior of the disk mass as a function of time, for the different surface
densities,  can be qualitatively understood in terms of viscous evolution 
by means of the analytic solutions of \citet[their Section~3.3]{lynden-bell1974}: 
for equally massive disks, the more compact the disk is
(i.e., the larger $\beta$), the more rapidly $M_{\mathrm{D}}$ reduces initially.
By a somewhat conservative assumption, as discussed above, 
disks that survive beyond $20\,\mathrm{Myr}$ 
are discarded and will not be given any further consideration.
This is the case, for example, for \emph{all} models with a photo-ionizing rate
characterized by  $f_{41}\le 1$ and the flattest initial surface density
($\beta=1/2$). Models of disks surviving less than $1\,\mathrm{Myr}$  will
also be discarded based on considerations on planet formation timescales,
as explained in Section~\ref{sec:LTM}.
\begin{deluxetable*}{crrrrrrrrrrrrr}
\tablecolumns{13}
\tablewidth{0pc}
\tablecaption{Lifetimes from Disk Models\label{tbl:tD}}
\tablehead{
\colhead{}    & \colhead{}    &
\multicolumn{11}{c}{$\tau_{\mathrm{D}}$\tablenotemark{a}} \\
\cline{3-13}
\colhead{}    & \colhead{}    &
\multicolumn{3}{c}{$\beta\tablenotemark{b}=1/2$} 
& \colhead{}    & 
\multicolumn{3}{c}{$\beta=1$}
& \colhead{}    & 
\multicolumn{3}{c}{$\beta=3/2$}\\ 
\cline{3-5} \cline{7-9} \cline{11-13}
\colhead{$M^{0}_{\mathrm{D}}/\Ms$}   & \colhead{$\nu_{1}$\tablenotemark{c}}&
\colhead{$f_{41}\tablenotemark{d}=10$} & \colhead{$100$} & \colhead{$1000$} &
\colhead{} & \colhead{$1$} & \colhead{$10$} & \colhead{$100$} &
\colhead{} & \colhead{$0.1$} & \colhead{$1$} & \colhead{$10$}
}
\startdata
$0.022$ & $4\times 10^{-6}$      & $10.8$ & $3.70$ & $1.31$  
                                               &  & $19.6$ & $10.8$ & $3.92$   
                                               &  & $3.35$ & $2.78$ & $2.21$ \\ 
$0.022$ & $8\times 10^{-6}$      & $10.3$ & $3.50$ & $1.24$
                                               &  & $12.6$ & $8.01$ & $3.75$
                                               &  & $1.84$ & $1.57$ & $1.25$ \\
$0.022$ & $1.6\times 10^{-5}$   & $9.95$ & $3.35$ & $1.16$
                                               &  & $7.75$ & $5.38$ & $3.12$
                                               &  & $0.98$ & $0.87$ & $0.74$ \\
\cline{2-13}
$0.044$ & $4\times 10^{-6}$      & $20.8$ & $7.05$ & $2.46$
                                               &  & $25.6$ & $16.2$ & $7.55$
                                               &  & $3.78$ & $3.20$ & $2.57$ \\
$0.044$ & $8\times 10^{-6}$     & $20.1$ & $6.72$ & $2.32$ 
                                              &  & $15.7$ & $10.9$ & $6.31$
                                              &  & $2.05$ & $1.78$ & $1.48$ \\
$0.044$ & $1.6\times 10^{-5}$  & $19.5$ & $6.47$ & $2.22$
                                              &  & $9.36$ & $6.90$ & $4.50$
                                              &  & $1.09$ & $0.98$ & $0.83$ \\
\cline{2-13}
$0.088$ & $4\times 10^{-6}$     & $40.5$ & $13.5$ & $4.66$
                                              &  & $31.8$ & $21.9$ & $12.7$
                                              &  & $4.21$ & $3.63$ & $3.00$ \\
$0.088$ & $8\times 10^{-6}$     & $39.3$ & $13.1$ & $4.43$
                                              &  &$19.0$ & $13.9$ & $9.07$
                                              &  & $2.26$ & $1.99$ & $1.69$ \\
$0.088$ & $1.6\times 10^{-5}$  & $38.5$ & $12.6$ & $4.25$
                                              &  & $11.0$ & $8.47$ & $5.97$
                                              &  & $1.20$ & $1.08$ & $0.93$
\enddata
\tablenotetext{a}{Time past which $M_{\mathrm{D}}\lesssim10^{-5}\,\Ms$, in units of Myr.}
\tablenotetext{b}{Initial ``slope'' of the disk's surface density.}
\tablenotetext{c}{Kinematic viscosity at $r_{1}=1\,\AU$ in units of 
$r^{2}_{1}\,\Omega_{1}=(G\,\Ms\,r_{1})^{1/2}$.}
\tablenotetext{d}{Rate of EUV ionizing photons emitted by the star in units of 
$10^{41}\,\mathrm{s^{-1}}$ (see Equation~\ref{eq:Speg}).}
\end{deluxetable*}

\begin{figure*}[t!]
\centering%
\resizebox{\figlen}{!}{%
\includegraphics[clip,bb=53 70 405 302]{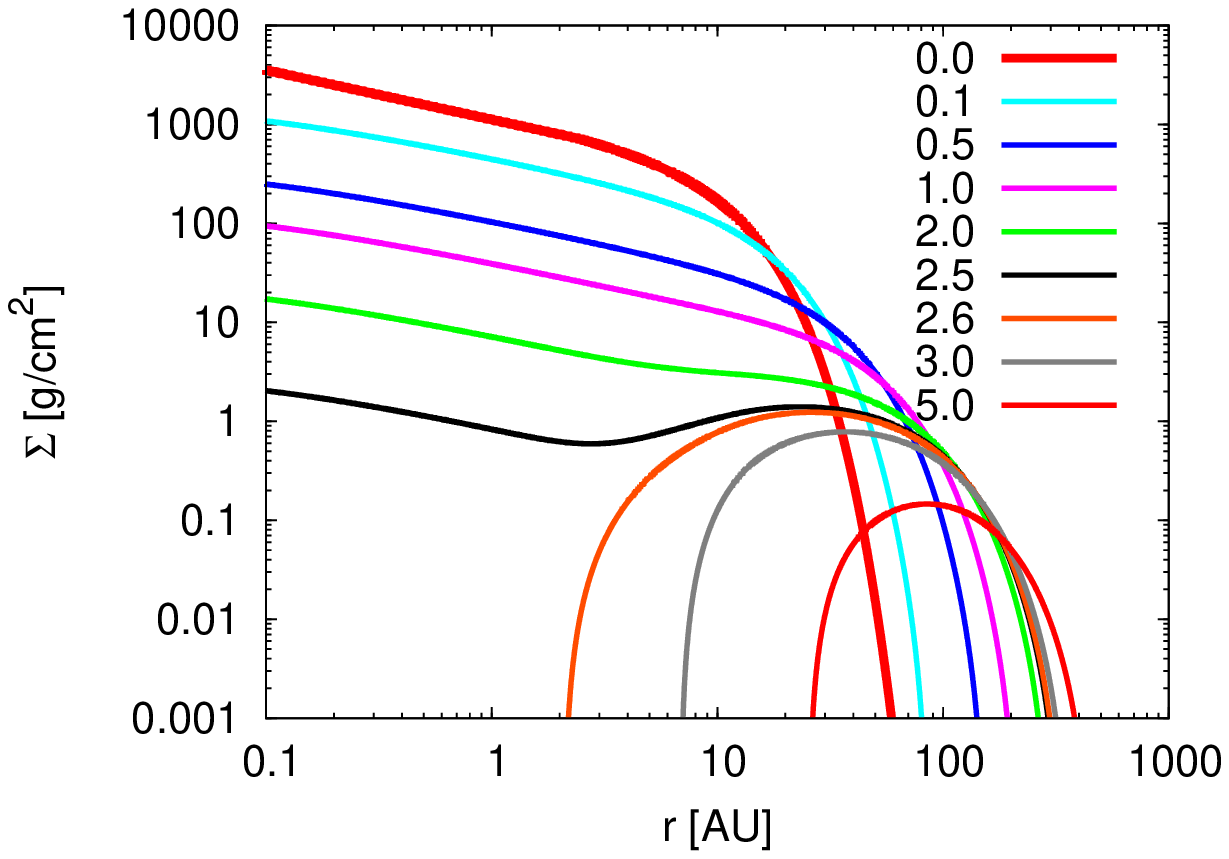}%
\includegraphics[clip,bb=53 70 405 302]{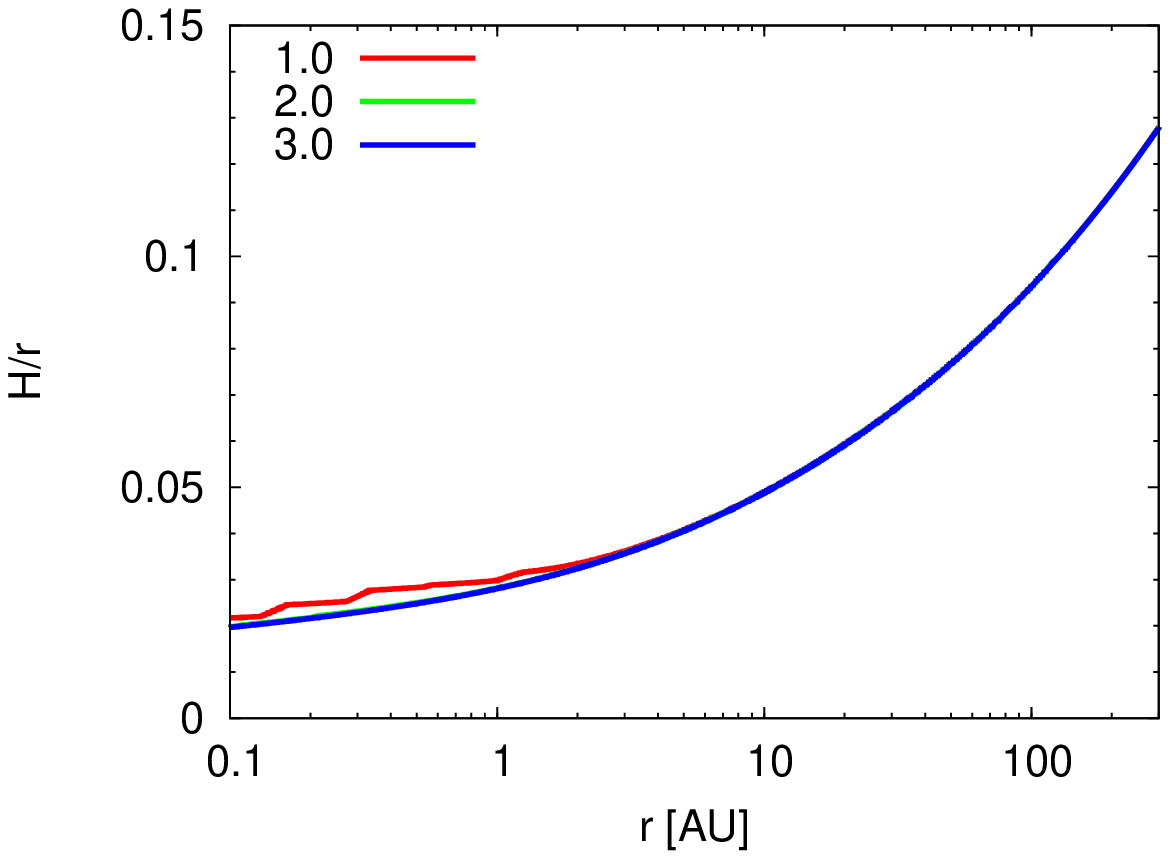}}
\resizebox{\figlen}{!}{%
\includegraphics[clip,bb=53 70 405 302]{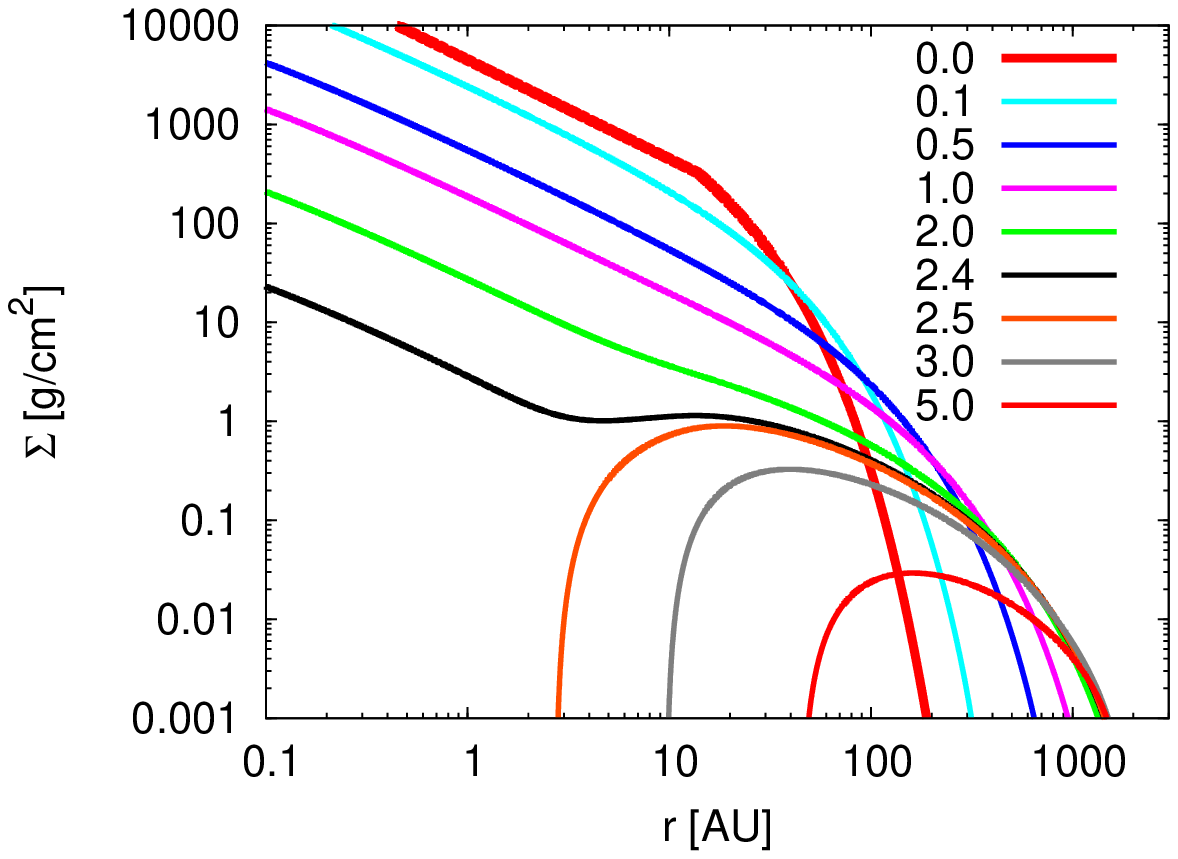}%
\includegraphics[clip,bb=53 70 405 302]{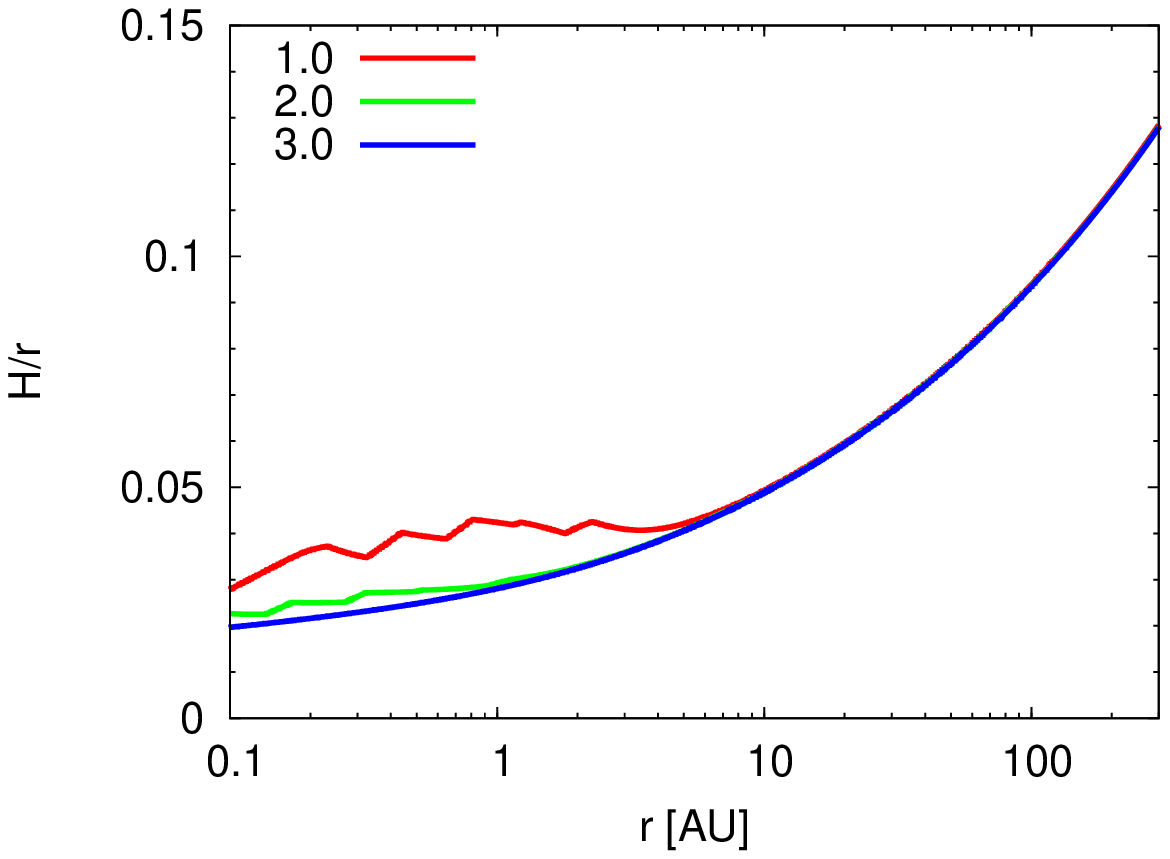}}
\resizebox{\figlen}{!}{%
\includegraphics[clip,bb=53 49 405 302]{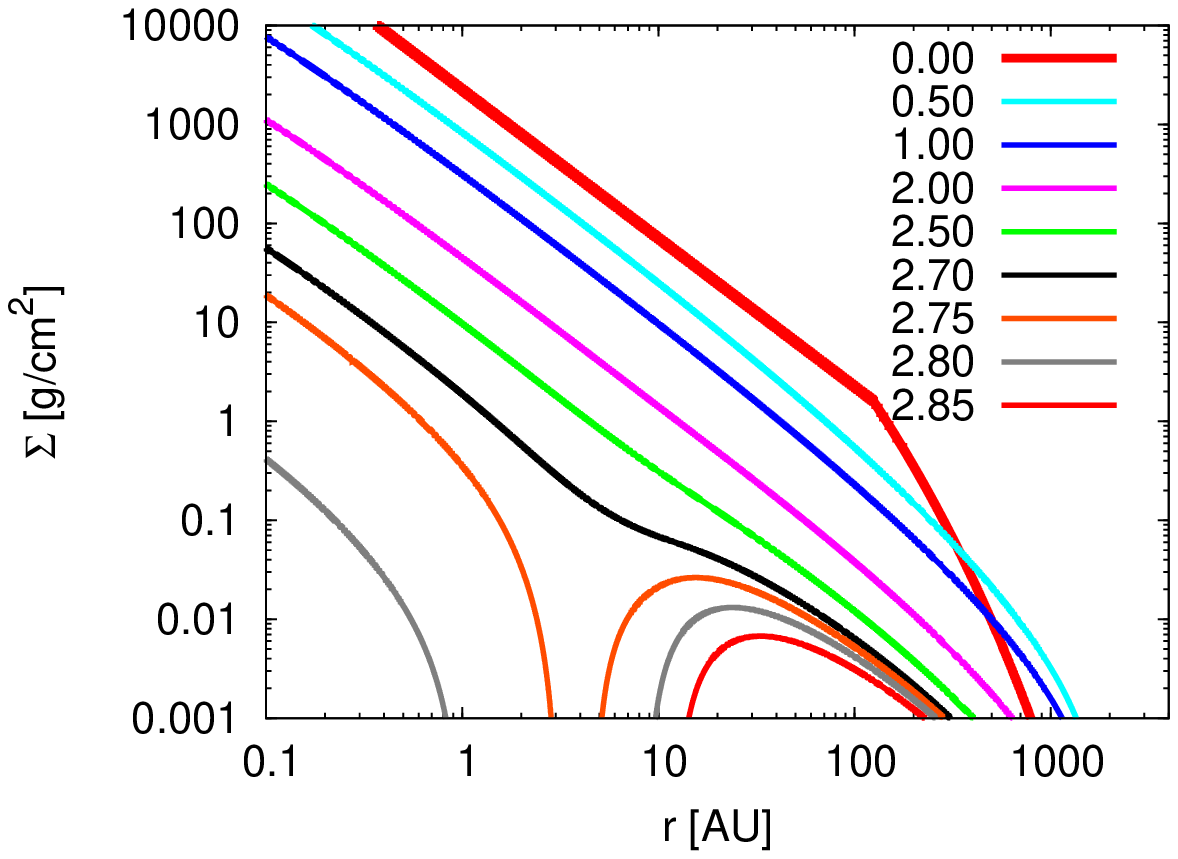}%
\includegraphics[clip,bb=53 49 405 302]{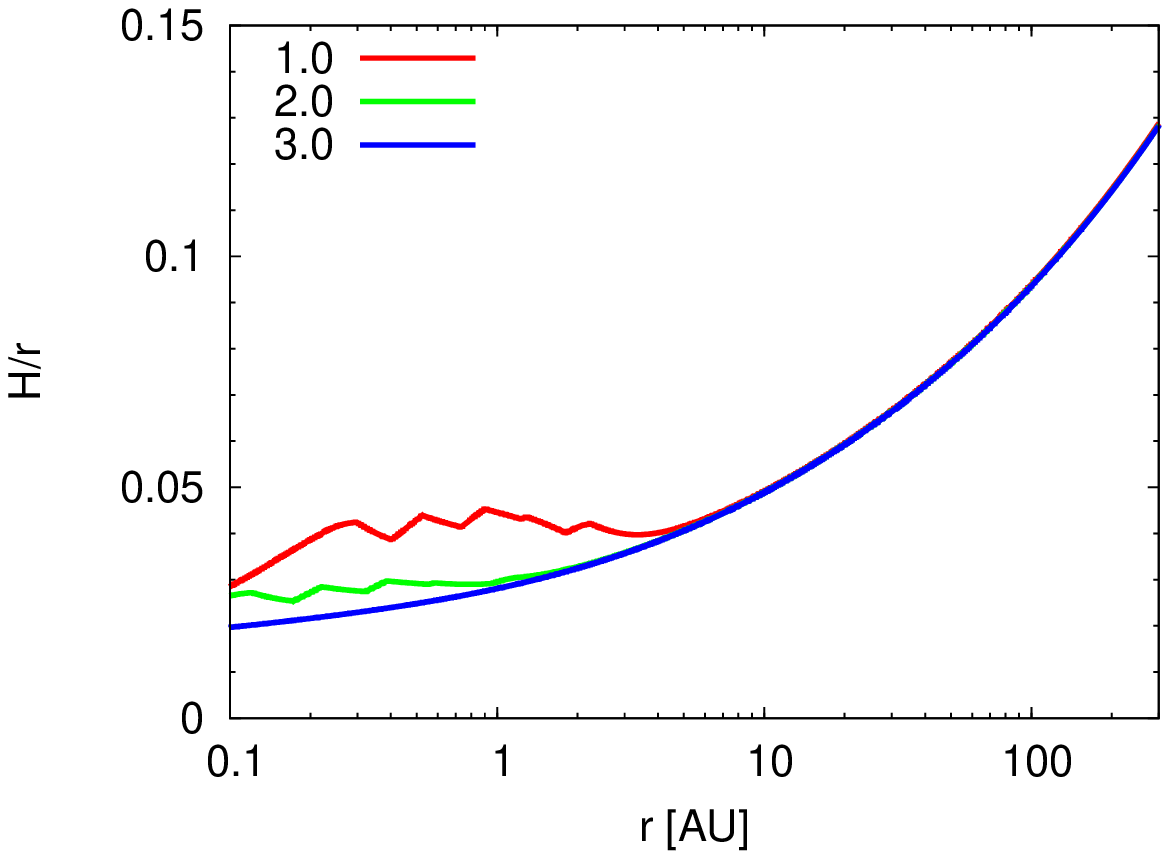}}
\caption{%
              Evolution of the disk's surface density (\textit{left})
              and disk's relative thickness (\textit{right})
              of disk models with
              $\nu_{1}=8\times 10^{-6}\,r^{2}_{1}\,\Omega_{1}$,
              $M^{0}_{\mathrm{D}}\simeq 0.022\,\Ms$, 
              $f_{41}=10$, and $\beta=1/2$, (\textit{top}), 
              $\nu_{1}=8\times 10^{-6}\,r^{2}_{1}\,\Omega_{1}$,
              $M^{0}_{\mathrm{D}}\simeq 0.088\,\Ms$, 
              $f_{41}=100$, and $\beta=1$ (\textit{center}), and
              $\nu_{1}=4\times 10^{-6}\,r^{2}_{1}\,\Omega_{1}$,
              $M^{0}_{\mathrm{D}}\simeq 0.044\,\Ms$, 
              $f_{41}=1$, and $\beta=3/2$ (\textit{bottom}).
              Times indicated in the legend are in Myr. 
             }
\label{fig:Svst}
\end{figure*}
A quantity of primary importance for planetary migration is the average 
surface density around the planet's orbit. 
In Figure~\ref{fig:Svst} (\textit{left panels}), the evolution of $\Sigma$ 
is shown for cases with different values of parameters $\beta$ and $f_{41}$. 
As anticipated at the end of Section~\ref{sec:DD}, once the accretion rate
drops below some threshold, photo-evaporation produces a gap in the surface
density at a radial distance of a few \AU. 
Then the disk inside the gap is removed by 
viscous diffusion on a timescale of order $r^{2}/(2\pi\nu)$ orbital periods.
We shall see in the Sections~\ref{sec:ToC} and \ref{sec:OMR}
that the disk's aspect ratio has also
a large impact on the rates and direction of migration. In the right panels of 
Figure~\ref{fig:Svst}, $H/r$ is plotted at reference times for the same
models as in the left panels. Once $\Sigma$ becomes small enough
and the viscous heating term $Q_{\nu}$ in Equation~(\ref{eq:EEq})
becomes unimportant, the disk temperature, and hence $H/r$
(see Section~\ref{sec:DT}), is dictated only by the stellar irradiation 
temperature (Equation~\ref{eq:Tirr}).

\begin{figure*}[t!]
\centering%
\resizebox{\figlen}{!}{%
\includegraphics[clip,bb=50 70 400 302]{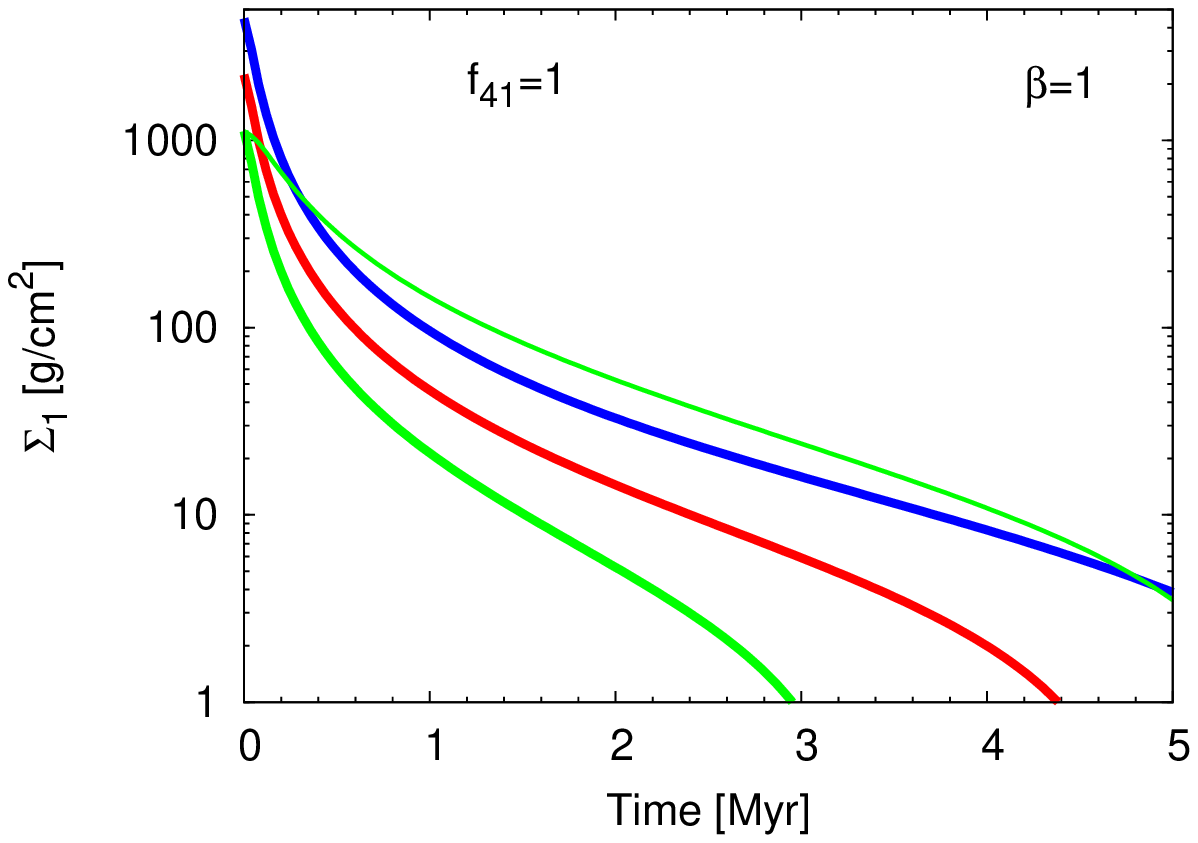}%
\includegraphics[clip,bb=50 70 400 302]{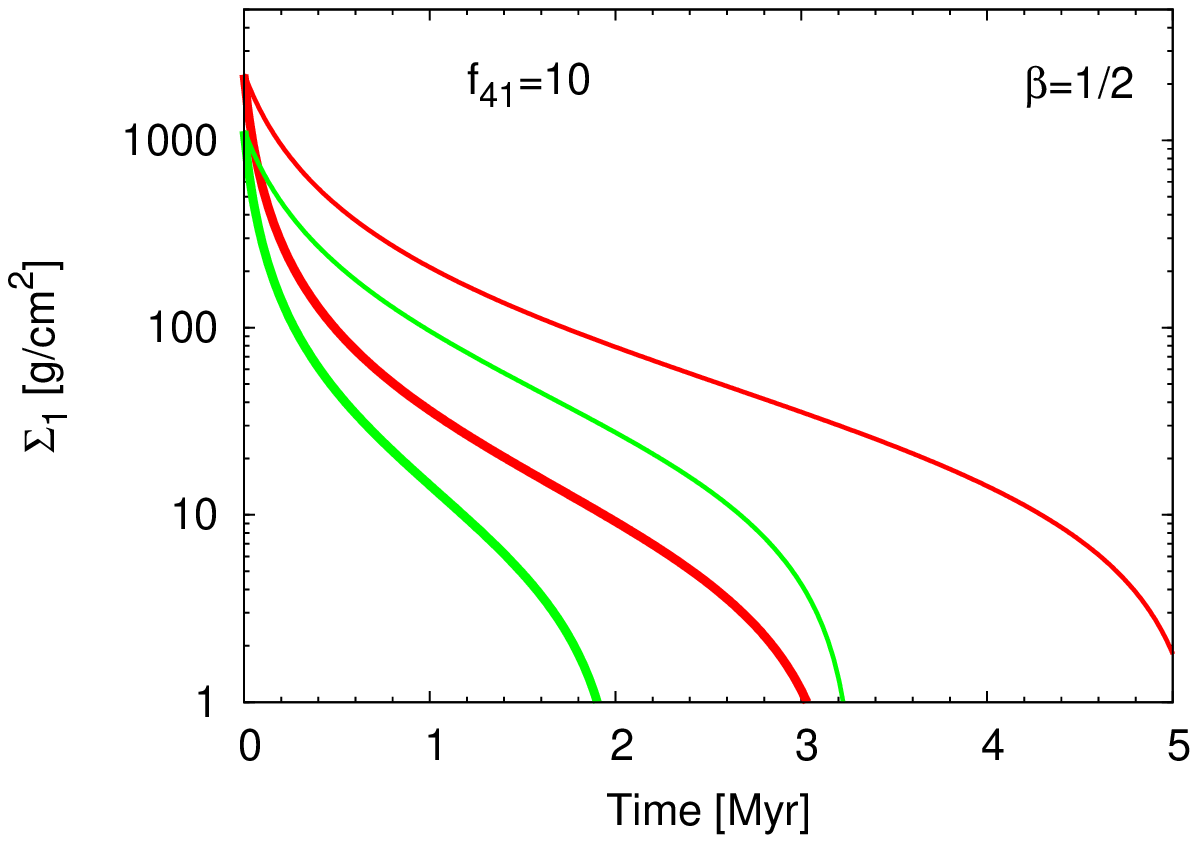}}
\resizebox{\figlen}{!}{%
\includegraphics[clip,bb=50 70 400 302]{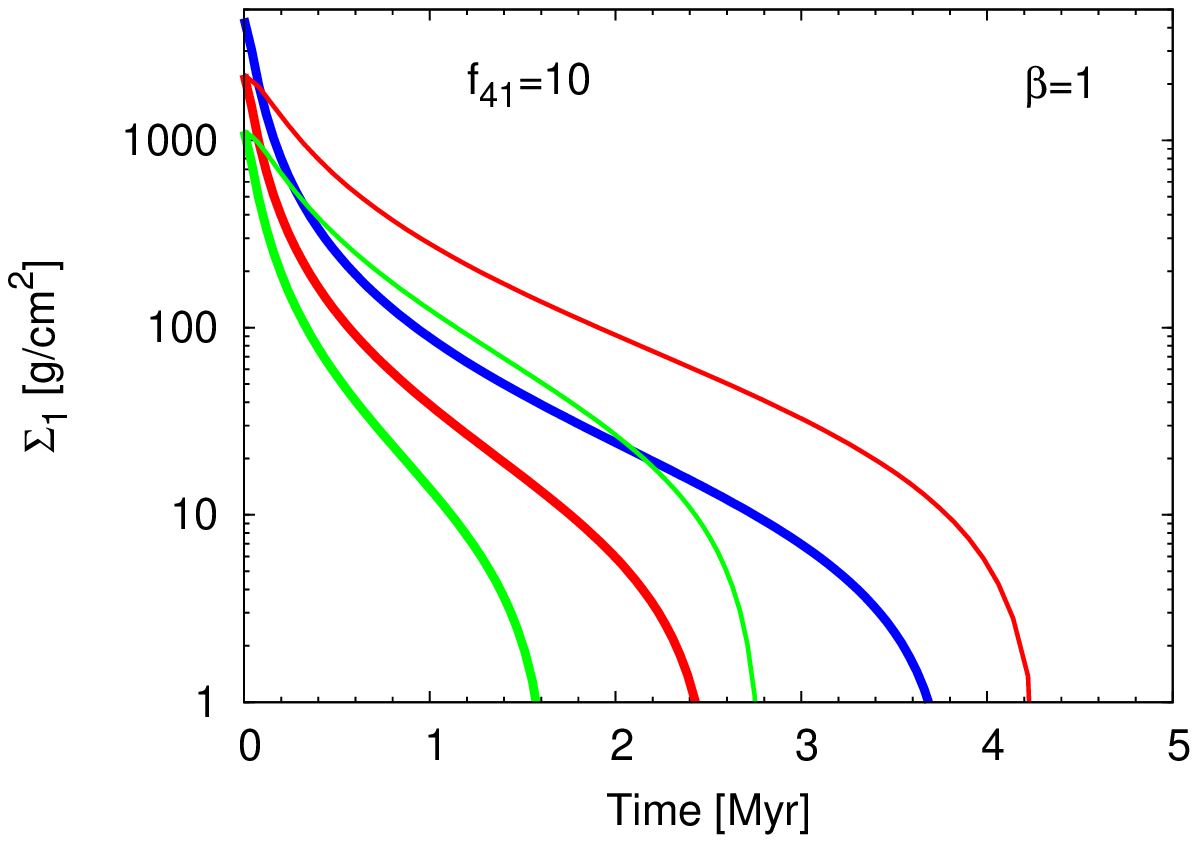}%
\includegraphics[clip,bb=50 70 400 302]{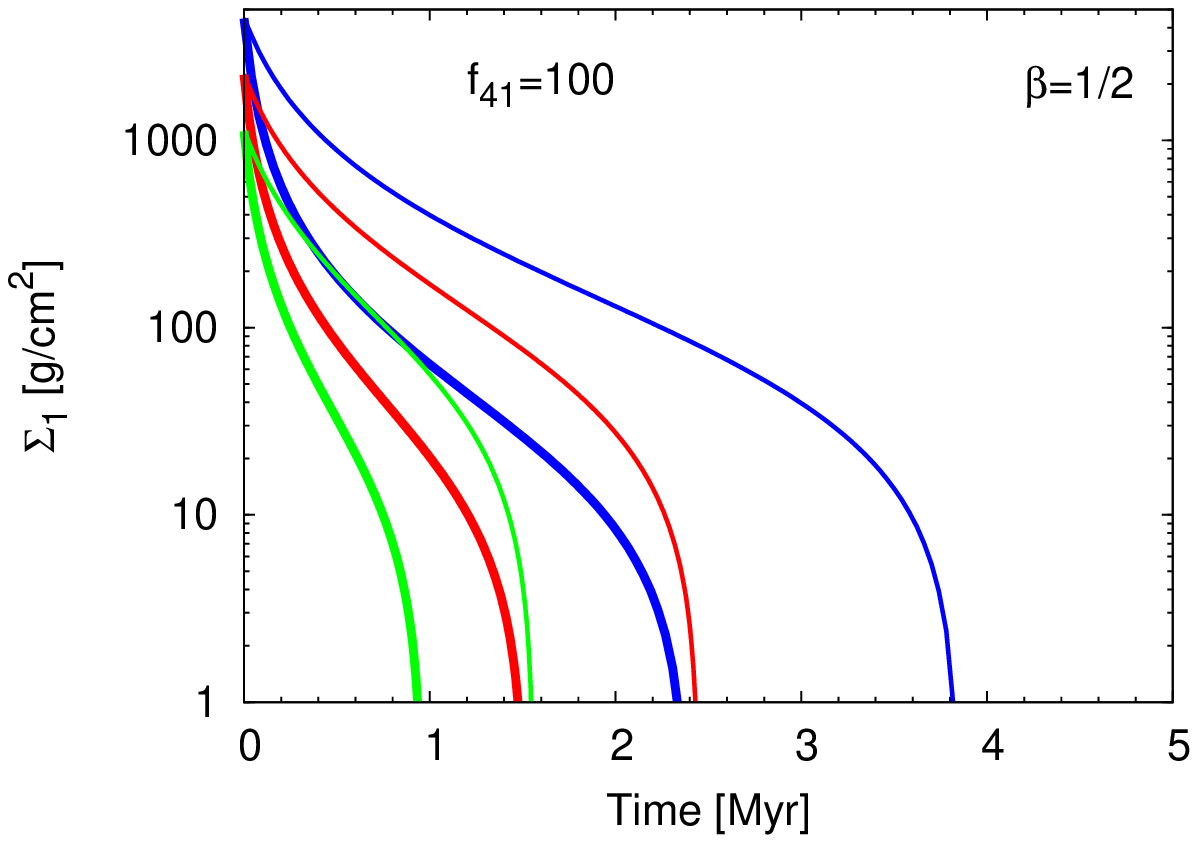}}
\resizebox{\figlen}{!}{%
\includegraphics[clip,bb=50 49 400 302]{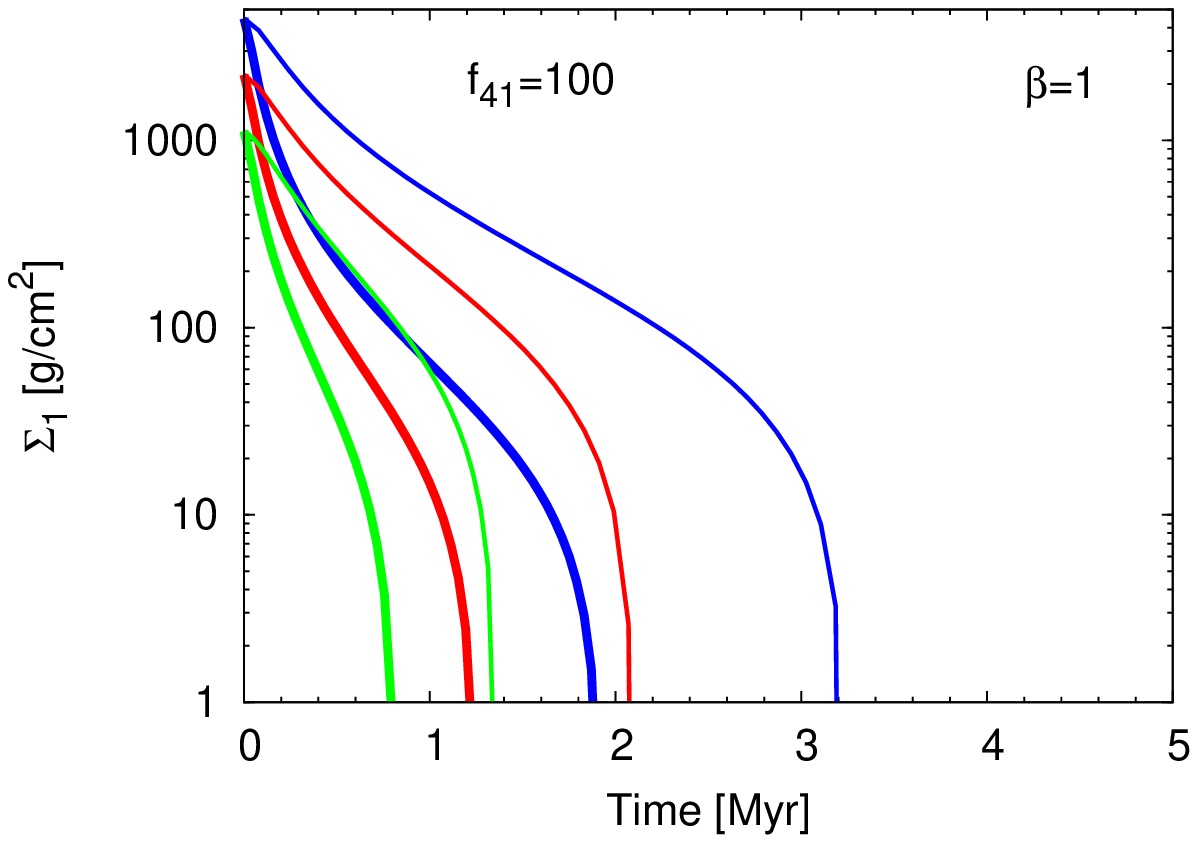}%
\includegraphics[clip,bb=50 49 400 302]{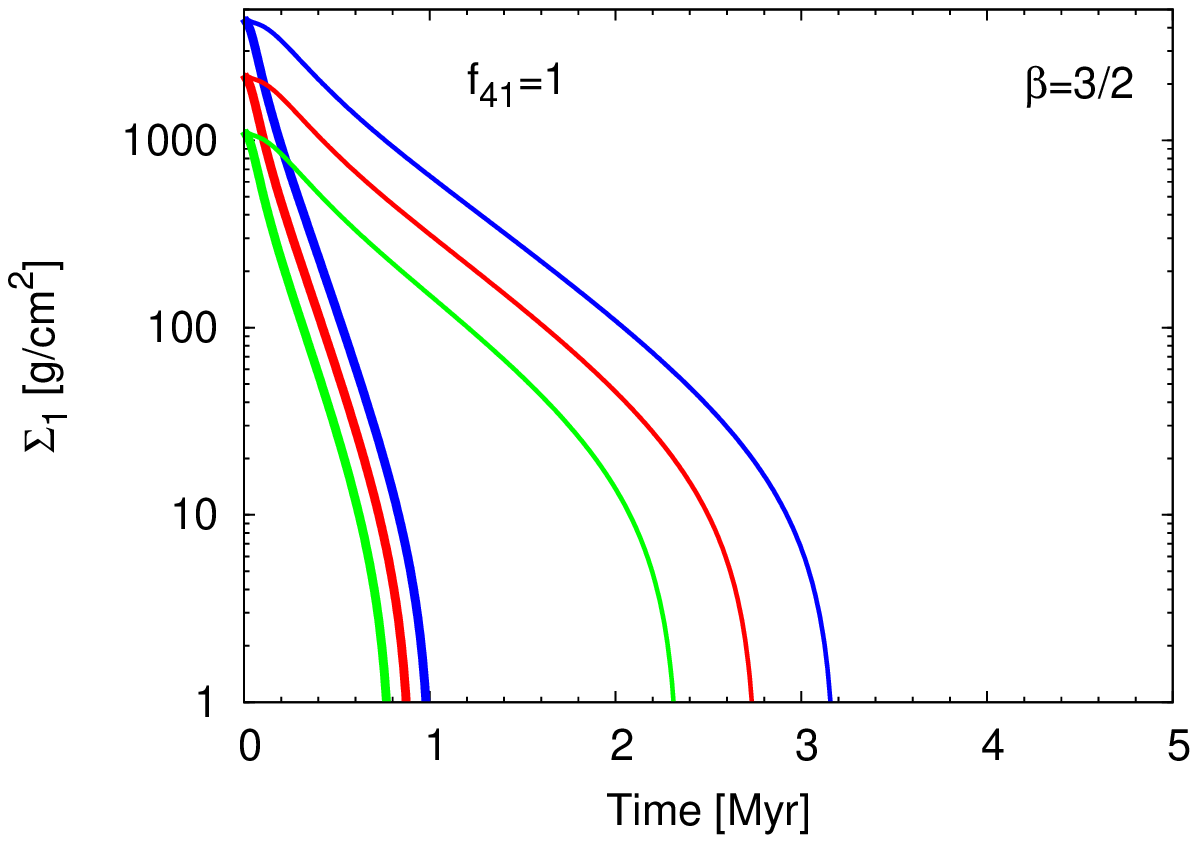}}
\caption{%
              Evolution of the disk's surface density at $1\,\AU$.   
              Thin and thick lines represent models with viscosity 
              $\nu_{1}=4\times 10^{-6}$ and 
              $1.6\times 10^{-5}\,r^{2}_{1}\,\Omega_{1}$, respectively.
              Different line colors
              correspond to different disk's initial masses.
              The value of the EUV ionizing photon rate, $f_{41}$, and of the
              initial surface density gradient, $\beta$, are given in each panel.
             }
\label{fig:S1}
\end{figure*}
\begin{figure*}[t!]
\centering%
\resizebox{\figlen}{!}{%
\includegraphics{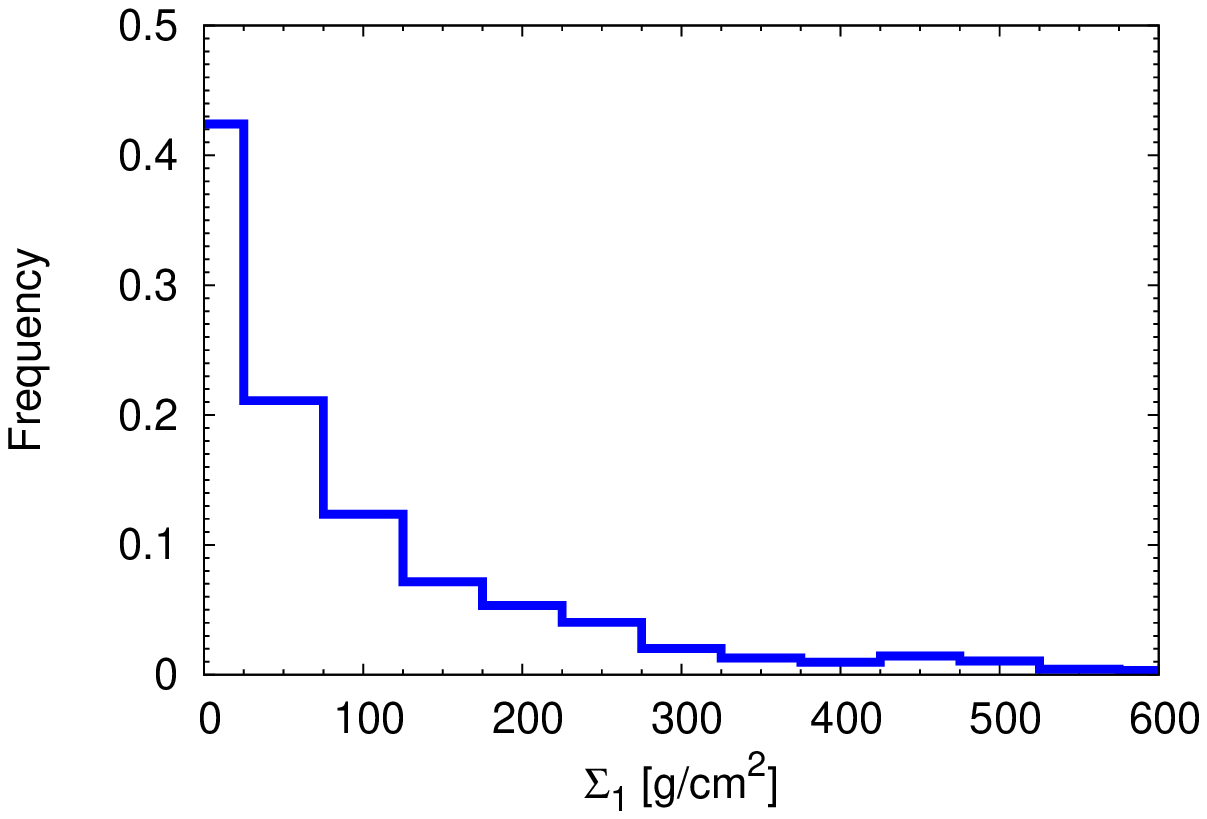}%
\includegraphics{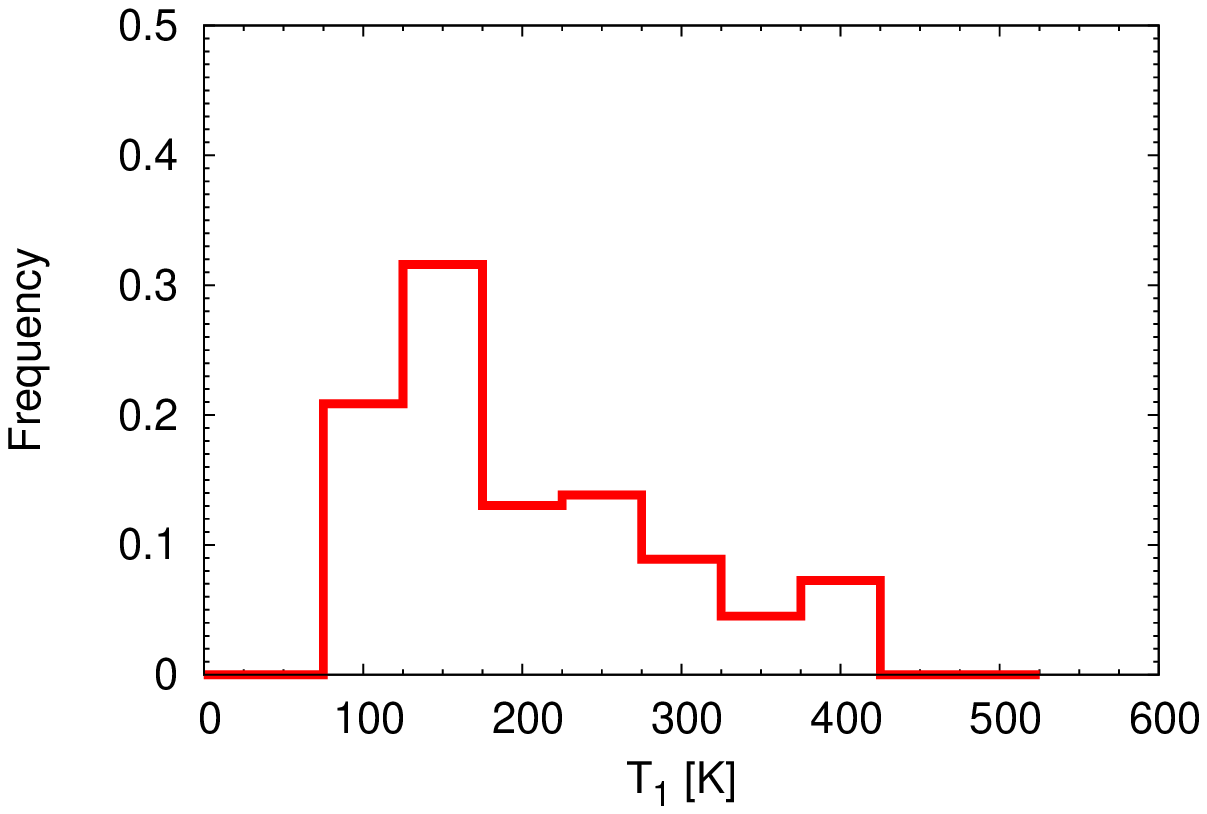}}
\caption{%
            Histograms of the surface density (\textit{left}) and mid-plane 
            temperature (\textit{right}) at $1\,\AU$, at time
            $t\approx 1\,\mathrm{Myr}$ for disk models whose lifetime 
            $\tau_{\mathrm{D}}$ is longer than $1$ and shorter than about 
            $20\,\mathrm{Myr}$. The histograms are based on over $2000$ 
            models using a random selection of $\nu_{1}$ and $f_{41}$, 
            for each pair $(M^{0}_{\mathrm{D}},\beta)$ listed in Table~\ref{tbl:tD}, 
            in the respective ranges indicated in the table.
             }
\label{fig:S1_hist}
\end{figure*}
The surface density at $1\,\AU$, $\Sigma_{1}$, vs.\  time is illustrated 
in Figure~\ref{fig:S1} for selected models from Table~\ref{tbl:tD}. 
Thin and thick lines refer to
smallest and largest values of $\nu_{1}$, respectively.  
To obtain a better statistical characterization of the disk density and 
temperature, 
for each pair of parameters $(M^{0}_{\mathrm{D}},\beta)$ listed in 
Table~\ref{tbl:tD},
$\nu_{1}$ and $f_{41}$ are varied randomly in the corresponding ranges
indicated in the table, for a total of more than $2000$ realizations. 
The histogram in Figure~\ref{fig:S1_hist} (\textit{left}) shows that
after $\sim 1\,\mathrm{Myr}$ the value of $\Sigma_{1}$ is 
$150\,\mathrm{g\,cm}^{-2}$, or less, in $\sim 80$\% of the models 
that may represent the solar nebula. 
The right panel of Figure~\ref{fig:S1_hist} illustrates
the occurrence frequency of the mid-plane temperature at $1\,\AU$. 
As a reference, the equilibrium temperature established by stellar
irradiation alone, i.e., Equation~(\ref{eq:Tsemp}), is $T_{1}\approx 100\,\mathrm{K}$.

\section{Tidal Interactions of Jupiter and Saturn with the Disk}
\label{sec:TI}

The evolution of the thermodynamical quantities (principally $\Sigma$, 
$T$, and $H/r$) of solar nebula models, obtained in the previous section,
can be used to evaluate the range of orbital migration of a pair of planets, 
over the disk lifetime, once appropriate migration rates are supplied. 
In this section we derive such rates.

\subsection{2D and 3D Hydrodynamical Calculations}
\label{sec:HC}

The migration of a Jupiter--Saturn pair in a gaseous disk is 
evaluated by using a combination of 2D and 3D hydrodynamical
calculations of tidal interactions between the planets and the disk.
We adopt a reference frame $\{O; r,\theta,\phi\}$ with
origin, $O$, fixed on the star, radius ranging from $0.25$ to $7\,\AU$, 
and azimuth varying between $0$ and $2\pi$. 
In the 2D disk approximation,  
the co-latitude angle $\theta$ is equal to $\pi/2$, whereas 
it varies from $\theta_{\mathrm{min}}$ to $\pi/2$ in a 3D disk.
In the latter case, the disk opening angle,  $\theta_{\mathrm{min}}$,
is such that the disk's vertical extent locally comprises at least three 
pressure scale-heights, $H$.
Mirror symmetry with respect to the $\theta=\pi/2$ plane is imposed
on account of the planets orbiting in this plane of symmetry.
The surface density $\Sigma$ has initially a dimensionless gradient 
$d\ln{\Sigma}/d\ln{r}=-1/2$.
We work in the assumption that the disk is locally isothermal, i.e.,
the temperature depends only on $r$, and that $H/r$ is a constant.
The gas pressure is therefore proportional either to $\Sigma/r$ (2D)
or to $\rho/r$ (3D), where $\rho$ is the mass density. 
It is further assumed that the kinematic viscosity of the disk, $\nu$, is 
constant throughout the disk.

The coordinate system rotates about the axis perpendicular to the planets'
orbital plane ($\theta=0$ axis) at a variable rotation speed, 
$\mathbf{\Omega}_{f}=\mathbf{\Omega}_{f}(t)$. 
Both $\mathbf{\Omega}_{f}$ and $\mathbf{\dot{\Omega}}_{f}$ are imposed
by the requirement that the (relative) azimuthal position of the interior planet, 
$\phi_{1}$,
remains constant in time and the (relative) angular velocity, $\dot{\phi}_{1}$,
is zero \citep[for details about the procedure, see][]{gennaro2005}.

Naming $\mathbf{r}_{\mathrm{1}}$ and $\mathbf{r}_{\mathrm{2}}$ 
the vector positions of the interior (Jupiter) and exterior planet (Saturn), 
respectively, the gravitational potential in the disk is
\begin{eqnarray}
\Phi &=&
-\frac{G\Ms}{r}
-\frac{G M_{\mathrm{1}}}{\sqrt{|\mathbf{r}-\mathbf{r}_{\mathrm{1}}|^{2}
+\varepsilon^{2}_{\mathrm{1}}}}
-\frac{G M_{\mathrm{2}}}{\sqrt{|\mathbf{r}-\mathbf{r}_{\mathrm{2}}|^{2}
+\varepsilon^{2}_{\mathrm{2}}}}
\nonumber\\%
& &
+\frac{G M_{\mathrm{1}}}{r^{3}_{\mathrm{1}}}\mathbf{r}\cdot\mathbf{r}_{\mathrm{1}}
+\frac{G M_{\mathrm{2}}}{r^{3}_{\mathrm{2}}}\mathbf{r}\cdot\mathbf{r}_{\mathrm{2}},
\label{eq:Phi}
\end{eqnarray}
which accounts for the contributions of non-inertial terms due to the
reference frame being centered on the star \citep[see][]{rnelson2000}.
The potential softening lengths $\varepsilon_{\mathrm{1}}$ and 
$\varepsilon_{\mathrm{2}}$ are set equal to $1/4$ 
(or $1/7$, in some calculations) times
the Hill radius, $R_{\mathrm{H}}$, of the corresponding planet.
It is worth stressing that the argument according to which $\varepsilon$
should be a fraction of the disk's scale-height, $H$, in the 2D 
geometry \citep[e.g.,][]{masset2002,mueller2012} applies to fully
embedded planets, when $\Rhill<H$. 
If $\Rhill\gtrsim H$, as in all our 2D calculations, 
the local disk scale-height depends also on the gravity of the planet itself 
\citep[e.g.,][]{gennaro2003b}. In such cases, one physical constrain on 
$\varepsilon$ is that it should be smaller than the radius over which
gas is effectively bound to (i.e., it rotates about) the planet 
\citep[$\sim\Rhill/3$, see, e.g.,][]{gennaro2003b}.

The Navier-Stokes equations that characterize the disk evolution are
solved by means of the finite-difference code described in 
\citet[][and references therein]{gennaro2005} with modifications 
detailed below. 
The disk is discretized in $678\times 16\times 700$ grid zones,
in $r$, $\theta$, and $\phi$, respectively 
(and $678\times 700$ in 2D).
Calculations were also performed at a higher resolution of
$1353\times 28\times 2096$.
Comparisons of the evolution of the planets' orbital elements
at these two resolutions yield good agreement.
We apply the wave-damping boundary
conditions of \citet{devalborro2006} within $r=0.3$ and beyond $r=6.65\,\AU$,
which are appropriate for planets far enough from the boundaries.

We implemented an orbital advection algorithm along the lines of the
FARGO algorithm of \citet{masset2000} \citep[see also][]{kley2009}. 
These types of algorithms exploit periodicity properties of the flow,
as those naturally occurring in the azimuthal direction of a disk.
\citep[Note that these algorithms can also be applied to local disk 
simulations, if periodicity is imposed at the patch boundaries, see][]{gammie2001}. 
As demonstrated by \citet{masset2000}, when the highest velocity component
is along the periodic direction,  in a 2D disk such techniques can increase 
the time step limit required by the Courant-Friedrichs-Lewy condition 
\citep[see, e.g.,][]{stone1992a}, relative to a standard advection scheme,
by factors $\sim10$, or larger. In a 3D disk, the gain may critically depend on 
the numerical resolution in the vertical ($\theta$) direction\footnote{%
Since a disk is typically thin, very high resolutions can be more 
easily achieved in the vertical direction. In a viscous disk, the
time step constraint required by the diffusion part of Navier-Stokes
equations scales as the square of the gird spacing. This requirement,
at high resolution and high viscosity, can severely reduce (or even nullify) 
the benefits of orbital advection.}.
The implementation requires care when handling the transport of quantities 
defined on staggered meshes. \citet{kley2009} use split cells, apply the
transport algorithm to each part, and then recombine the partial information 
to reconstruct the full transport.
Unlike them, we define the auxiliary variables required in 
the procedure \citep[see][for details]{masset2000} on the same staggered 
meshes as the transport quantities are defined, wherever they are necessary. 
This approach requires more copies of the standard auxiliary variables 
to be defined, and hence more memory storage, but offers the advantage 
that the advection of all hydrodynamical quantities can be performed 
in a single step.
Contrary to the implementations of both \citet{masset2000} and \citet{kley2009}, 
the algorithm used here avoids any directional bias, maintaining the 
full symmetry of the advection scheme, by using a sequence that alternates 
the transport among directions \citep[see][]{stone1992a}.

The equations of motion of two planets orbiting in a disk around a star, 
written in a reference frame rotating at variable angular speed, are
\begin{eqnarray}
\mathbf{\ddot{r}}_{\mathrm{1}}&=&
-\frac{G(\Ms+M_{\mathrm{1}})}{r^{3}_{\mathrm{1}}}\mathbf{r}_{\mathrm{1}}%
-\frac{GM_{\mathrm{2}}}{r^{3}_{\mathrm{2}}}\mathbf{r}_{\mathrm{2}}%
-\frac{GM_{\mathrm{2}}}{r^{3}_{\mathrm{12}}}\mathbf{r}_{\mathrm{12}}%
+\mathbf{\mathcal{A}}_{\mathrm{1}}-\mathbf{\mathcal{A}}_{s}
\nonumber\\%
& &
-\mathbf{\Omega}_{f}\mathbf{\times}(\mathbf{\Omega}_{f}\mathbf{\times}\mathbf{r}_{\mathrm{1}})%
-2\,\mathbf{\Omega}_{f}\mathbf{\times}\mathbf{\dot{r}}_{\mathrm{1}}%
-\mathbf{\dot{\Omega}}_{f}\mathbf{\times}\mathbf{r}_{\mathrm{1}}
\label{eq:accj}\\
& & \nonumber\\
\mathbf{\ddot{r}}_{\mathrm{2}}&=&
-\frac{G(\Ms+M_{\mathrm{2}})}{r^{3}_{\mathrm{2}}}\mathbf{r}_{\mathrm{2}}%
-\frac{GM_{\mathrm{1}}}{r^{3}_{\mathrm{1}}}\mathbf{r}_{\mathrm{1}}%
+\frac{GM_{\mathrm{1}}}{r^{3}_{\mathrm{12}}}\mathbf{r}_{\mathrm{12}}%
+\mathbf{\mathcal{A}}_{\mathrm{2}}-\mathbf{\mathcal{A}}_{s}
\nonumber\\%
& &
-\mathbf{\Omega}_{f}\mathbf{\times}(\mathbf{\Omega}_{f}\mathbf{\times}\mathbf{r}_{\mathrm{2}})%
-2\,\mathbf{\Omega}_{f}\mathbf{\times}\mathbf{\dot{r}}_{\mathrm{2}}%
-\mathbf{\dot{\Omega}}_{f}\mathbf{\times}\mathbf{r}_{\mathrm{2}},
\label{eq:accs}
\end{eqnarray}
where
$\mathbf{r}_{\mathrm{12}}=\mathbf{r}_{\mathrm{1}}-\mathbf{r}_{\mathrm{2}}$.
Note that, since the origin of the coordinate system is on the star, 
Equations~(\ref{eq:accj}) and (\ref{eq:accs}) include the forces 
per unit mass exerted on the star by the planets. 
The gravitational acceleration terms imposed by the disk, 
$\mathbf{\mathcal{A}}_{\mathrm{1}}$,
$\mathbf{\mathcal{A}}_{\mathrm{2}}$,
and $\mathbf{\mathcal{A}}_{s}$, are defined by Equations~(8)
and (9) of \citet{gennaro2005}
and updated every hydrodynamical time step, $\Delta t$.
Equations~(\ref{eq:accj}) and (\ref{eq:accs}) are integrated numerically
over $\Delta t$
by means of a high-order Gragg-Bulirsch-Stoer extrapolation algorithm
with order and step-size control \citep{hairer1993}. The algorithm
chooses automatically a suitable order at each (internal) step, 
which basically depends on the required tolerance of the solution 
error. We set a relative tolerance of $10^{-9}$ and an absolute tolerance 
of $10^{-14}$.

\subsection{Torque Calculations and Outward Migration}
\label{sec:ToC}

The basic mechanism that may allow a pair of resonant-orbit
planets to experience a positive torque exerted by a gaseous disk
and migrate outward was first described by \citet{masset2001}.
Labeling with subscripts $1$ and $2$ the inner and outer 
planet, respectively, in order for this mechanism to be active, 
the following conditions must be fulfilled:
\begin{itemize}
\item[\textit{\i})] the planet-to-star mass ratios ($q_{i}=M_{i}/\Ms$) must be such 
that $q_{1}>q_{2}$; 
\item[\textit{\i\i})] the separation of the semimajor axes 
$\Delta a=a_{2}-a_{1}$ must be such that
$\Delta a=b(R_{\mathrm{H},1}+R_{\mathrm{H},2})$, where
$b\lesssim 4.5$ (as we shall discuss below);
\item[\textit{\i\i\i})] $q_{2}$ must be large enough to open a gap,
or partial gap, in the density distribution by tidal torques.
\end{itemize}

\begin{figure}[t!]
\centering%
\resizebox{\linewidth}{!}{%
\includegraphics{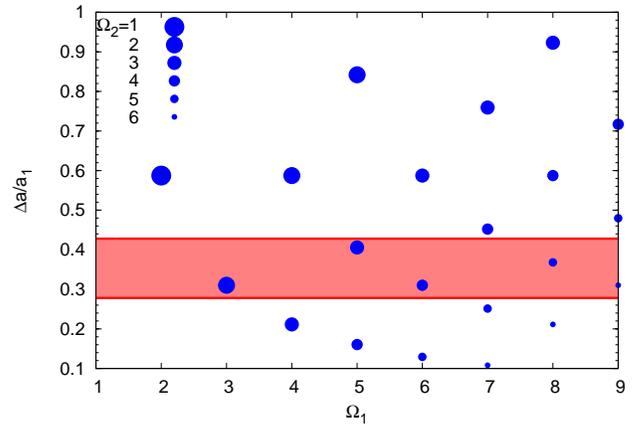}}
\caption{%
              Fractional difference, $\Delta a/a_{1}=a_{2}/a_{1}-1$, 
              between the semimajor axes of
              the interior ($a_{1}$) and exterior ($a_{2}$) planets,
              as a function of the mean motion of the interior planet
              $\Omega_{1}$ (in scaled units). Different symbol
              sizes refer to different mean motions of the exterior
              planet, $\Omega_{2}$ (see the legend). The region of the graph
              below the shaded area is unstable due to planet--planet
              interaction. The Hill stability criterion adopted in this figure 
              is given by Equation~(23) of \citet{gladman1993}. 
              The region above the shaded area does not fulfill
              condition \textit{\i\i}) for gap overlap (see also Figures~\ref{fig:zoom}
              and \ref{fig:dtdm}). Resonant orbits falling in the shaded area 
              may activate outward migration.
             }
\label{fig:reso}
\end{figure}
\begin{figure*}[t!]
\centering%
\resizebox{0.5384\linewidth}{!}{%
\includegraphics[clip,bb=0 25 561 337]{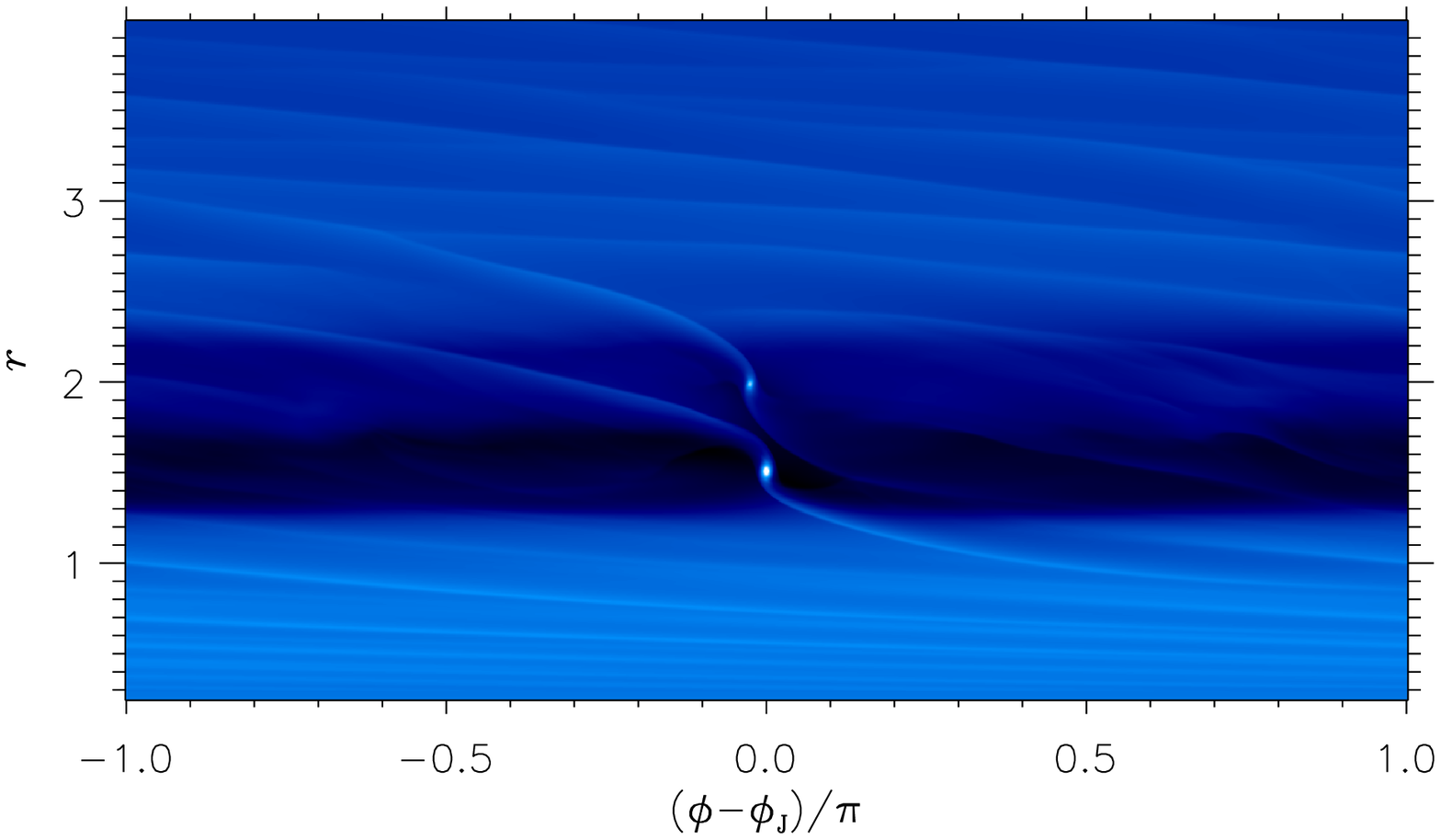}}%
\resizebox{0.4616\linewidth}{!}{%
\includegraphics[clip,bb=0 25 481 337]{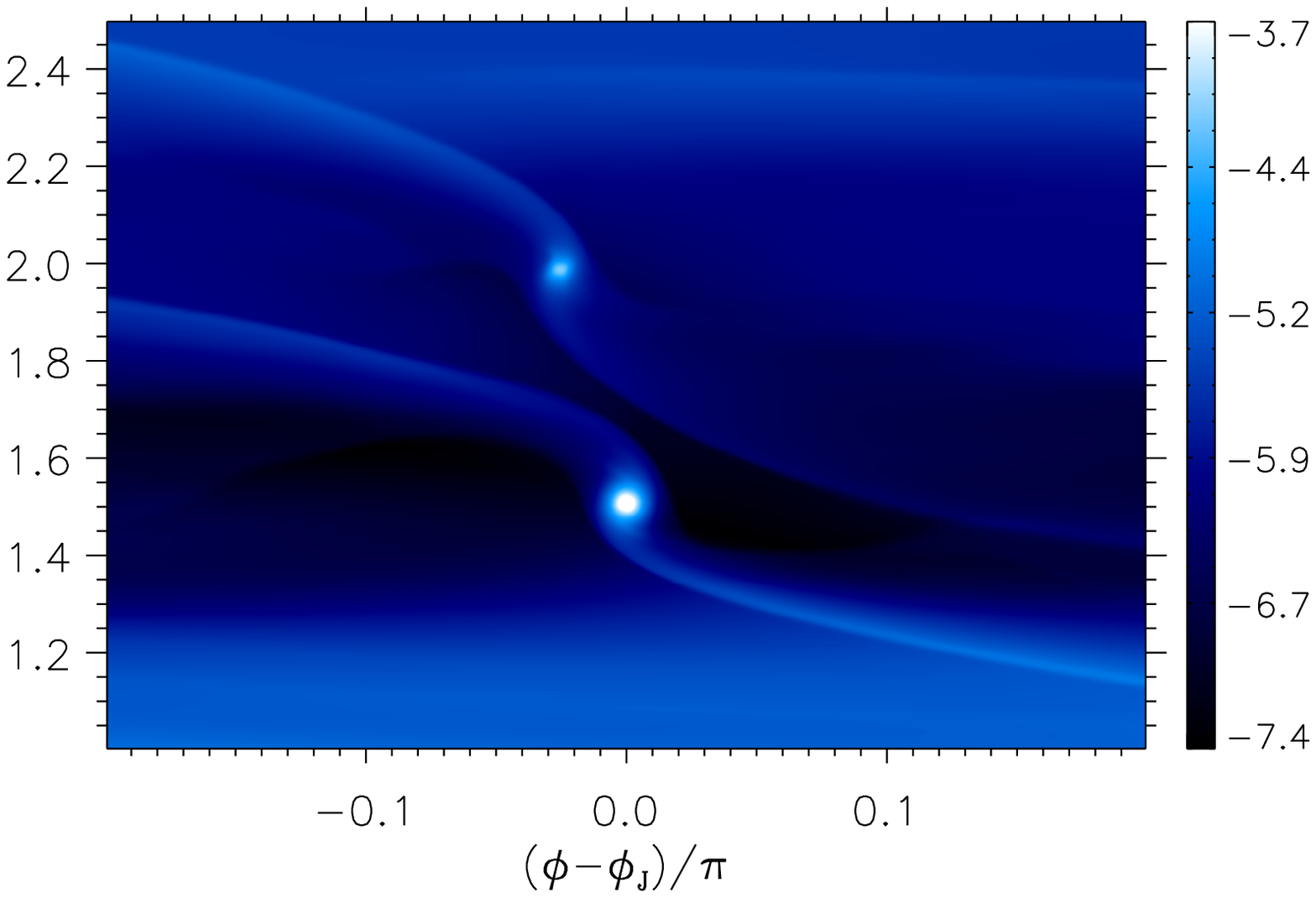}}
\resizebox{0.5384\linewidth}{!}{%
\includegraphics[clip,bb=0 00 561 317]{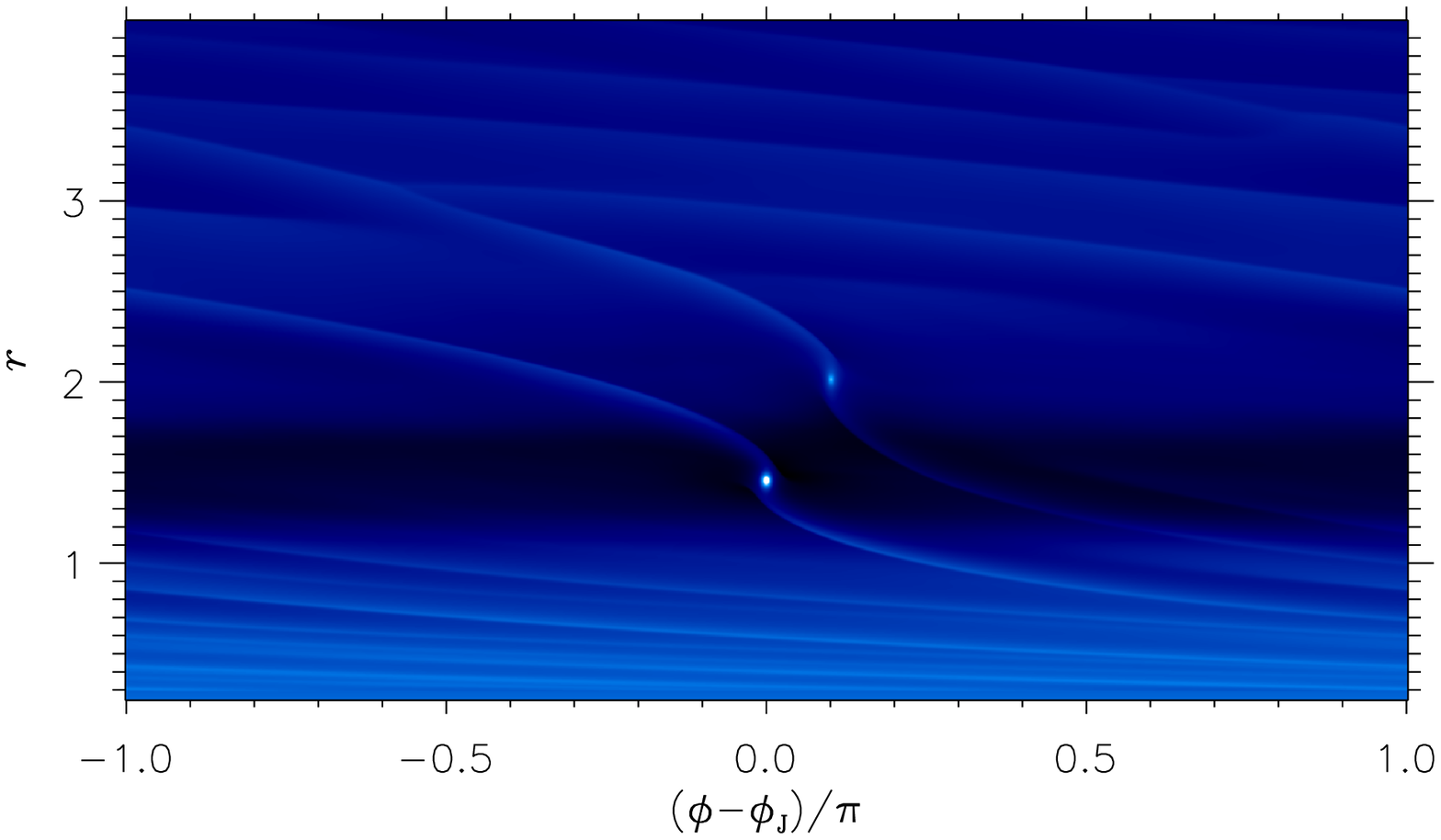}}%
\resizebox{0.4616\linewidth}{!}{%
\includegraphics[clip,bb=0 00 481 317]{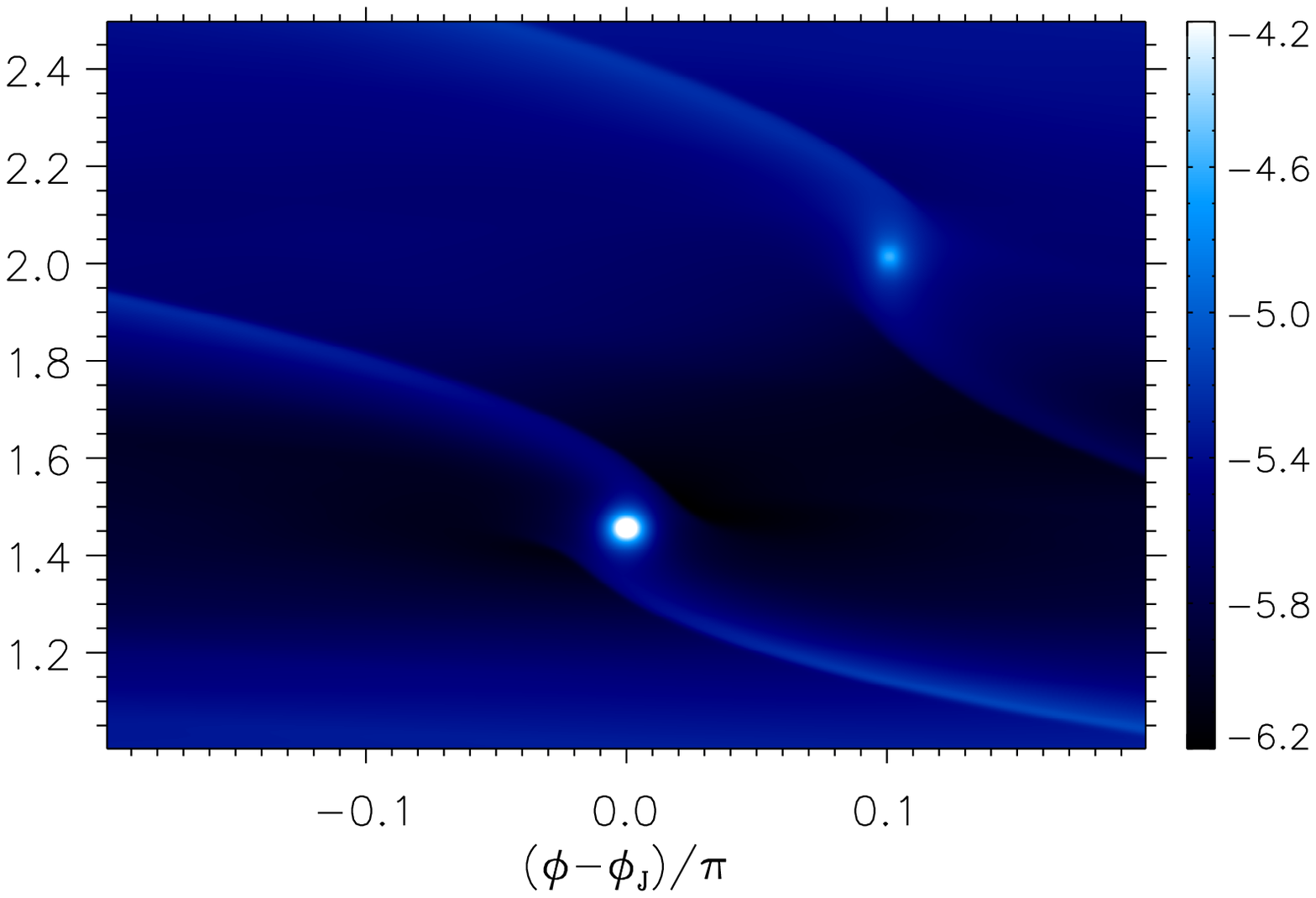}}
\resizebox{\linewidth}{!}{%
\includegraphics[clip,bb=-35 -15 501 216]{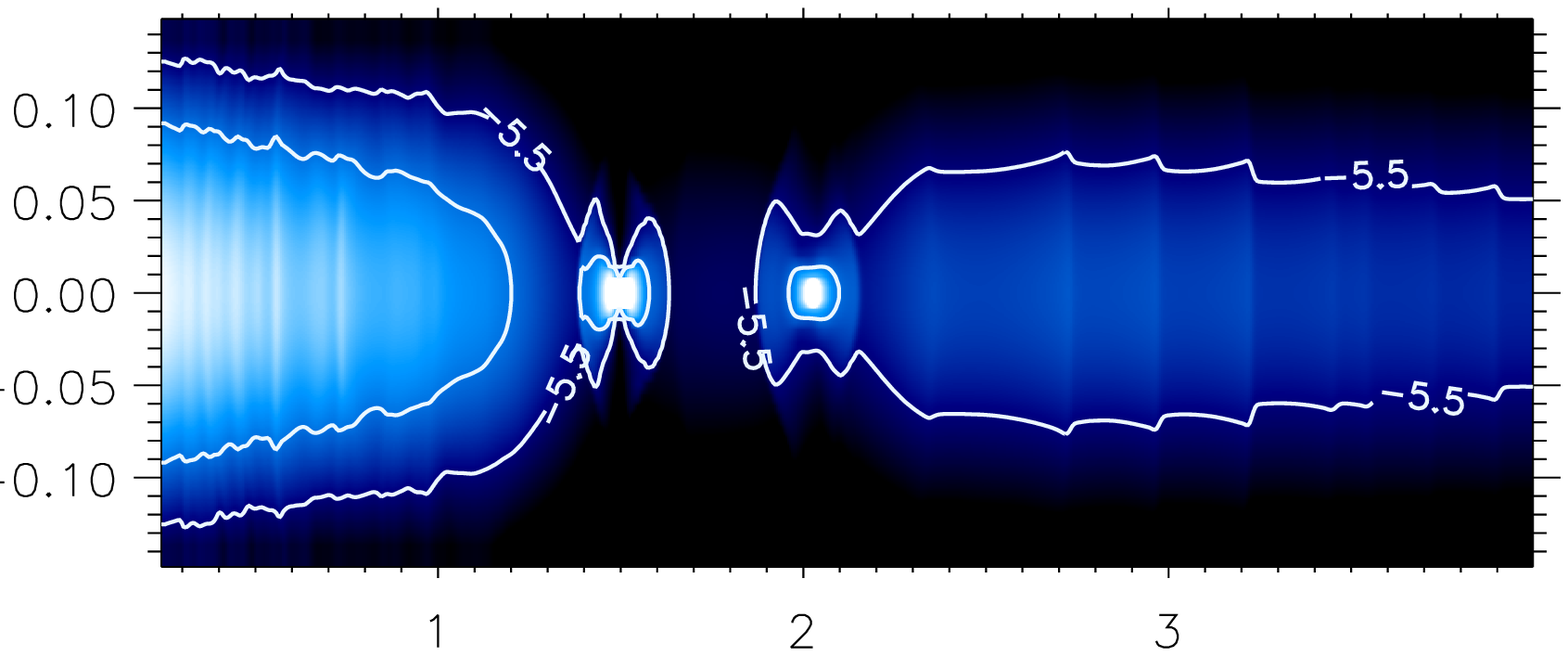}%
\includegraphics[clip,bb= 45 -15 561 216]{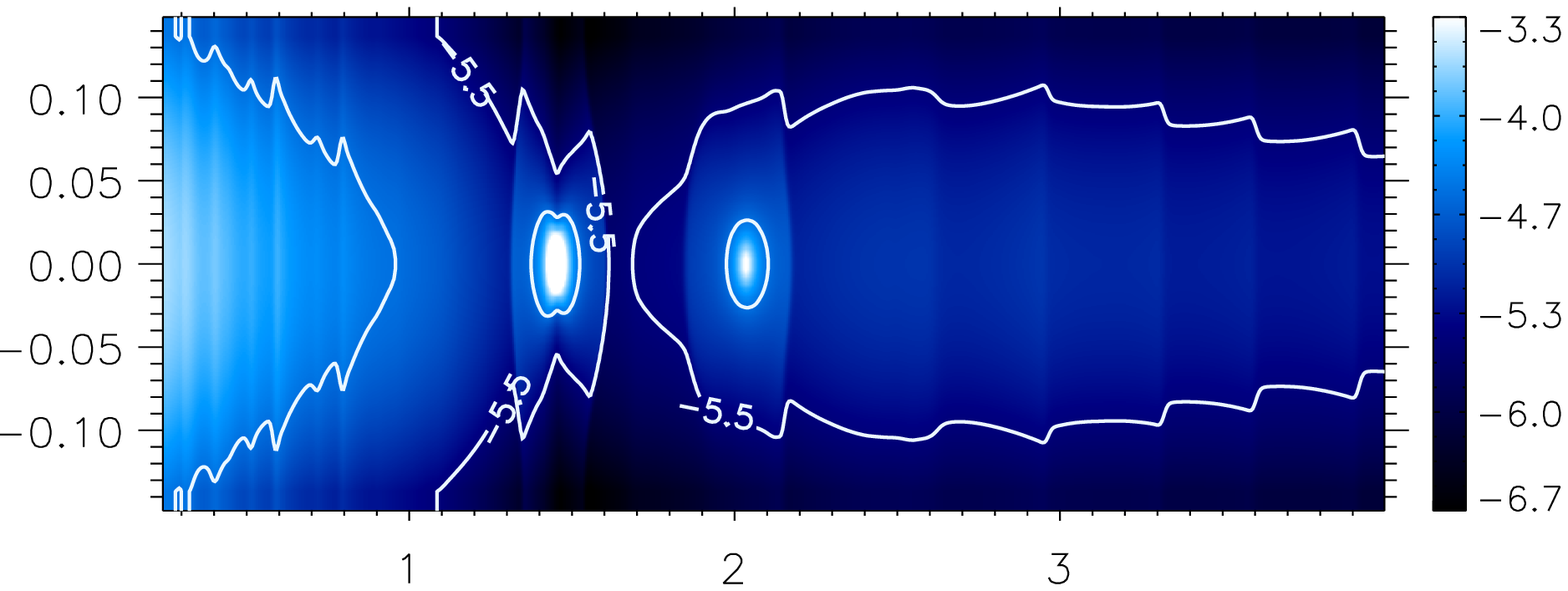}}
\caption{%
             Surface density distribution obtained from 3D calculations of
             a disk with $H/r=0.04$ (\textit{top}) and  $0.07$ (\textit{middle}).
             The angle $\phi_{\mathrm{J}}$ is the azimuth of the interior planet,
             Jupiter.
             The turbulence parameter $\alpha_{\mathrm{t}}$ is $0.005$
             in both cases. A zoom of the region around the planets is shown
             in the right panels. The color scale 
             is logarithmic and given 
             in units of $\Ms\,r^{-2}_{1}$, where $r_{1}$ indicates the radius $r=1$
             (i.e., the unit of length).
             The bottom panels show the vertical stratification of the mass density,
             $\rho$, at the disk azimuth $\phi_{\mathrm{J}}=\phi_{\mathrm{S}}$ for 
             $H/r=0.04$ (\textit{left}) and $0.07$ (\textit{right}). The angle 
             $\vartheta=\pi/2-\theta$ is the disk's latitude and the color
             scale is in units of $\Ms\,r^{-3}_{1}$.
             }
\label{fig:zoom}
\end{figure*}
Condition \textit{\i\i}) above implies that 
$\Delta a/a_{1}=(\sqrt[3]{q_{1}}+\sqrt[3]{q_{2}})/(\sqrt[3]{3}/b-\sqrt[3]{q_{2}})$.
However, Hill stability for close planets on circular orbits also imposes 
that $\Delta a/a_{1}\gtrsim 2.40\, \sqrt[3]{q_{1}+q_{2}}$ 
\citep[this inequality strictly applies in the limit of vanishing masses
and absence of gas, see][]{gladman1993}, 
whose right-hand side is about equal to $0.26$
for $q_{1}=M_{\mathrm{1}}/M_{s}=9.8\times 10^{-4}$ and 
$q_{2}=M_{\mathrm{2}}/M_{s}=2.9\times 10^{-4}$,
hence $b\gtrsim 2$ 
\citep[or $2.2$, adopting a more precise determination of the Hill stability criterion, see]%
[for details. See also Figure~\ref{fig:reso}]{gladman1993}. 
Conditions \textit{\i})  and \textit{\i\i}) suggest that, for a 
Jupiter--Saturn pair, the first encountered first-order mean motion resonance in which 
the mechanism may be activated is the 3:2 (the second-order 5:3 commensurability 
being another possibility), as indicated in Figure~\ref{fig:reso}.
In principle, configurations external, but sufficiently near, to resonances may 
also promote outward migration, as we shall see in Section~\ref{sec:OMR}. 
It is worth stressing here that simple capture in a mean motion resonance does
not imply outward migration \citep[see, e.g.,][]{zhang2010}, as we shall see in 
Section~\ref{sec:21}.

Figure~\ref{fig:zoom} (\textit{top} and \textit{middle}) illustrates the surface density 
perturbed by resonant-orbit planets, derived from 3D calculations for disks of different 
thicknesses (see figure's caption). The bottom panels of the figure show the mass
density in a vertical slice of the disk, while the planets are aligned with the star.
The exterior planet opens a partial gap in the case of a thinner disk, 
but it does so to a lesser extent in the other case (see \textit{bottom panels}).

\begin{figure*}[t!]
\centering%
\resizebox{\figlen}{!}{%
\includegraphics{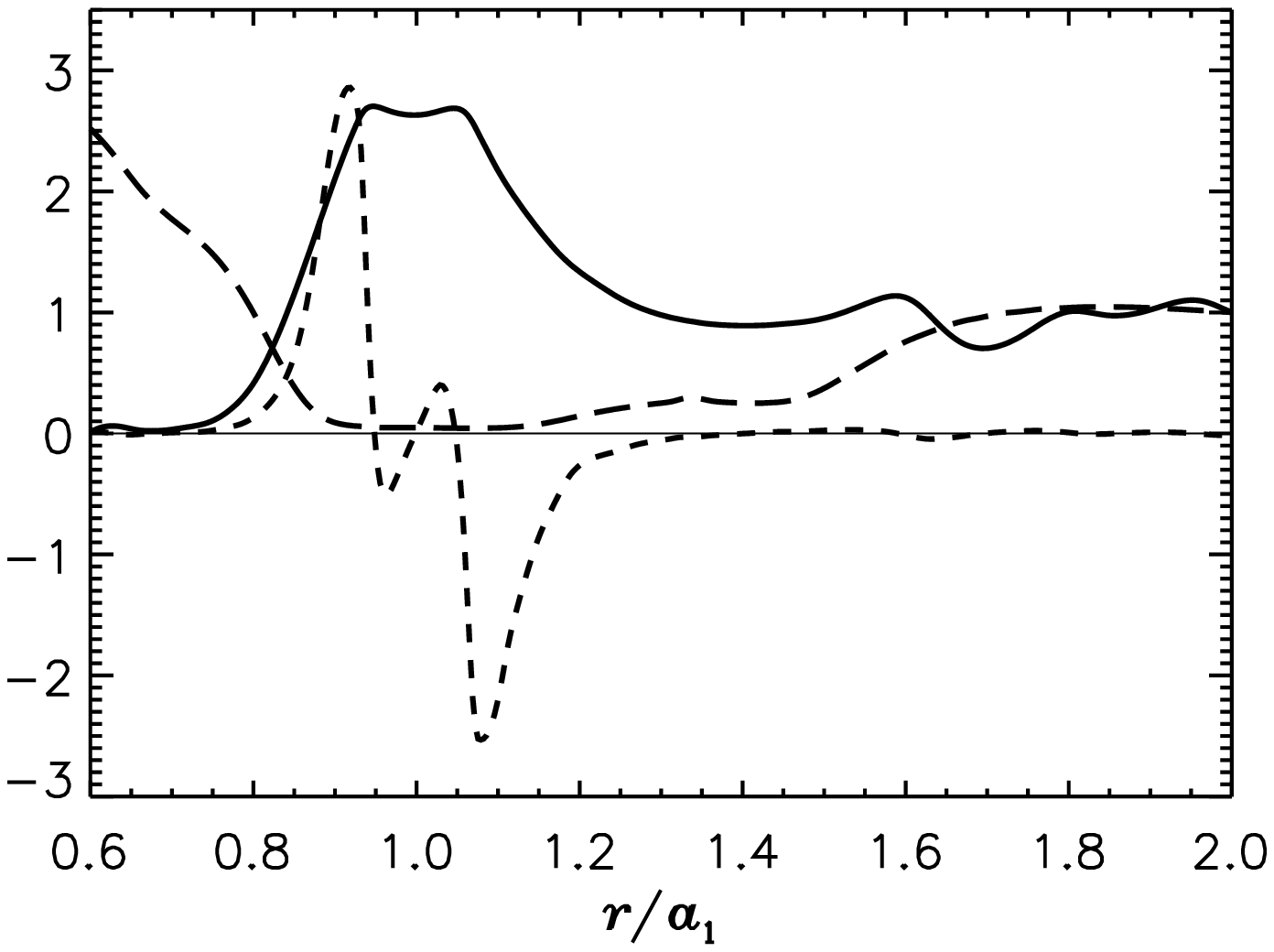}%
\includegraphics{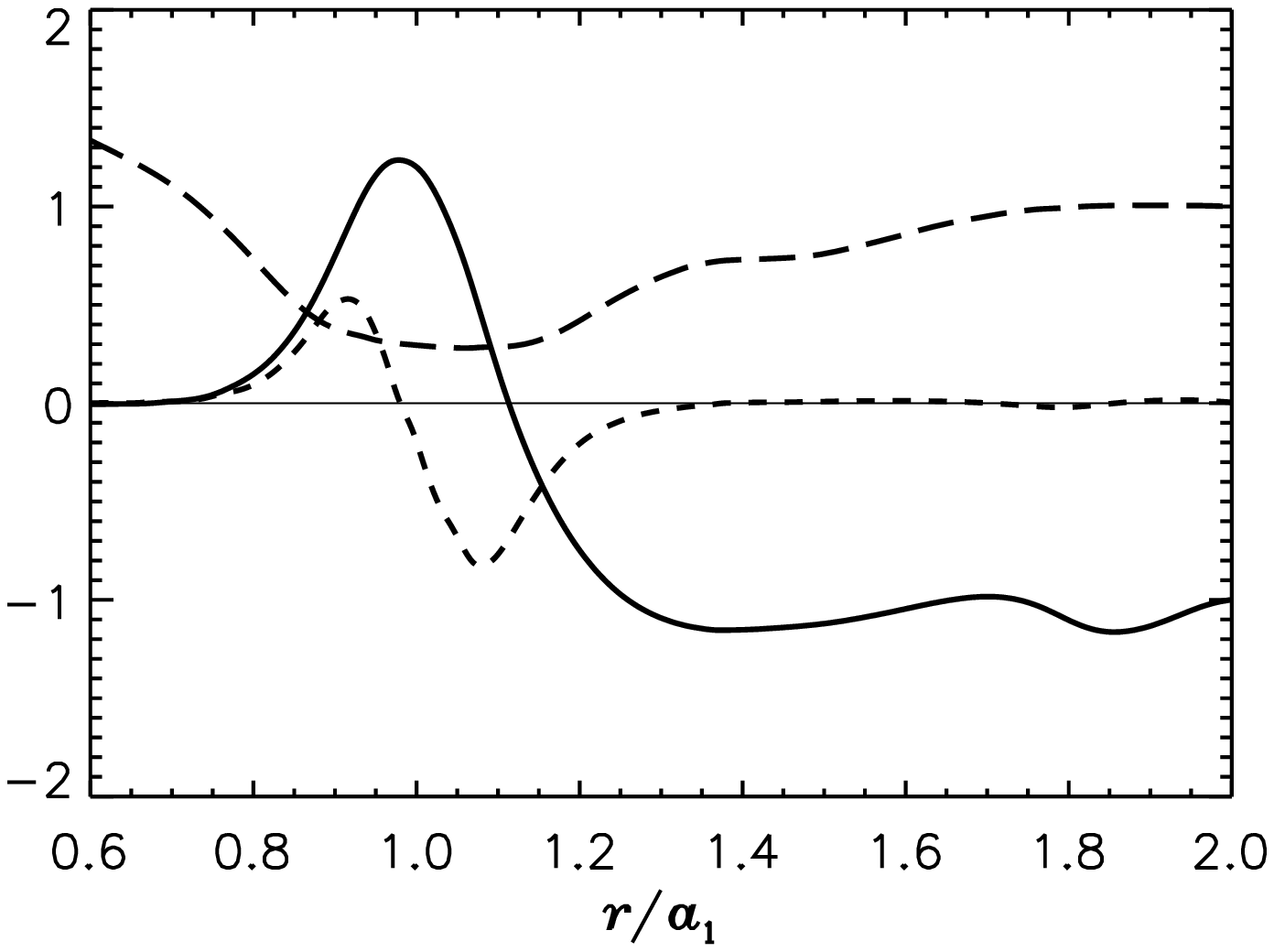}}
\caption{%
              Azimuthally averaged surface density (\textit{long-dashed line}),
              torque per unit disk mass (\textit{short-dashed line}) exerted on 
              the inner 
              planet, and cumulative torque (Equation~\ref{eq:Tcum}) acting 
              on the inner planet (\textit{solid line}) for the same models as in
              Figure~\ref{fig:zoom} (cases with $H/r=0.04$ and $0.07$ on the 
              left and right, respectively). 
              Quantity $d\mathcal{T}/dM$ 
              is normalized to $10^{3} G\Ms(M_{1}/\Ms)^{2}/a_{1}$ ($a_{1}$ is
              the semimajor axis of the interior planet).
              The surface density and cumulative torque are normalized by their 
              absolute values at $r/a_{1} = 2$.
              The peaks occurring at the planets' orbital radii 
              (due to mass accumulation within the Roche lobe) have been 
              removed from the surface density profile.
              Results displayed here were obtained from 3D calculations and 
              averaged over a few orbital periods of the outer planet.
             }
\label{fig:dtdm}
\end{figure*}
The occurrence of outward migration of resonant-orbit planets can be 
intuitively understood from Figure~\ref{fig:dtdm}, which reports on the 
results obtained from the same calculations as in Figure~\ref{fig:zoom},  
where Saturn is caught in a 2:3 mean motion resonance with Jupiter. 
The azimuthally averaged surface 
density in normalized units (\textit{long-dashed lines}) indicates that Saturn 
has cleared a partial gap, whose inner part  overlaps with the outer part of 
the gap opened by Jupiter.
The short-dashed lines in the figure represent torque
density distributions (\citealp{gennaro2008}, hereafter 
\citetalias{gennaro2008}) due to Jupiter, $d\mathcal{T}/dM$. 
These functions yield the total torque when integrated over the disk mass.
The torque exerted on the interior planet peaks at 
$a_{1}\pm R_{\mathrm{H,1}}$ and is mostly comprised in a radial region 
of average width $\sim 3.5\,R_{\mathrm{H,1}}$ on either side of the orbit
(note that $R_{\mathrm{H,1}}>H$ in the case displayed in the left panel 
and $R_{\mathrm{H,1}}\approx H$ in the other case).
But since gas depletion due to gap formation may extend somewhat beyond 
this distance (see long-dashed lines in Figure~\ref{fig:dtdm}), one may allow 
for a maximum value of the factor $b$ in condition \textit{\i\i}) above between
$4$ and $5$. 
It is also important to notice that the orbital eccentricity of a planet acts to widen
and  smooth out gap edges \citep[see, e.g., Figure~2 of][]{gennaro2006}, 
which may also affect somewhat the factor $b$.

The solid lines in Figure~\ref{fig:dtdm} represent the cumulative torque, 
which is defined as
\begin{equation}
\mathcal{T}_{\mathrm{CM}}(r)=%
2\pi\int_{0}^{r}\frac{d\mathcal{T}}{dM}\langle\Sigma\rangle r' dr'.
\label{eq:Tcum}
\end{equation}
Looking at $\mathcal{T}_{\mathrm{CM}}$ in the left panel of Figure~\ref{fig:dtdm}, 
it appears clear that the positive torque exerted by the disk interior of Jupiter's 
orbit is larger than that exerted by the disk exterior of the orbit, principally 
because Saturn has lowered the density there. The right panel illustrates
the situation for a thicker disk, in which gas depletion operated by 
the exterior planet is not sufficient to reverse the sign of the torque.
\begin{figure}[t!]
\centering%
\resizebox{\linewidth}{!}{%
\includegraphics[clip, bb=15 52 481 205]{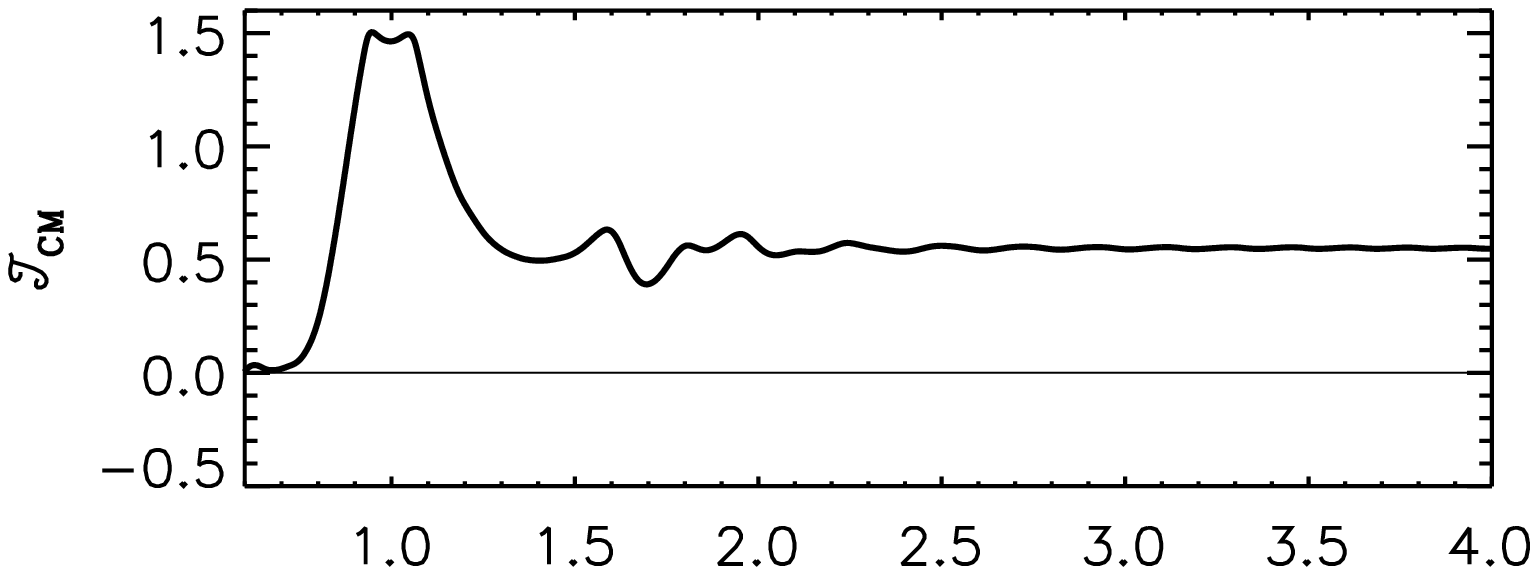}}
\resizebox{\linewidth}{!}{%
\includegraphics[clip, bb=15 52 481 205]{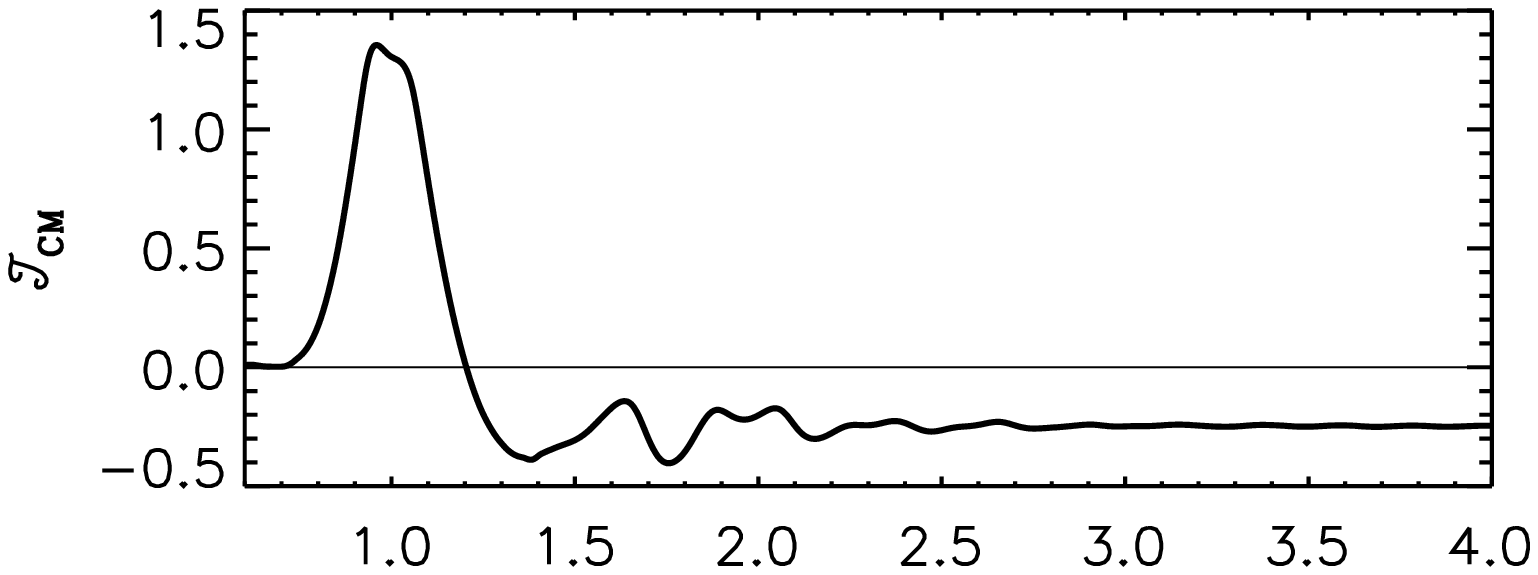}}
\resizebox{\linewidth}{!}{%
\includegraphics[clip, bb=15 10 481 205]{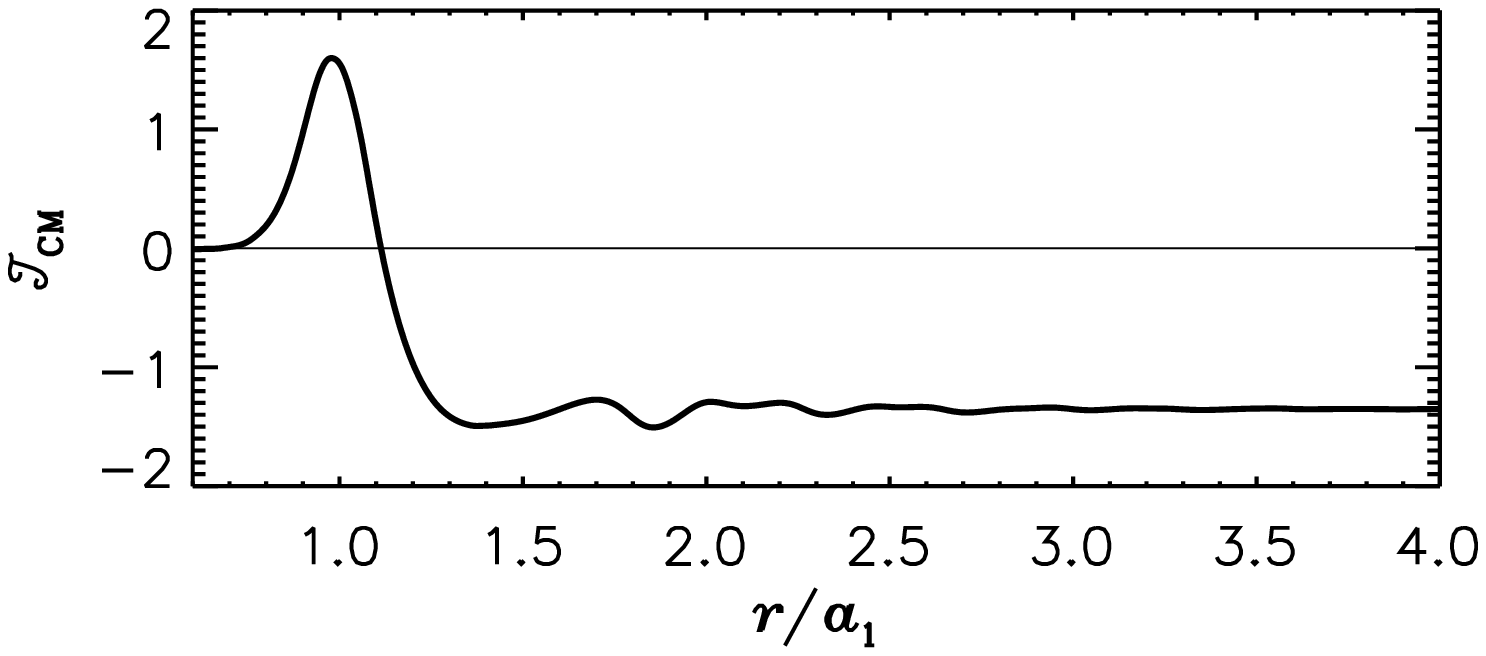}}
\caption{%
              Cumulative torque, Equation~(\ref{eq:Tcum}), exerted on the interior 
              planet by a 3D disk for which the turbulence parameter is
              $\alpha_{\mathrm{t}}=0.005$ and the thickness is $H/r=0.04$
              (\textit{top}), $0.05$  (\textit{center}),  and $0.07$ (\textit{bottom}). 
              Quantity $\mathcal{T}_{\mathrm{CM}}$ is normalized to 
              $10^{-3}G M^{2}_{\mathrm{s}}(M_{1}/\Ms)^{2}/a_{1}$, 
              where $a_{1}$ is the semimajor axis of the interior planet.
             }
\label{fig:tcum}
\end{figure}
Figure~\ref{fig:tcum} allows for a comparison of the cumulative torque 
for three different values of the disk thickness: $H/r=0.04$ (\textit{top}), $0.05$ 
(\textit{center}), and $0.07$ (\textit{bottom}) (see figure's caption for further details).
A closer inspection of $d\mathcal{T}/dM$ and $\mathcal{T}_{\mathrm{CM}}$ in 
Figure~\ref{fig:dtdm} (\textit{left}) and of the cumulative torque in Figure~\ref{fig:tcum} 
(\textit{top}) indicates that the total (positive) torque is basically driven by Lindblad 
resonances and that corotation torques are unimportant ($\mathcal{T}_{\mathrm{CM}}$ 
does not vary significantly over the radial width of the corotation region), as also 
argued by \citet{morbidelli2007}.

It thus appears from Figures~\ref{fig:dtdm} and \ref{fig:tcum}  
that the discriminant factor for outward, stalled, or inward migration
is the depth (and width) of the outer planet's gap. 
If the outer planet opened a very deep and sufficiently wide gap, 
the inner planet would be subjected only to a one-sided Lindblad (positive) 
torque exerted by the interior disk that, to within a factor of order unity, 
can be written as \citep[see][and referenced therein]{lubow2010}
\begin{equation}
\mathcal{T}_{\mathrm{OS}}\sim%
a^{4}\,\Omega^{2}\,\Sigma\left(\frac{\Mp}{\Ms}\right)^{2}%
\left(\frac{a}{\widetilde{\Delta}}\right)^{3},
\label{eq:Tos}
\end{equation}
where $\widetilde{\Delta}=\max{(H,R_{\mathrm{H}})}$. 
The concept of one-sided torque is very useful to evaluate the presence 
of a tidally-induced gap, i.e., condition \textit{\i\i\i}) above. 
To first-order approximation, a density depletion begins to form when 
the one-sided torque exceeds the viscous torque (see Section~\ref{sec:DD}), 
$\mathcal{T}_{\mathrm{OS}}\gtrsim\mathcal{T}_{\nu}$, which yields a
simple order-of-magnitude condition
$q^{2}\gtrsim 3\pi\alpha_{\mathrm{t}}(H/a)^{2}(\widetilde{\Delta}/a)^{3}$
\citep[see also][]{papaloizou1984,ward2000},
or
\begin{equation}
g=\frac{q}{\sqrt{3\pi\alpha_{\mathrm{t}}}}\left(\frac{a}{H}\right)%
\left(\frac{a}{\widetilde{\Delta}}\right)^{3/2} \gtrsim 1.
\label{eq:gcon}
\end{equation}
This conditions should be regarded as a measure of how much 
the density along the planet's orbit is depleted, hence it can be 
considered a condition for gas depletion. 
A condition for tidal truncation (gap formation) is then $g\gg1$
\citep[see also][]{lin1986b}.
If predictions from the inequality~(\ref{eq:gcon}) are compared 
with results from direct 3D calculations 
(\citetalias{gennaro2008}, Figures~6 and 8), one finds that $g\approx 1$ 
corresponds to a $\sim 20$\% density depletion (relative to the unperturbed 
state, i.e., with no planet), and $g\approx 2.7$ to $\sim 60$\% depletion. 
If applied to Saturn in 
the disks of Figure~\ref{fig:dtdm}, $g\approx 3.4$ for a density depletion 
of roughly $75\%$ (\textit{left panels}) and again $g\approx 1$ for a
density drop of $\sim 20$\% (\textit{right panels}).

\subsection{Orbital Migration Rates}
\label{sec:OMR}

\begin{figure}[t!]
\centering%
\resizebox{\linewidth}{!}{%
\includegraphics{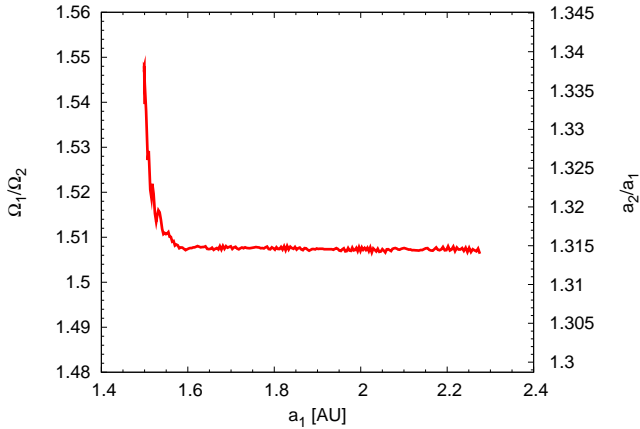}}
\caption{%
              Ratio of orbital frequencies of the inner to the outer planet showing
              convergence toward the 3:2 commensurability and subsequent
              resonant-orbit migration.
              The direction of migration of both planets is outward. 
              The labels on the right vertical axis are \textit{an approximation} 
              of the semimajor axis ratio, $a_{2}/a_{1}$.
             }
\label{fig:oo}
\end{figure}

In order to derive migration rates, we shall assume that the orbits of
the interior and exterior planets, Jupiter and Saturn respectively, are 
in the 3:2 mean motion resonance and that this resonance is maintained 
during migration. Results from a calculation that support this assumption
are plotted in Figure~\ref{fig:oo}. All calculations resulting in outward 
migration behave similarly, but there are also instances in which the
resonance is broken (see below). 
Since the ratio $a_{\mathrm{2}}/a_{\mathrm{1}}$ is supposed to be a constant, 
we concentrate on the migration rate of the interior planet, 
Jupiter. We seek an expression for $\dot{a}$ of the form
\begin{equation}
\frac{d a}{d t}=\dot{a}_{\mathrm{ref}}\,%
k_{1}(\Sigma)\,k_{2}(a)\,k_{3}(g),
\label{eq:dotaJ0}
\end{equation}
where $\dot{a}_{\mathrm{ref}}$ is a reference migration speed and
$k_{1}$, $k_{2}$, and $k_{3}$ are dimensionless functions. 
To derive such expression, we employ results from both 2D and 
3D calculations. In this section, the planets are fully formed since 
the beginning of the calculations and non-accreting.

\begin{figure*}[t!]
\centering%
\resizebox{\figlen}{!}{%
\includegraphics{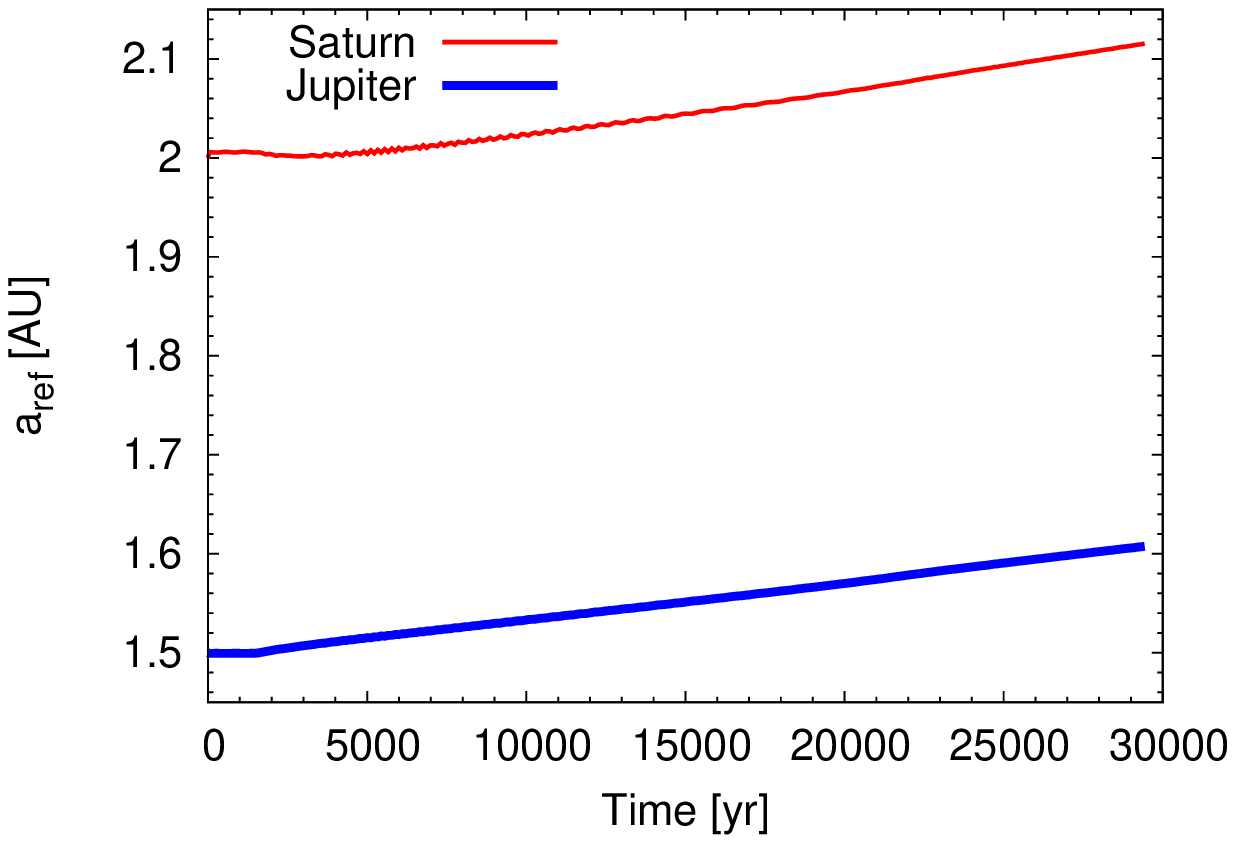}%
\includegraphics{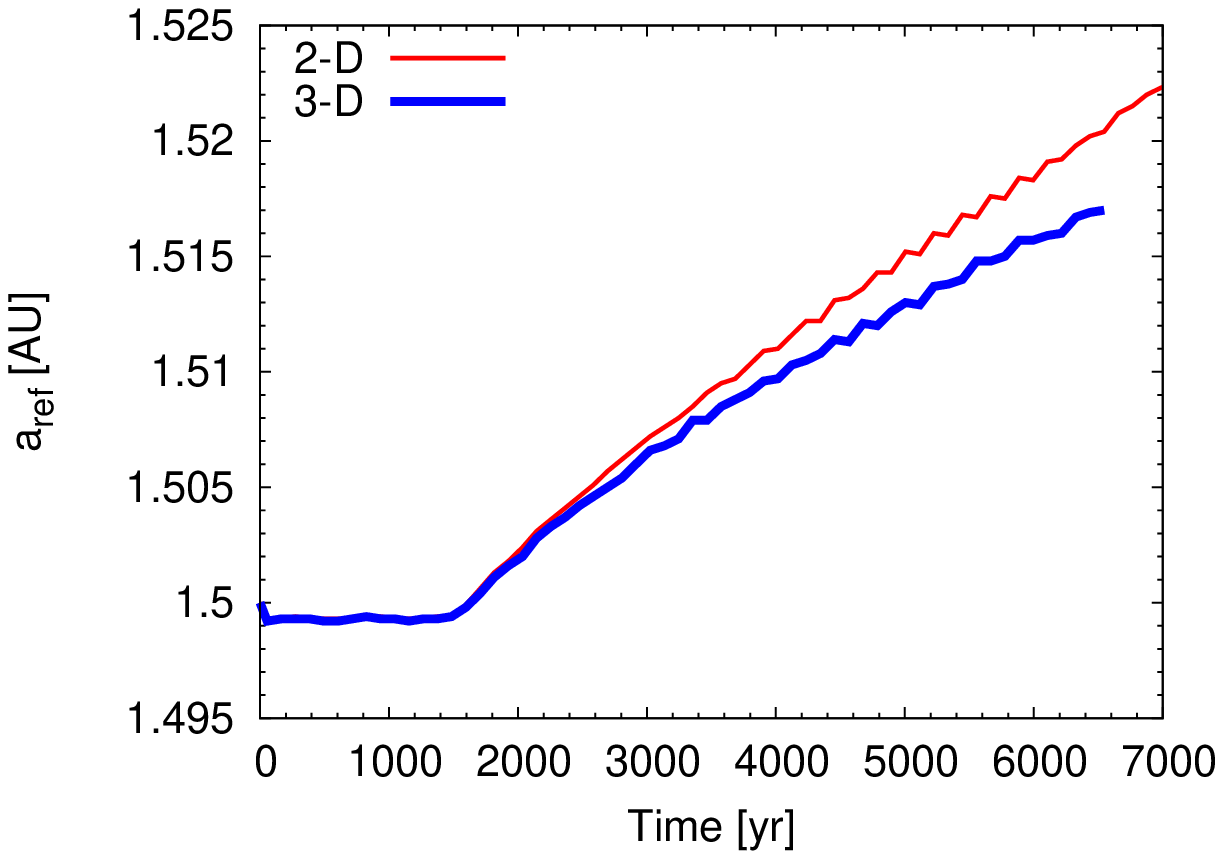}}
\resizebox{\figlen}{!}{%
\includegraphics{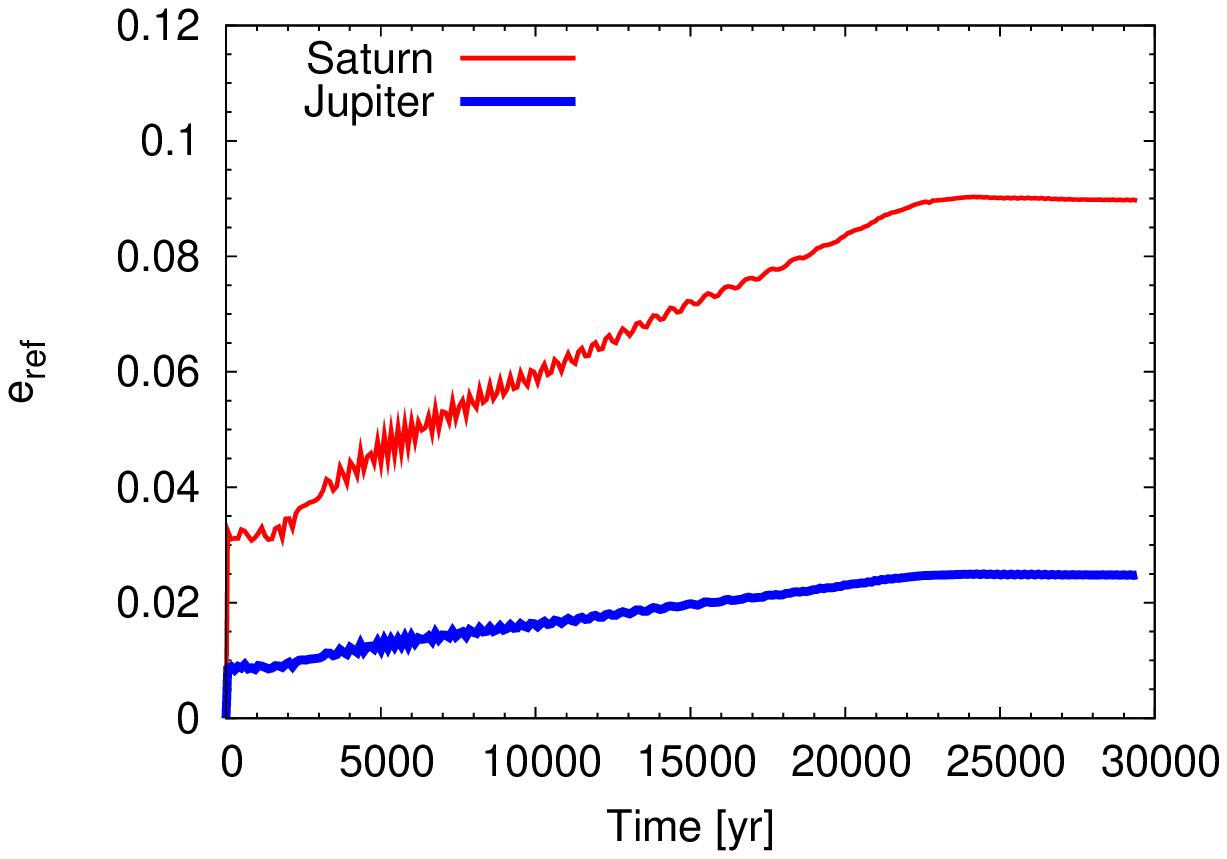}%
\includegraphics{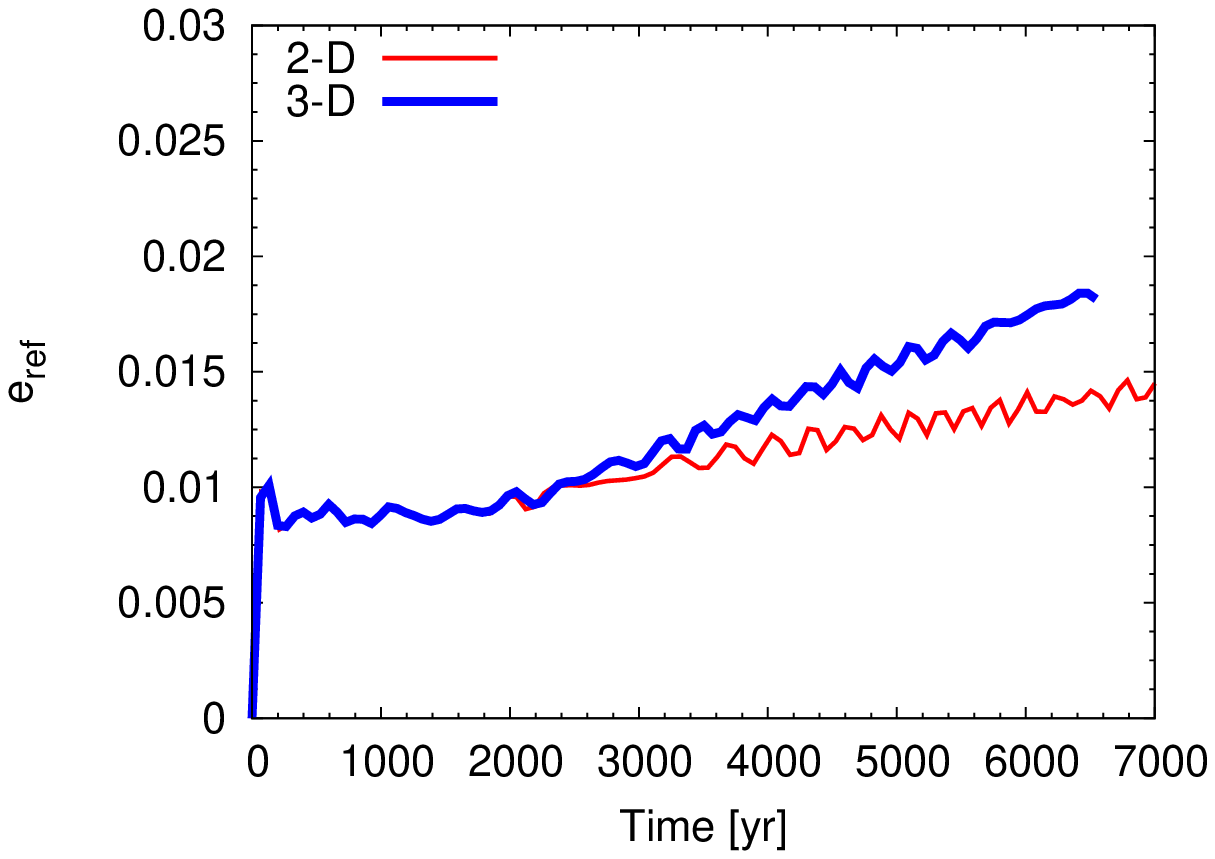}}
\caption{%
              \textit{Left:}
              semimajor axis (\textit{top}) and eccentricity (\textit{bottom}) evolution 
              of a Jupiter--Saturn system locked in the 3:2 mean motion resonance,
              for the reference model. The planets evolve in a disk
              with  $H/r=0.04$, $\alpha_{\mathrm{t}}=0.005$, and an initial
              $\Sigma$ at $1\,\AU$ of $50\,\mathrm{g\,cm}^{-2}$. During the
              first $1500$ years, the planets interact with each other and with
              the star, but do not ``feel'' the disk. 
              Elliptical (osculating) orbital elements are obtained applying
              Gauss perturbation equations \citep[e.g.,][]{beutler2005} in a 
              non-rotating frame (including perturbations from the other planet) 
              and then averaging over $60$ and $40$ orbital periods of the inner
              and outer planet, respectively.
              \textit{Right:}
              comparison between evolutions of the semimajor axis (\textit{top}) 
              and eccentricity (\textit{bottom}) of Jupiter,
              obtained for the reference model, in a 2D and 3D disk. 
             }
\label{fig:aref}
\end{figure*}
In Figure~\ref{fig:aref} (\textit{top-left panel}), the semimajor axis evolution 
of the reference model is shown for both planets (see figure's caption for details).
The disk has an aspect ratio $H/r=0.04$ and turbulence parameter
$\alpha_{\mathrm{t}}=0.005$.
At $r=1\,\AU$, the surface density is $\Sigma_{1}=50\,\mathrm{g\,cm}^{-2}$
at time $t=0$. This value is chosen from the disk evolution calculations
discussed in Section~\ref{sec:MR}, which show that $\gtrsim 50$\% of disks 
have $\Sigma_{1}\lesssim 50\,\mathrm{g\,cm}^{-2}$ after $\sim 1\,\mathrm{Myr}$.
Following \citet{pierens2011}, we set $a_{\mathrm{1}}=1.5\,\AU$ and 
$a_{\mathrm{2}}=2\,\AU$ at $t=0$ (see caption of Figure~\ref{fig:aref} for
further details), slightly outside the 3:2 mean motion resonance
(see Figure~\ref{fig:oo}). 
In the top-right panel of the Figure, a comparison between
2D and 3D calculation results is presented. 
The 3D migration rate is about $25$\% 
smaller than the 2D one, which correction we apply to all 2D
results. Note that this comparison is carried out at the same numerical
resolution in the $r--\phi$ plane (see Section~\ref{sec:HC}).
The parameter that may produce the largest differences
between 2D and 3D outcomes is the disk thickness.  In this case,
however, we rely only on 3D calculations to approximate 
Equation~(\ref{eq:dotaJ0}). Similar plots are shown in the bottom panels 
of Figure~\ref{fig:aref}, but for the orbital eccentricity. For the duration
of the evolution we consider, the eccentricity of Jupiter does not exceed 
$\sim 0.03$ in any of the models discussed in this section. 
The orbital eccentricity of the exterior planet grows larger, 
to values $e_{2}\sim 0.1$ \citep[see also][]{pierens2011}.

As mentioned above, Figures~\ref{fig:oo} and \ref{fig:aref} indicate
that outward migration can be activated also if orbital configurations
are external, but somewhat near, the 3:2 commensurability. This
effect is related to the gap widths and the extent to which density
perturbations compound. In this context, orbital eccentricity may play
some important role, since it affects the shape of a gap \citep{gennaro2006}.

In Section~\ref{sec:ToC}, we argued that the torque exerted on Jupiter
would tend to the one-sided Lindblad torque $\mathcal{T}_{\mathrm{OS}}$
(Equation~\ref{eq:Tos}),
if Saturn (i.e., the exterior planet) carved a very deep and wide gap in the disk.
In the opposite limit of very large disk thickness, $H$, and/or viscosity parameter, 
$\alpha_{\mathrm{t}}$, neither Jupiter nor Saturn would be capable of
depleting the disk significantly and therefore it is expected that 
the torque exercised upon the interior planet will be of \textit{type~I}
\citep{ward1986,lin1986a}
\begin{equation}
\mathcal{T}_{\mathrm{I}}\sim%
-a^{4}\,\Omega^{2}\,\Sigma\left(\frac{a}{H}\right)^{2}%
\left(\frac{\Mp}{\Ms}\right)^{2},
\label{eq:TI}
\end{equation}
where, again, we neglect a factor (typically) of order unity in front 
of the right hand side. Since both
$\mathcal{T}_{\mathrm{OS}}$ and $\mathcal{T}_{\mathrm{I}}$ are linear
in $\Sigma$, and since in the limit of zero orbital eccentricity
\begin{equation}
\frac{da}{dt}=\frac{2\mathcal{T}}{a\Omega\Mp},
\label{eq:dadt_circ}
\end{equation}
one obvious guess is to approximate $k_{1}$ as a linear function.
\begin{figure*}[t!]
\centering%
\resizebox{\figlen}{!}{%
\includegraphics{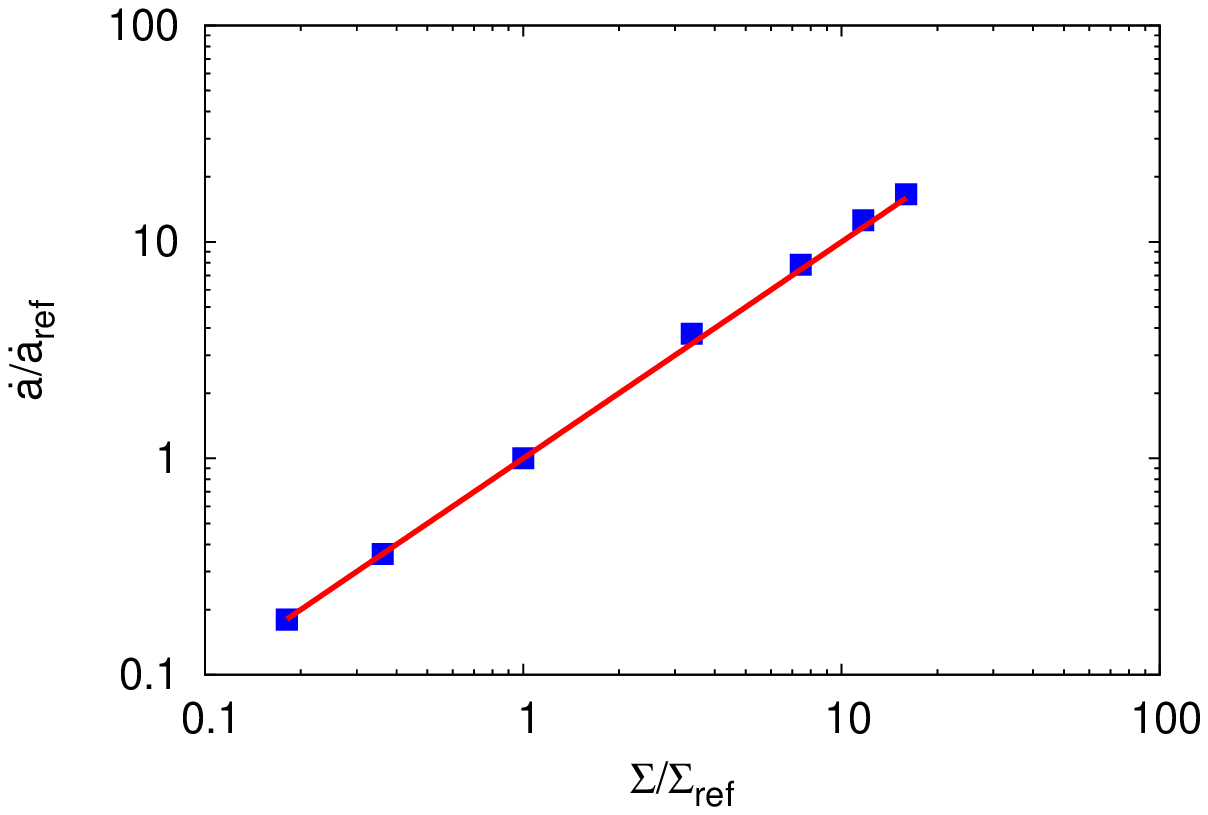}%
\includegraphics{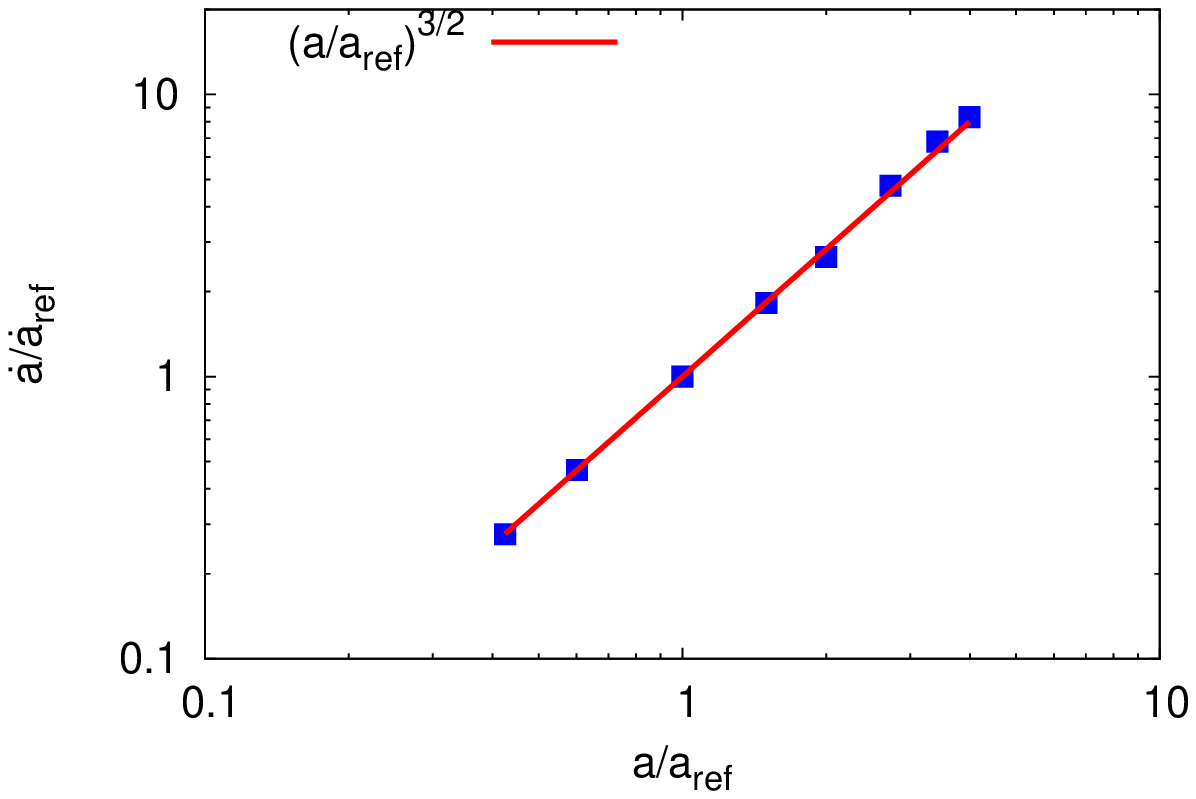}}
\caption{%
             Normalized migration speed $\dot{a}/\dot{a}_{\mathrm{ref}}$ vs.\
             the normalized (azimuthally averaged) surface density  
             $\Sigma/\Sigma_{\mathrm{ref}}$ (\textit{left}) and vs.\
             $a/a_{\mathrm{ref}}$ (\textit{right}). The surface density is sampled
             at $a_{1}-5.5\,R_{\mathrm{H},1}$. 
              All the calculations are run for an evolution time between $10^{4}$ 
              and $\gtrsim 2\times 10^{4}$ years. 
              To determine $\dot{a}$, first $a(t)$ is averaged in time over several tens 
              of years and then a linear fit to the data is performed using a base
              time span of one to several thousand years.
             }
\label{fig:adep1}
\end{figure*}
In the left panel of Figure~\ref{fig:adep1}, the migration velocity $\dot{a}$ of
the inner planet from calculations (symbols), normalized to the velocity from 
the reference model (Figure~\ref{fig:aref}), is plotted against the value 
of the normalized $\Sigma$. 
Here the value of the surface density is that at a distance of $5.5\,R_{\mathrm{H},1}$ 
from the (inner) planet's orbit and interior to it.
A function proportional to $\Sigma$ (\textit{solid line}) appears a reasonable 
approximation of  $k_{1}$ over the range of densities shown in the Figure. 
Hence, we will assume that $k_{1}(\Sigma)=\Sigma/\Sigma_{\mathrm{ref}}$,
where the density is sampled as stated above.

\begin{figure}[h!]
\centering%
\resizebox{\linewidth}{!}{%
\includegraphics{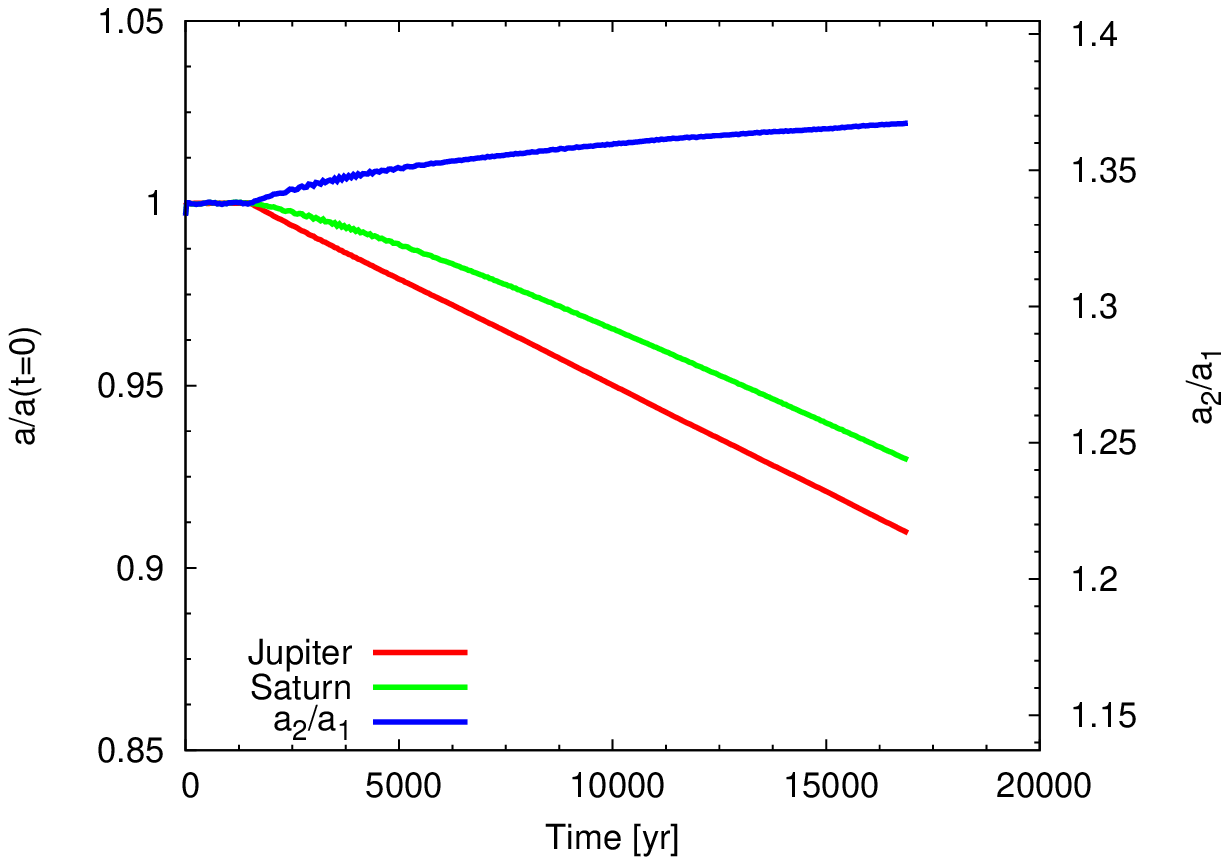}}
\caption{%
              Semimajor axes vs.\ time of a Jupiter--Saturn type system
              evolving in a 3D disk with $H/r=0.07$, $\alpha_{\mathrm{t}}=0.005$, 
              and an initial $\Sigma$ at $1\,\AU$ of $50\,\mathrm{g\,cm}^{-2}$.
              Each semimajor axis is normalized to its initial value. 
              The planets are subjected to disk torques after the first $1500$ years. 
              The planets undergo divergent inward migration.
             }
\label{fig:outmig}
\end{figure}
We note in passing that if the interior, and hence the exterior, planet is subjected
to a torque of the type given in Equation~(\ref{eq:TI}), the resonance may be broken since
the inner, more massive, planet may drift inward at larger speed than the outer,
less massive, planet. This is indeed observed in some calculations of relatively 
thick disks, as illustrated in Figure~\ref{fig:outmig} (see figure's caption for further details).

The form of function $k_{2}$ can be guessed following a similar 
line of argument. In their natural units of $a^{2}\Omega^{2}$ (times a mass), 
both torques $\mathcal{T}_{\mathrm{OS}}$ (Equation~\ref{eq:Tos}) and 
$\mathcal{T}_{\mathrm{I}}$ (Equation~\ref{eq:TI})
scale as $a^{2}\Sigma$ (the dependence on $H/a$ is considered later), 
which can be seen as a measure of the local disk mass. Therefore, if
$\Sigma$ was constant, $\dot{a}\propto a^{2}$ in units of $a\Omega$ 
(see Equation~\ref{eq:dadt_circ}) and thus $\dot{a}$ would scale as $a^{3/2}$.
In the right panel of Figure~\ref{fig:adep1}, the migration speed of
the inner planet from calculations (symbols), normalized to the reference 
migration speed, is plotted as a function of $a$, normalized to the semimajor 
axis in the reference model, $a_{\mathrm{ref}}$. Also in this case, the approximation
seems satisfactory (\textit{solid line}) and so we shall assume that 
$k_{2}(a)=(a/a_{\mathrm{ref}})^{3/2}$.

In Section~\ref{sec:ToC}, it was anticipated that the depth and width of
the density depletion produced by the exterior planet play a fundamental
role in determining magnitude and direction of the interior planet's migration,
and hence of the pair as a whole. Quantity $g$ (Equation~\ref{eq:gcon}) can
be used as a proxy to discriminate among the various situations, i.e.,
different combinations of  $H/r$ and $\alpha_{\mathrm{t}}$ (and $q$) that
may affect $\Sigma$. 
On account of the compact orbital configuration, since $q_{1}>q_{2}$
we have that  $g_{1}>g_{2}$ (unless $H/r$ and/or $\alpha_{\mathrm{t}}$
are rapidly varying with disk radius). If $g_{2}\gg 1$, then $g_{1}\gg 1$, and
the interior planet is likely subjected to a torque whose limit is the one-sided 
Lindblad torque in Equation~(\ref{eq:Tos}) and the two planets migrate outward. 
Otherwise, if  $g_{1}\ll 1$, then $g_{2}\ll 1$, and
migration will be dictated by a type~I torque (Equation~\ref{eq:TI})
and be directed inward. 

\begin{figure}[t!]
\centering%
\resizebox{\linewidth}{!}{%
\includegraphics{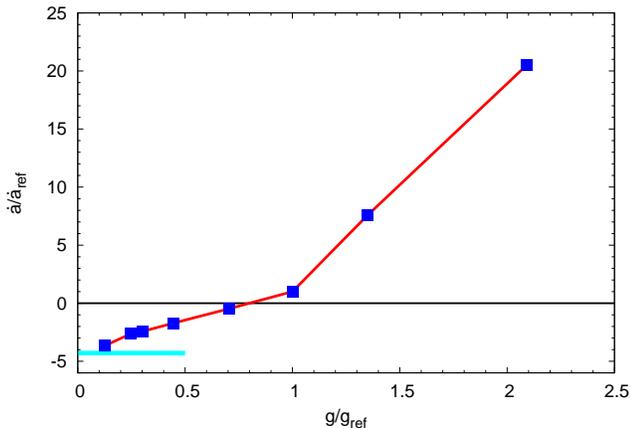}}
\caption{%
              Normalized migration speed $\dot{a}/\dot{a}_{\mathrm{ref}}$
              for various values of the ratio $g/g_{\mathrm{ref}}$, calculated
              for the exterior planet, according to Equation~(\ref{eq:ggref}).
             All models are run for an evolution time of between $6\times 10^{3}$ and 
             $1.2\times 10^{4}$ years. The thick horizontal line segment represents the
             type~I migration (see Equation~\ref{eq:TI})
             that the inner planet would be subjected to
             if the disk had a relative thickness $H/r=0.1$. 
             }
\label{fig:adepg}
\end{figure}
The migration velocity of the inner planet should depend 
on both $g_{1}$ and $g_{2}$. 
However, in the present context $g_{2}$ should have the larger
impact of the two and therefore, for the sake of simplicity, 
we assume that function $k_{3}$ depends only on $g_{2}$. In particular,
referring to the value of $g_{2}$ in the reference model (Figure~\ref{fig:aref})
as $g_{\mathrm{ref}}$ and indicating $g_{2}$ simply as $g$, $k_{3}$ will
be approximated as a function of $g/g_{\mathrm{ref}}$. 
Since it is unclear how the agreement between 2D and 3D calculations varies 
as a function of the disk thickness (it should likely worsen as $H$ increases), 
we only use 3D calculations to find an approximation to function $k_{3}$.
Figure~\ref{fig:adepg} shows the ratio $\dot{a}/\dot{a}_{\mathrm{ref}}$ obtained
from calculations for various values of the ratio $g/g_{\mathrm{ref}}$.
The thick horizontal line in the plot indicates the type~I migration speed
that would apply to the inner planet if $H/r=0.1$, corresponding to the value
used for left-most data point on the graph.
For reference, the normalized 
migration speed corresponding to the one-sided torque in Equation~(\ref{eq:Tos}) 
would be $\sim 90$.
The broken line is a
linear interpolation of the data, which will be used as a representation of $k_{3}$.

\subsection{Approximation of the 3:2 Resonant-orbit Migration Velocity}
\label{sec:ROM}

Summarizing the results of Section~\ref{sec:OMR}, we write the
migration speed of the interior planet as
\begin{equation}
\frac{d a}{d t}=\dot{a}_{\mathrm{ref}}%
\left(\frac{\Sigma}{\Sigma_{\mathrm{ref}}}\right)%
\left(\frac{a}{a_{\mathrm{ref}}}\right)^{3/2}%
k_{3}(g/g_{\mathrm{ref}}),
\label{eq:dotaJ}
\end{equation}
where the dimensionless function $k_{3}$ is obtained via linear interpolation
of the numerical data in Figure~\ref{fig:adepg}, 
i.e., the thick solid line in the Figure. 
The reference values in Equation~(\ref{eq:dotaJ}) are taken from
the reference model, corrected for 3D effects,
as discussed in Section~\ref{sec:OMR}:
$\dot{a}_{\mathrm{ref}}= 2.7\times 10^{-6}\,\AU\,\mathrm{yr}^{-1}$ for 
$\Sigma_{\mathrm{ref}}=42\,\mathrm{g\,cm}^{-2}$ and $a_{\mathrm{ref}}=1.57\,\AU$.
Recall that both $\Sigma$ and  $\Sigma_{\mathrm{ref}}$ are evaluated at
a distance of $5.5\,R_{\mathrm{H},1}$ interior of the inner planet's orbit.
The argument of function $k_{3}$ is given by
\begin{equation}
\frac{g}{g_{\mathrm{ref}}}=\sqrt{\frac{\alpha_{\mathrm{t,ref}}}{\alpha_{\mathrm{t}}}}%
\left(\frac{H_{\mathrm{ref}}}{H}\right)
\left(\frac{R_{\mathrm{H}}}{\widetilde{\Delta}}\right)^{3/2},
\label{eq:ggref}
\end{equation}
where $\alpha_{\mathrm{t,ref}}=0.005\sqrt{1\,\AU/a_{\mathrm{ref}}}$, 
$(H/a)_{\mathrm{ref}}=0.04$, and
$\widetilde{\Delta}=\max{(H,R_{\mathrm{H}})}$.
Recall that here parameters $g$ and $g_{\mathrm{ref}}$ are
computed from Equation~(\ref{eq:gcon}) applied to the exterior planet.
For a mass ratio $q$ different from that of the reference model, 
the right-hand side of Equation~(\ref{eq:ggref}) should be multiplied 
by $q/q_{\mathrm{ref}}$.
Equation~(\ref{eq:dotaJ}), without the correction due to 3D effects, 
is also in reasonable 
agreement with the results presented by \citet{pierens2011} in their Figure~21, 
for a disk of $0.4\,M_{\mathrm{J}}$ within $1.5\,\AU$ 
($\Sigma\sim 500\,\mathrm{g\,cm}^{-2}$ at $1\,\AU$).
As explained above,
the exterior planet's orbit may not always be resonant with that of the
interior planet, when migration is inward (see Figure~\ref{fig:outmig}).
Nonetheless, for the outer planet's orbit we set $a_{2}=(3/2)^{2/3}a_{1}$.

Equation~(\ref{eq:dotaJ}) predicts stalling points 
(where $k_{3}$, and hence $\dot{a}$, is $\approx 0$)
at $g_{0}\approx 0.8\,g_{\mathrm{ref}}$, which will be regarded as a
nominal value. 
But notice that since the approximation to $k_{3}$ is sampled at a
limited number of points, its zero could be located 
at a somewhat different abscissa, yet it is located between 
$0.7\,g_{\mathrm{ref}}$ and $g_{\mathrm{ref}}$. 
In principle, there could be multiple stalling points in a disk 
(if disk properties are not monotonic), which would 
represent locations of stable equilibrium since they are convergent radii 
for the planets' semimajor axes. The argument, however, is based on the 
assumption that the 3:2 commensurability is preserved even for inward migration.
This is not true in general (as illustrated, for example, in Figure~\ref{fig:outmig}), 
but it further requires that the inward migration of the exterior planet be
faster than that of the interior planet and that the condition for gap overlap
(see condition \textit{\i\i}) in Section~\ref{sec:ToC}) be satisfied.
When these conditions are not met,
the pair of planets can migrate past a stalling point, toward the star. 
This may happens, for example, if disk temperature and viscosity 
are low enough so that both planets open up a gap and drift according to 
\textit{type~II} migration before capture into resonance occurs.

\begin{figure}[t!]
\centering%
\resizebox{\linewidth}{!}{%
\includegraphics{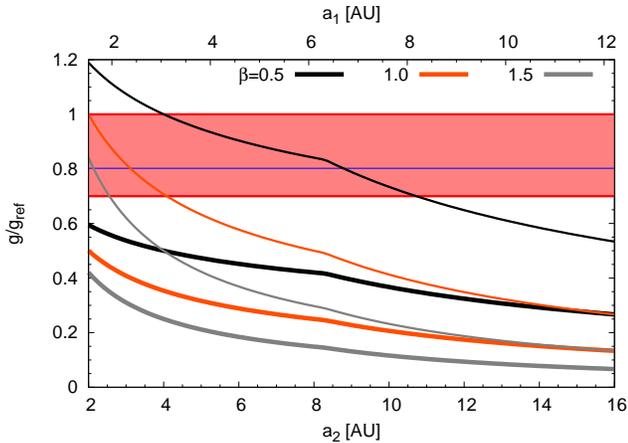}}
\caption{%
             Ratio $g/g_{\mathrm{ref}}$ (Equation~\ref{eq:ggref}) as a function 
             of the exterior (\textit{bottom axis}) and interior (\textit{top axis}) planet's
             orbital radius. Here, $H/r$ is approximated by Equation~(\ref{eq:HRsemp}).
             The shaded area indicates the region where $\dot{a}_{1}$ changes sign:
             $\dot{a}_{1}>0$ ($\dot{a}_{1}<0$) above (below) the shaded area.  
             The horizontal line within the boundaries of the shaded area is the nominal 
             value for stalling migration (see the text). The various curves represent 
             $g/g_{\mathrm{ref}}$ for different ``slope'' parameters, $\beta$ (see the legend).
             Thin and thick curves correspond to $\nu_{1}=4\times 10^{-6}$ and 
             $1.6\times 10^{-5}\,r^{2}_{1}\,\Omega_{1}$, respectively.
             }
\label{fig:gg0}
\end{figure}
Some constraints on the range of outward migration predicted by Equation~(\ref{eq:dotaJ})
can be derived for the disk models discussed in Section~\ref{sec:MR}. 
The disk thickness, $H$, is affected by internal (viscous) heating typically 
over the first few million years of evolution (see Figure~\ref{fig:Svst}, \textit{left}).
If we neglect that source of heating, $H$ can be approximated by using 
Equation~(\ref{eq:HRsemp}) and then Equation~(\ref{eq:ggref}) can provide
a rough measure of a disk's radial range over which the planets
can drift away from the star, as shown in Figure~\ref{fig:gg0}. The addition of
internal heating would raise the value of $H$, hence reducing the ratio
$g/g_{\mathrm{ref}}$, which suggests that outward migration may not be
activated in the warm interiors of a young disk. 
The radial region over which a curve extends above or, to some degree, inside
the shaded area is favorable to outward migration. The intersection of a curve
with the horizontal line in the shaded area gives the nominal radius of the stalling
point of each planet (see above).
According to the plot, outward migration of a Jupiter-mass planet locked 
in the 3:2 mean motion resonance with a Saturn-mass planet cannot 
proceed beyond $\sim 7\,\AU$ (\textit{top x-axis}). 
The numerical experiments discussed 
in Section~\ref{sec:LTM} are broadly consistent with this prediction.

\begin{figure}[t!]
\centering%
\resizebox{\linewidth}{!}{%
\includegraphics{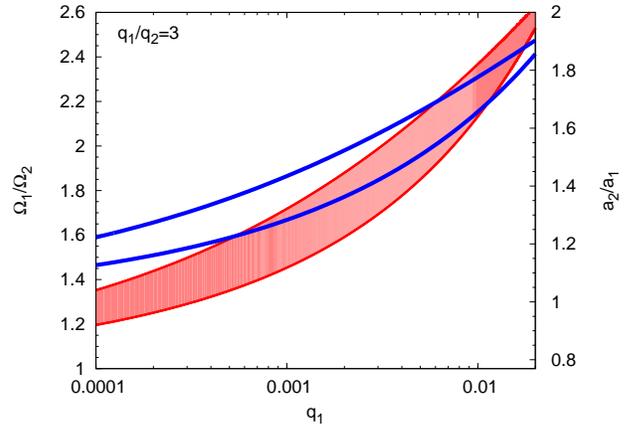}}
\caption{%
             Shaded area represents orbital frequency ratios,
             as a function of the interior planet mass,
             which may possibly activate outward migration of a pair of planets 
             whose mass ratio is such that $q_{1}/q_{2}=M_{1}/M_{2}=3$. 
             The upper boundary of the shaded area is given by condition 
             \textit{\i\i}) of Section~\ref{sec:ToC}. The lower boundary is given 
             by a Hill stability criterion for circular orbit planets 
             \citep[Equation~23 of ][]{gladman1993}, in absence of gas.
             It is assumed that parameter $g$, defined in Equation~(\ref{eq:gcon})
             and applied to the exterior planet, is large enough to allow for sufficient
             gas depletion. The area bounded by the thicker lines represents this
             condition in terms of the ratio $a_{2}/a_{1}$ (labeled on the right vertical axis).
             }
\label{fig:mass_con}
\end{figure}
The validity of Equation~(\ref{eq:dotaJ}) for planet-to-star mass ratios
different from those adopted here ($q_{1}=M_{1}/\Ms=9.8\times 10^{-4}$ 
and $q_{2}=M_{2}/\Ms=2.9\times 10^{-4}$) was not investigated. 
Parameter $g$ (Equation~\ref{eq:gcon}) is $\propto q_{2}$ if $\Rhill<H$ 
and $\propto \sqrt{q_{2}}$ otherwise.
Therefore, a faster outward migration might be expected as $q_{2}$
increases, provided that the ratio $q_{1}/q_{2}$ stays roughly constant.
Condition \textit{\i\i}) of Section~\ref{sec:ToC}, along with the requirement
of Hill stability, suggests that, as $q_{2}$ increases beyond $\sim 0.001$, 
the 2:1, rather than the 3:2, commensurability may be available 
to activate outward migration for a ratio $q_{1}/q_{2}\approx 3$, 
in accord with the findings of \citet{crida2009}. 
This is schematically illustrated in Figure~\ref{fig:mass_con} (see figure's
caption for further details). 
As the ratio $q_{1}/q_{2}$ approaches $1$, the shaded area in the graph 
shifts upward.

\section{Long-Term 3:2 Resonant-orbit Migration of Jupiter and Saturn}
\label{sec:LTM}

The results from the disk models of Section~\ref{sec:MR} can be combined
with the results of Section~\ref{sec:ROM} to study the migration of a resonant-orbit
pair of planets with masses corresponding to those of Jupiter and Saturn.
Here we shall assume that, by a time $\tau_{p}$, both planets have fully formed, 
i.e., they have reached their final masses, and their orbits have become locked 
in the 3:2 mean motion resonance.
Planet formation calculations of a giant planet via core nucleated accretion
\citep[e.g.,][]{hubickyj2005,alibert2005b,lissauer2009,naor2010,mordasini2011} 
indicate that the formation time of Jupiter is greater than $\sim 1\,\mathrm{Myr}$,
although this timescale is affected by the formation of the solid core and thus
by the distance where the core forms. The formation of Saturn, the exterior planet, 
seemingly takes somewhat longer \citep[see, e.g.,][]{pollack1996,dodson2008,benvenuto2009}. 
In addition to the formation time, there is the time required by convergent orbital migration 
to bring the planets into mean motion resonance (which is also included in $\tau_{p}$).
Assuming that $\tau_{p}$ is determined mainly by the formation time and 
for lack of better constraints, we choose three reference times $\tau_{p}$ of 
$1$, $2$ and $3\,\mathrm{Myr}$ \citep[e.g.,][]{kenyon2009,bromley2011}.

It is important to bear in mind that longer timescales are possible, whereas
it is unclear if shorter timescales are feasible. In fact, there are also 
observational constraints suggesting that Jupiter formed after several million
years \citep[see][and references therein]{scott2006}.

The disk evolution models presented in Section~\ref{sec:MR} are recalculated, starting
from time  $\tau_{p}$ and using Equations~(\ref{eq:dotaJ}) and (\ref{eq:ggref}), 
to integrate the orbital radius of the interior planet.
The orbit of the exterior planet is constrained by the 2:3 resonant-orbit requirement
with the interior planet (see Section~\ref{sec:ROM}). 
In particular, the disk models provide the quantities 
$\Sigma$, $\alpha_{\mathrm{t}}$, and $H$, 
used in Equations~(\ref{eq:dotaJ}) and (\ref{eq:ggref}), as they evolve over 
time.
The effects of the torques exerted by the planets on the disk are not taken 
into account in the 1D models because, for current purposes, the feedback 
of the tidal field on the disk and its effects on the migration rates 
are included in the 2D and 3D calculations discussed above.
The mass of both planets is constant, but we will contemplate the impact of gas 
accretion in Section~\ref{sec:PPoA}.
Since $\tau_{p}\ge1\,\mathrm{Myr}$, useful disk models are those for which
$\tau_{\mathrm{D}}>1\,\mathrm{Myr}$ (and $\lesssim 20\,\mathrm{Myr}$).

\begin{figure*}[t!]
\centering%
\resizebox{\linewidth}{!}{%
\includegraphics[clip,bb=55 70 400 302]{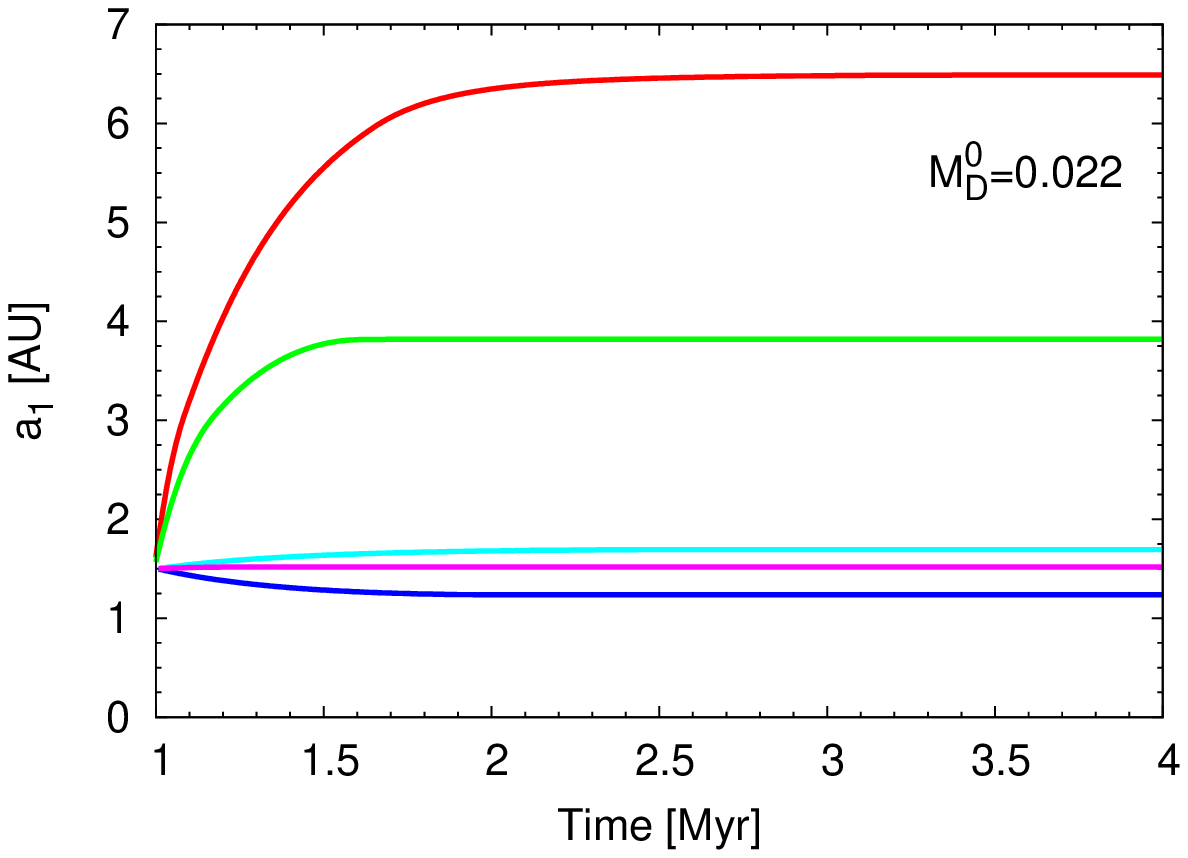}%
\includegraphics[clip,bb=80 70 400 302]{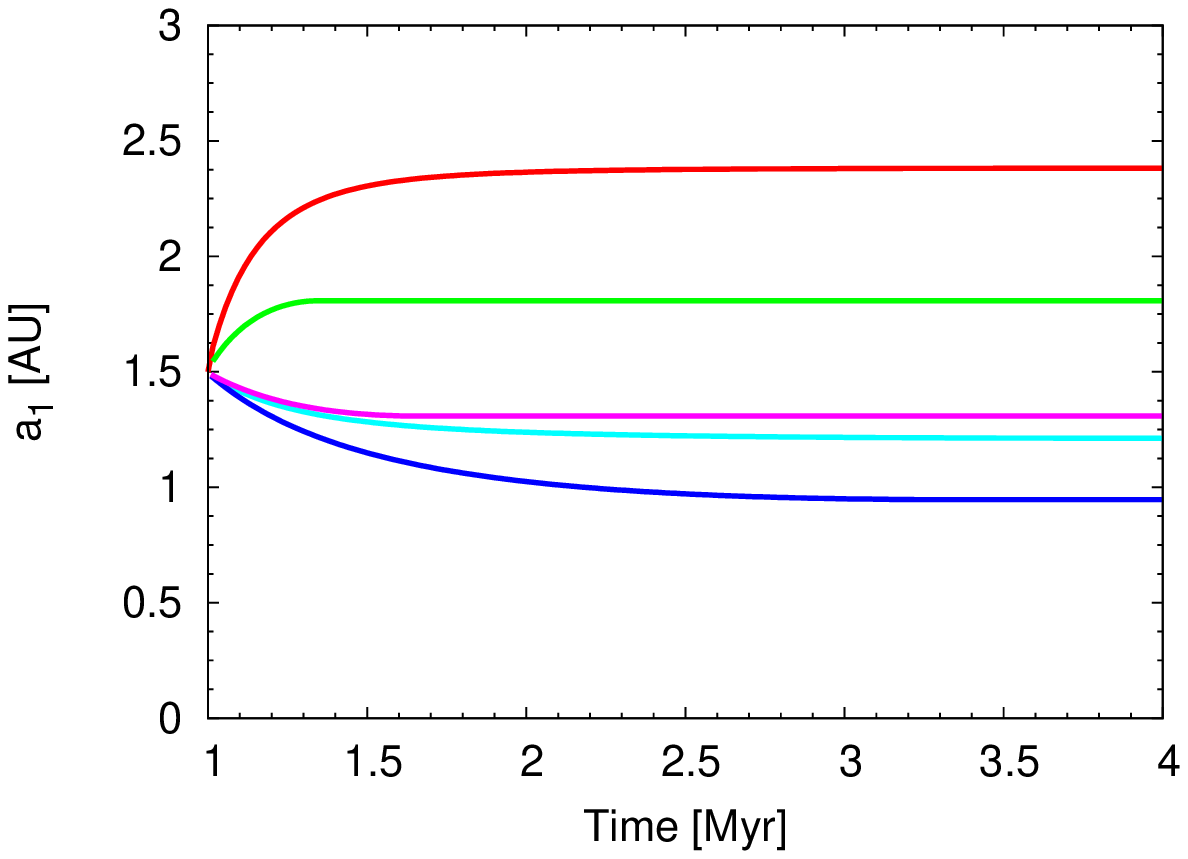}%
\includegraphics[clip,bb=80 70 400 302]{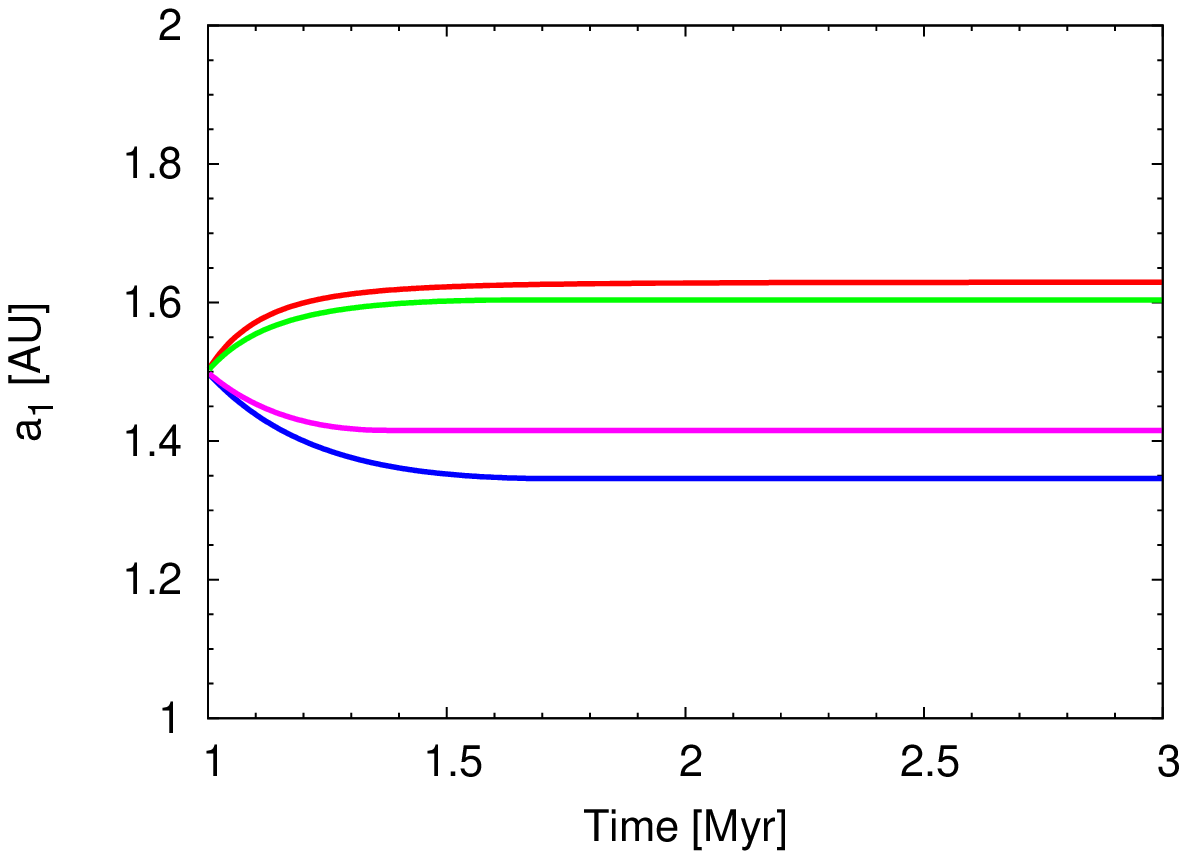}}
\resizebox{\linewidth}{!}{%
\includegraphics[clip,bb=55 70 400 302]{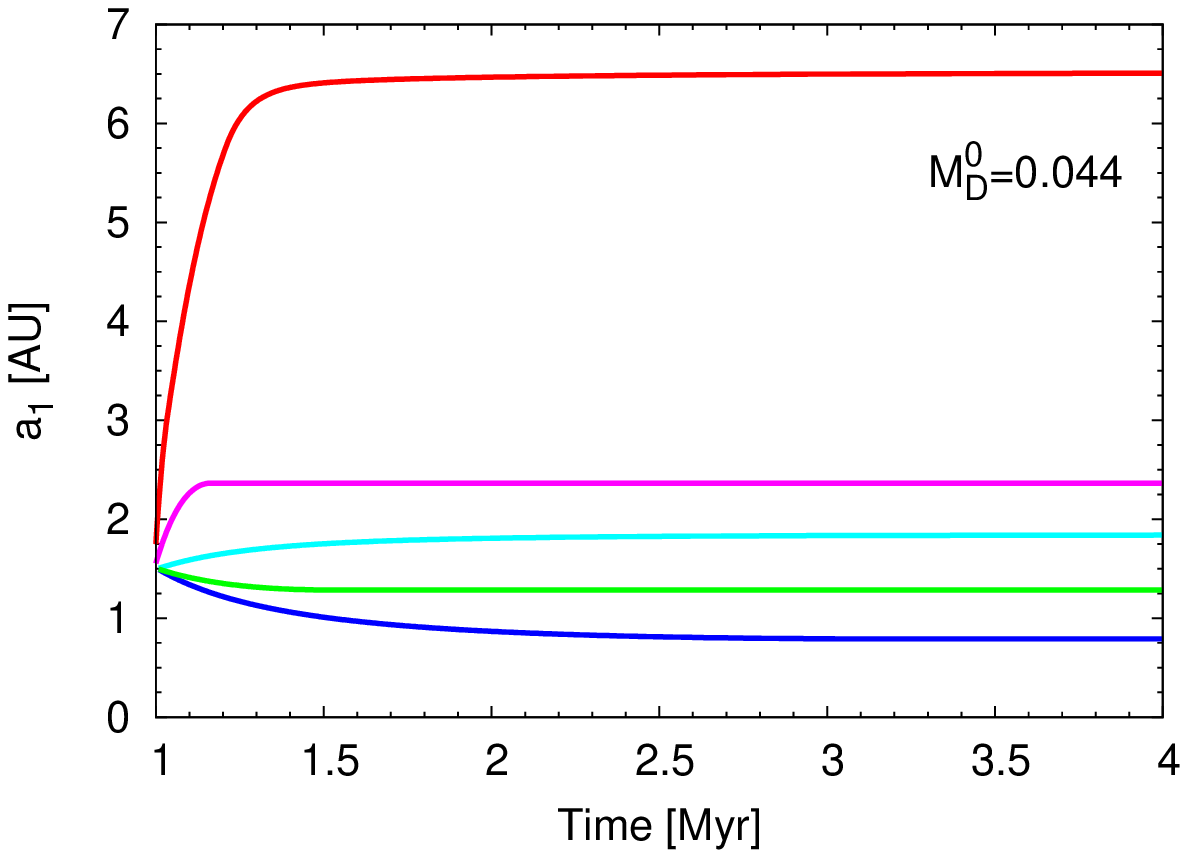}%
\includegraphics[clip,bb=80 70 400 302]{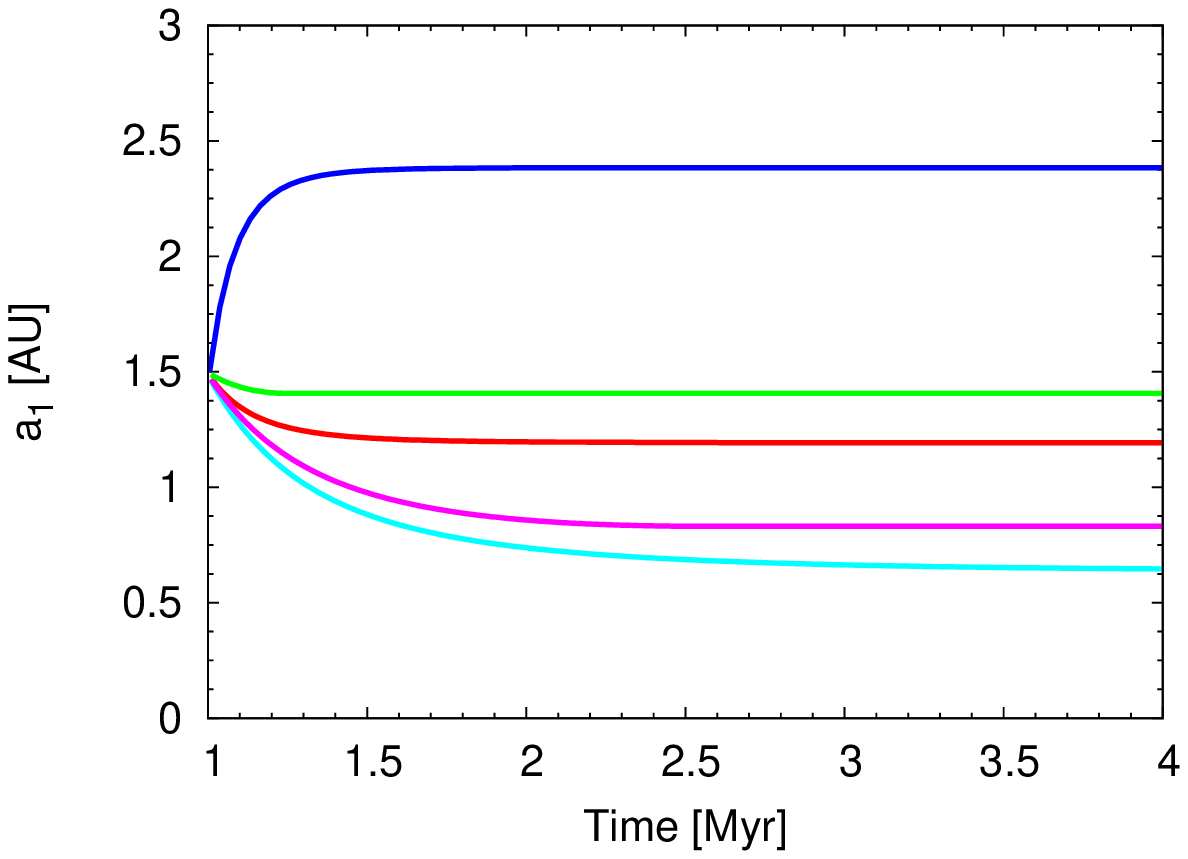}%
\includegraphics[clip,bb=80 70 400 302]{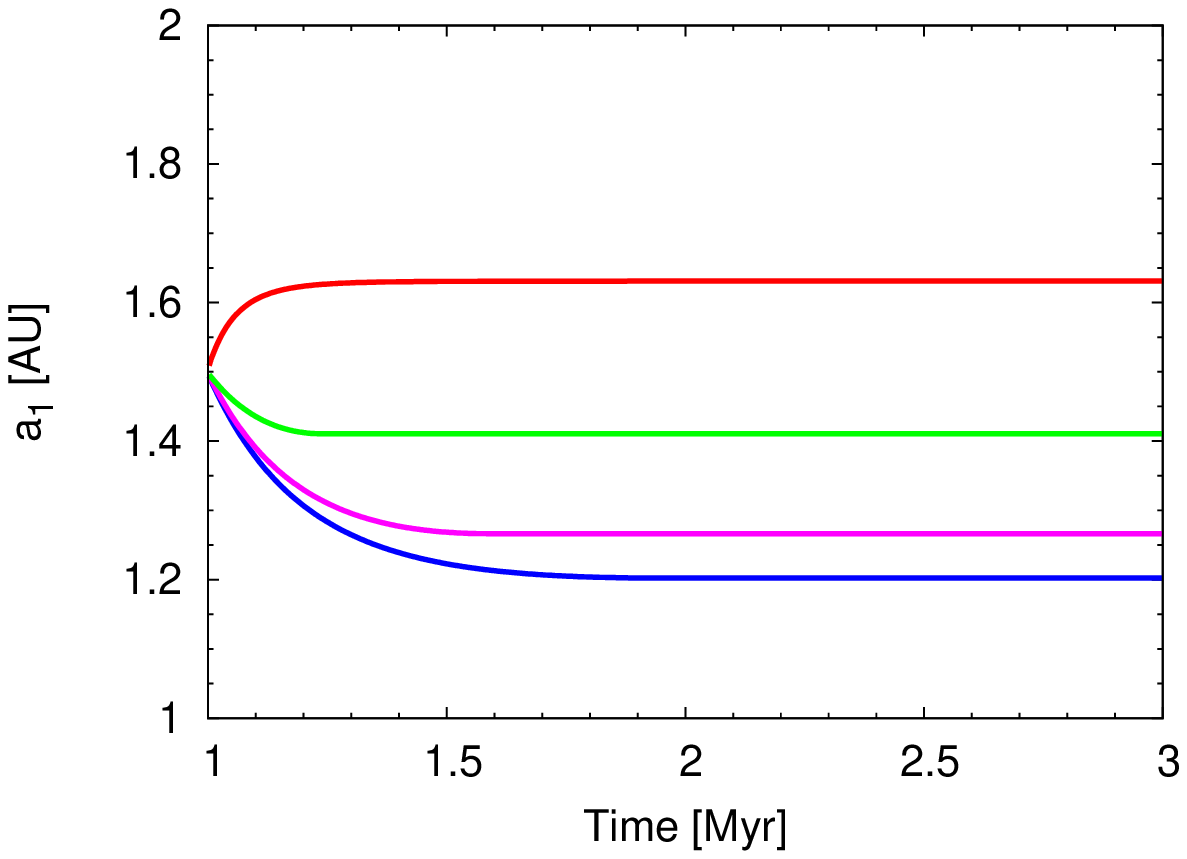}}
\resizebox{\linewidth}{!}{%
\includegraphics[clip,bb=55 49 400 302]{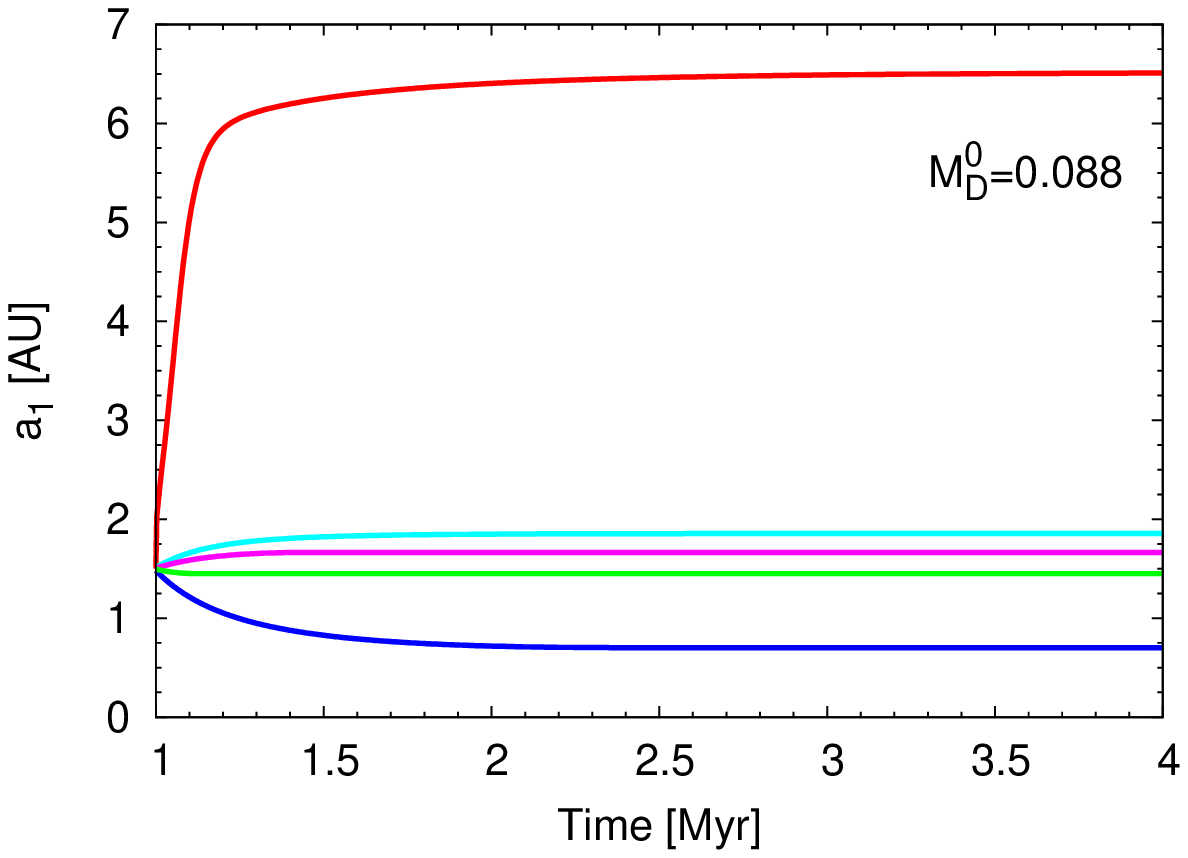}%
\includegraphics[clip,bb=80 49 400 302]{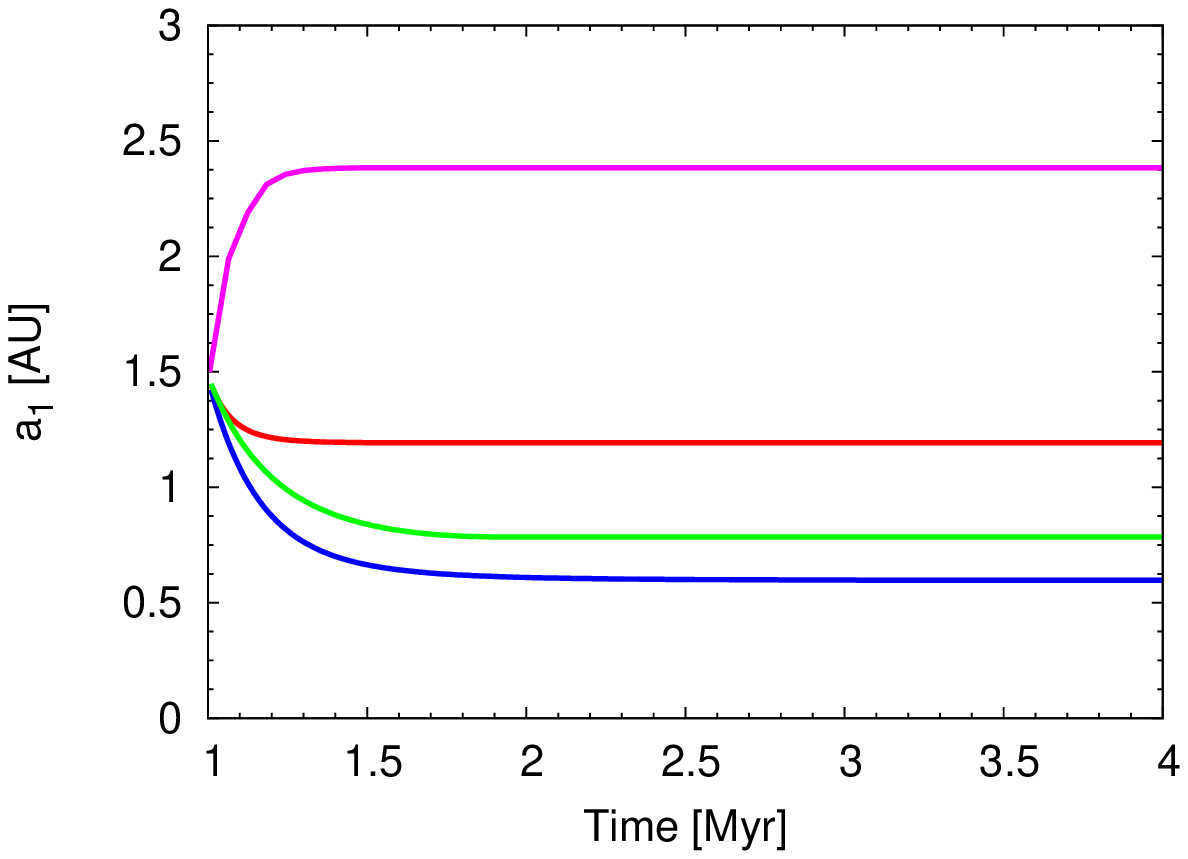}%
\includegraphics[clip,bb=80 49 400 302]{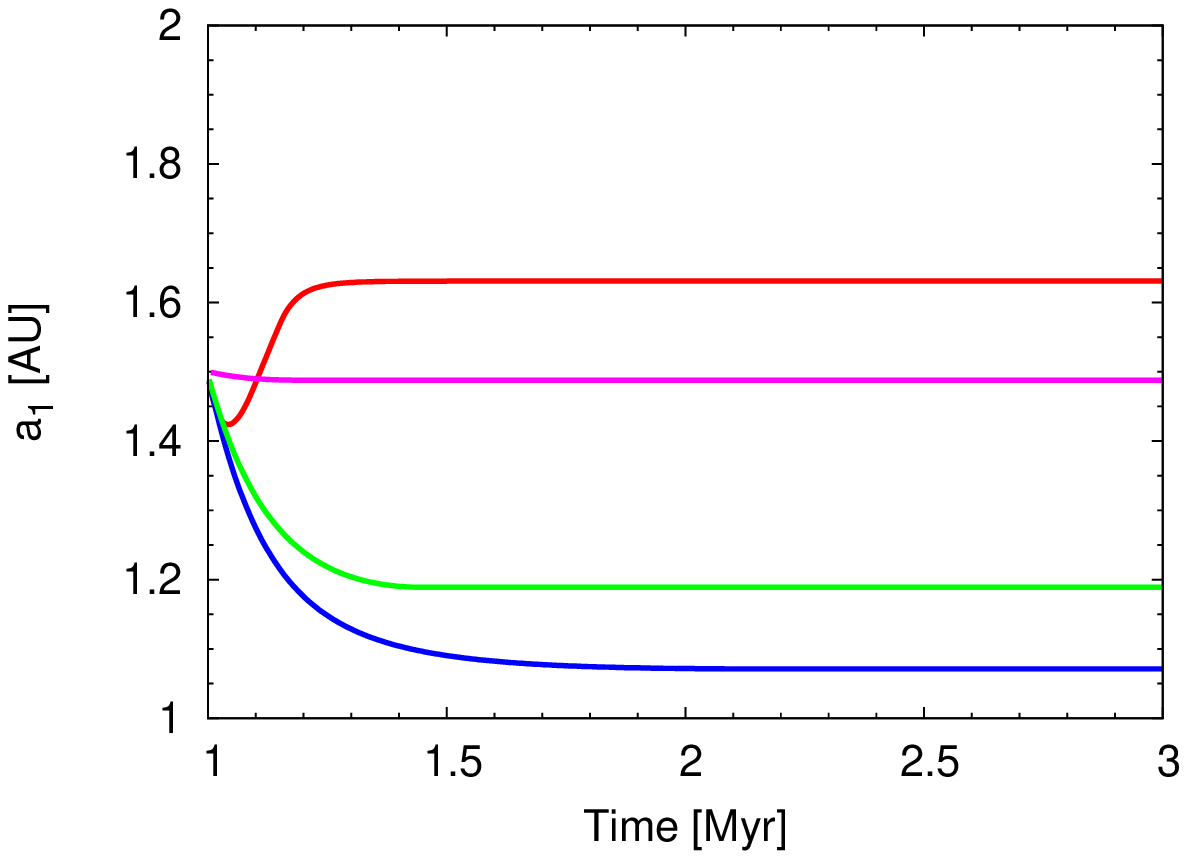}}
\caption{%
             Evolution of the orbital radius ($a_{1}$) of the interior planet computed 
             by integrating Equation~(\ref{eq:dotaJ}) along with the disk models of 
             Section~\ref{sec:MR} and assuming a formation timescale
             $\tau_{p}=1\,\mathrm{Myr}$ (see the text).
             The initial disk mass, in units of $\Ms$, is indicated in the top-right
             corner of the left panels. The initial ``slope'' parameter of $\Sigma$ is
             $\beta=1/2$ (\textit{left}), $1$ (\textit{center}), and $3/2$ (\textit{right}).
             The curves in each panel refer to different values of $\nu_{1}$ and 
             $f_{41}$ (see Table~\ref{tbl:ainf_1} to identify each curve).
             A complete list of the asymptotic values, $a_{\infty}$, is reported in 
             Table~\ref{tbl:ainf_1}. It is assumed that $a_{2}=(3/2)^{2/3}a_{1}$.
             }
\label{fig:a1vst}
\end{figure*}
\begin{deluxetable*}{crrrrrrrrrrrrr}
\tablecolumns{13}
\tablewidth{0pc}
\tablecaption{Asymptotic Orbital Radii for $\tau_{p}=1\,\mathrm{Myr}$\label{tbl:ainf_1}}
\tablehead{
\colhead{}    & \colhead{}    &
\multicolumn{11}{c}{$a_{\infty}$\tablenotemark{a}} \\
\cline{3-13}
\colhead{}    & \colhead{}    &
\multicolumn{3}{c}{$\beta\tablenotemark{b}=1/2$} 
& \colhead{}    & 
\multicolumn{3}{c}{$\beta=1$}
& \colhead{}    & 
\multicolumn{3}{c}{$\beta=3/2$}\\ 
\cline{3-5} \cline{7-9} \cline{11-13}
\colhead{$M^{0}_{\mathrm{D}}/\Ms$}   & \colhead{$\nu_{1}$\tablenotemark{c}}&
\colhead{$f_{41}\tablenotemark{d}=10$} & \colhead{$100$} & \colhead{$1000$} &
\colhead{} & \colhead{$1$} & \colhead{$10$} & \colhead{$100$} &
\colhead{} & \colhead{$0.1$} & \colhead{$1$} & \colhead{$10$}
}
\startdata
$0.022$ & $  4\times 10^{-6}$     & $6.49$ & $3.82$ & $1.50$   
                                                &  & $2.38$ & $2.35$ & $1.81$   
                                                &  & $1.63$ & $1.63$ & $1.60$ \\ 
$0.022$ & $  8\times 10^{-6}$     & $1.70$ & $1.52$ & $1.50$
                                                &  & $1.21$ & $1.30$ & $1.50$
                                                &  & $1.35$ & $1.41$ & $1.50$ \\
$0.022$ & $1.6\times 10^{-5}$    & $1.23$ & $1.50$ & $1.50$
                                                &  & $0.95$ & $1.31$ & $1.50$
                                                &  & \nodata& \nodata&\nodata \\
\cline{2-13}
$0.044$ & $  4\times 10^{-6}$     & $6.51$ & $6.50$ & $2.37$ 
                                                &  &  \nodata & $2.38$ & $2.37$
                                                &  & $1.63$ & $1.63$ & $1.63$ \\
$0.044$ & $  8\times 10^{-6}$     & $1.84$ & $1.71$ & $1.50$ 
                                                &  & $1.19$ & $1.20$ & $1.30$
                                                &  & $1.20$ & $1.27$ & $1.41$ \\
$0.044$ & $1.6\times 10^{-5}$    & $0.79$ & $1.28$ & $1.50$
                                                &  & $0.64$ & $0.83$ & $1.41$
                                                &  & $1.50$ &  \nodata &  \nodata \\
\cline{2-13}
$0.088$ & $  4\times 10^{-6}$     & \nodata &  $6.51$ & $6.50$
                                                &  & \nodata &  \nodata & $2.38$
                                                &  & $1.63$ & $1.63$ & $1.63$ \\
$0.088$ & $  8\times 10^{-6}$     &  \nodata&  $1.86$ & $1.66$
                                                &  & $1.19$ & $1.19$ & $1.19$
                                                &  & $1.07$ & $1.10$ & $1.19$ \\
$0.088$ & $1.6\times 10^{-5}$    & \nodata &  $0.70$ & $1.45$
                                                &  & $0.60$ & $0.60$ & $0.79$
                                                &  & $1.49$ & $1.50$ & \nodata
\enddata
\tablenotetext{a}{Asymptotic value of the semimajor axis, in \AU, of the interior planet's orbit.}
\tablenotetext{b}{Initial ``slope'' of the disk's surface density.}
\tablenotetext{c}{Kinematic viscosity at $r_{1}=1\,\AU$ in units of 
$r^{2}_{1}\,\Omega_{1}=(G\,\Ms\,r_{1})^{1/2}$.}
\tablenotetext{d}{Rate of EUV ionizing photons emitted by the star in units of 
$10^{41}\,\mathrm{s^{-1}}$ (see Equation~\ref{eq:Speg}).}
\end{deluxetable*}
The evolution of $a_{1}$, the interior planet's orbital radius, is illustrated
in Figure~\ref{fig:a1vst} for selected cases (see figure's caption for details). 
The complete list of the asymptotic values of $a_{1}$, $a_{\infty}$,
is reported in Table~\ref{tbl:ainf_1} for $\tau_{p}=1\,\mathrm{Myr}$.
In some cases, the orbital radius remains nearly unchanged.
This happens as a result of $\Sigma$ (around $r\sim a_{1}$) 
being too low, at time $t=1\,\mathrm{Myr}$, 
for any significant amount of angular momentum to be transferred to/from the disk.
Obviously, this result holds for $\tau_{p}>1\,\mathrm{Myr}$.

The outcomes of Figure~\ref{fig:a1vst}  and Table~\ref{tbl:ainf_1} can be
interpreted with the aid of Figure~\ref{fig:gg0}. 
For the highest viscosity regime, outward migration is not activated, 
and both planets migrate inward regardless of other disk parameters.
At the intermediate viscosity, $\dot{a}_{1}$ may be positive, but the nominal
stalling radius is within $2\,\AU$, hence the interior planets may not proceed
beyond this distance. The lowest viscosity regime offers the widest range of
outward migration, resulting in nominal stalling radii of $\approx 6.5\,\AU$
($\beta=1/2$), $\approx 2.4\,\AU$ ($\beta=1$), and $\approx 1.6\,\AU$
($\beta=3/2$). Because of the variation of $\nu$ with
radius, as the density steepens, the nominal stalling radius moves inward.
At even smaller viscosity, migration of the resonant pair may proceed to
larger distances, but in this case there are at least two possible issues 
that may arise. One is related to the disk lifetime, which increases as 
$\nu$ decreases (see Table~\ref{tbl:tD}). The other is related to the
mode of migration of the exterior planet prior to resonance capture,
which may transition to type~II at low enough viscosity, hence
convergent migration toward the interior planet may be compromised
(see Section~\ref{sec:21}).

Figure~\ref{fig:gg0} can also assist in extending the results illustrated 
in Figure~\ref{fig:a1vst},  and reported Table~\ref{tbl:ainf_1}, to different 
initial orbital radii. Assuming that viscous heating does not represent
a major source term in the energy budget, Equation~(\ref{eq:EEq}),
a pair of planets that become locked in the 3:2 mean motion resonance
inside the stalling radii (of each planet) will migrate toward those locations. 
Whether or not the planets may reach those radii depends on the
gas density level in the disk. 
At $\nu_{1}=4\times 10^{-6}\,r^{2}_{1}\,\Omega_{1}$, 
models for which $a_{\infty}\approx 6.5\,\AU$ for $\beta=1/2$, 
or $\approx 2.4\,\AU$ for $\beta=1$, or $\approx 1.6\,\AU$ for $\beta=3/2$,
have reached their stalling radii.
If the pair becomes locked into resonance
outside the stalling radii, the planets will converge toward, or transit across,
them depending on the disk conditions (see discussion in Section~\ref{sec:ROM}).

\begin{figure*}[t!]
\centering%
\resizebox{\figlen}{!}{%
\includegraphics{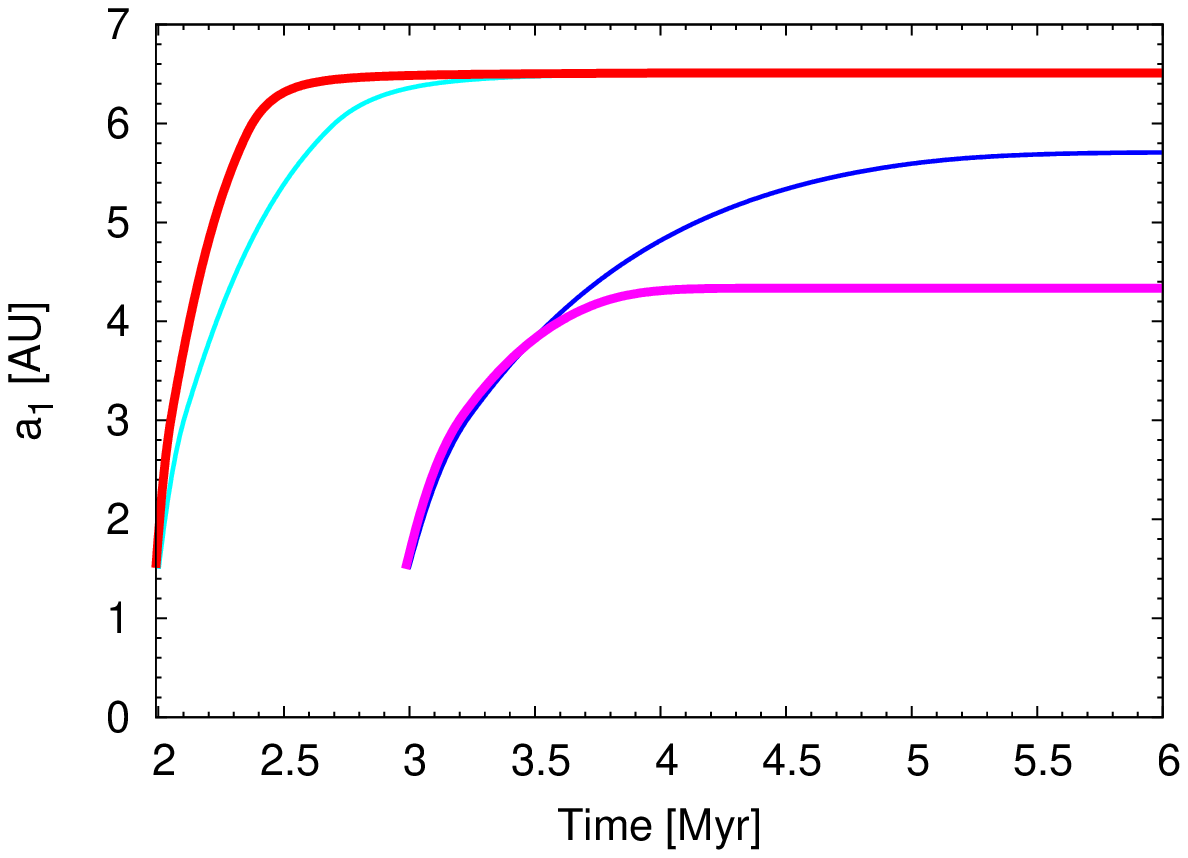}%
\includegraphics{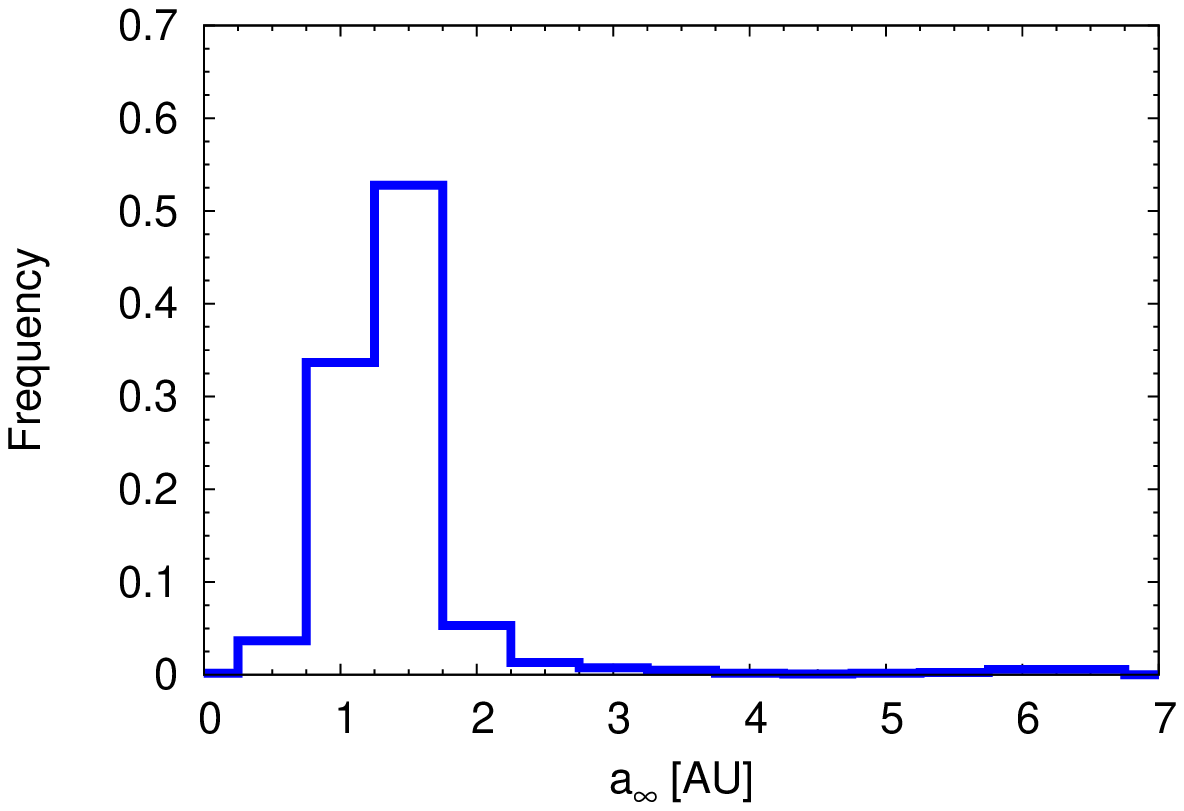}}
\caption{%
             \textit{Left:}
             orbital radius of the interior planet vs.\ time, assuming
             a formation timescale
             $\tau_{p}=2\,\mathrm{Myr}$ and $3\,\mathrm{Myr}$.
             Thin and thick curves represent, respectively, models with 
             an initial disk mass 
             $M^{0}_{\mathrm{D}}=0.044$ ($f_{41}=10$) and $0.088\,\Ms$
             ($f_{41}=100$). In all cases, parameters are $\beta=1/2$ and 
             $\nu_{1}=4\times 10^{-6}\,r^{2}_{1}\,\Omega_{1}$.
             For $\tau_{p}=2$ and $3\,\mathrm{Myr}$,
             these are the models that show the longest range of outward
             migration among those listed in Table~\ref{tbl:ainf_1}.
             \textit{Right:}
             histogram of the asymptotic orbital radius of the interior planet
             obtained by varying randomly the parameters $\nu_{1}$, $f_{41}$,
             and $\tau_{p}$. See the text for further details.
             }
\label{fig:a1_tau}
\end{figure*}
Among the sets of parameters listed in Table~\ref{tbl:ainf_1},
only five are compatible with the outward migration of the interior
planet beyond $\sim 5\,\AU$ (when resonance locking occurs
at $\sim 1\,\AU$). Only two sets of parameters remain
compatible with this requirement if the formation timescale is
$\tau_{p}=2\,\mathrm{Myr}$, and only one 
if $\tau_{p}=3\,\mathrm{Myr}$, as illustrated in the left panel of
Figure~\ref{fig:a1_tau} (see figure's caption for further details). 

In order to derive a distribution of the asymptotic orbital radii
of the interior planet, $a_{\infty}$, for each pair of values 
$(M^{0}_{\mathrm{D}},\beta)$ listed in Table~\ref{tbl:ainf_1},
$\nu_{1}$ and $f_{41}$ are varied randomly between
the corresponding minimum and maximum values reported
in the table, and $\tau_{p}$ is varied randomly between
$1$ and $3\,\mathrm{Myr}$. A total of over $1200$ models 
were computed and the histogram of the results is shown 
in the right panel of Figure~\ref{fig:a1_tau}.
Overall, there is a $97$\% probability that the interior
planet will achieve an asymptotic radius $a_{\infty}\lesssim 3\,\AU$
and a $98$\% probability that $a_{\infty}\lesssim 4\,\AU$.

\section{Gas Accretion and Planet Growth}
\label{sec:PPoA}

As mentioned in Section~\ref{sec:ToC}, the first condition necessary
to activate the outward migration mechanism of resonant-orbit planets
is that the interior planet's mass must exceed that of the exterior planet.
In the limit of equal mass planets, one expects the outer Lindblad 
torque exerted on the exterior planet to overcome the inner Lindblad 
torque exerted on the interior planet \citep[see, e.g.,][]{morbidelli2007}. 
Several outcomes are then possible, including breaking of the resonance, 
scattering, and inward type~II migration of both planets.

The neglect of gas accretion, especially on the exterior planet, represents 
possibly the most serious limitation of this mechanism.  
Hydrodynamical calculations
can provide maximum, or disk-limited, gas accretion rates for such planets. 
Although they do not necessarily represent the actual accretion rates, 
formation models of Jupiter 
\citep{lissauer2009} indicate that, once runaway accretion begins, a giant 
planet does grow at a disk-limited accretion rate. 
In a disk with $H/r\lesssim 0.05$ and $\alpha_{\mathrm{t}}\lesssim 0.005$,
an \emph{isolated} Saturn-mass planet may accrete gas at a rate a few 
times as large as that of a Jupiter-mass planet 
at the same location in the disk \citep[see][and references therein]{lissauer2009}.
In a disk with $H/r\gg 0.05$ or $\alpha_{\mathrm{t}}\gg 0.005$,
these rates would be comparable \citepalias{gennaro2008}.
Even assuming the same accretion rate for both planets, $\dot{M}_{p}$,
the initial growth time, $M_{p}/\dot{M}_{p}$, of Saturn is shorter
than that of Jupiter hence Saturn may approach the mass of Jupiter 
more or less quickly, depending on the local values of $\Sigma$ and $H/r$.
Furthermore, there is no obvious reason as to why Jupiter and Saturn should
accrete gas at a disk-limited rate and then suddenly stop accreting despite
the continuing supply of gas from the disk.

\begin{figure}[t!]
\centering%
\resizebox{\linewidth}{!}{%
\includegraphics{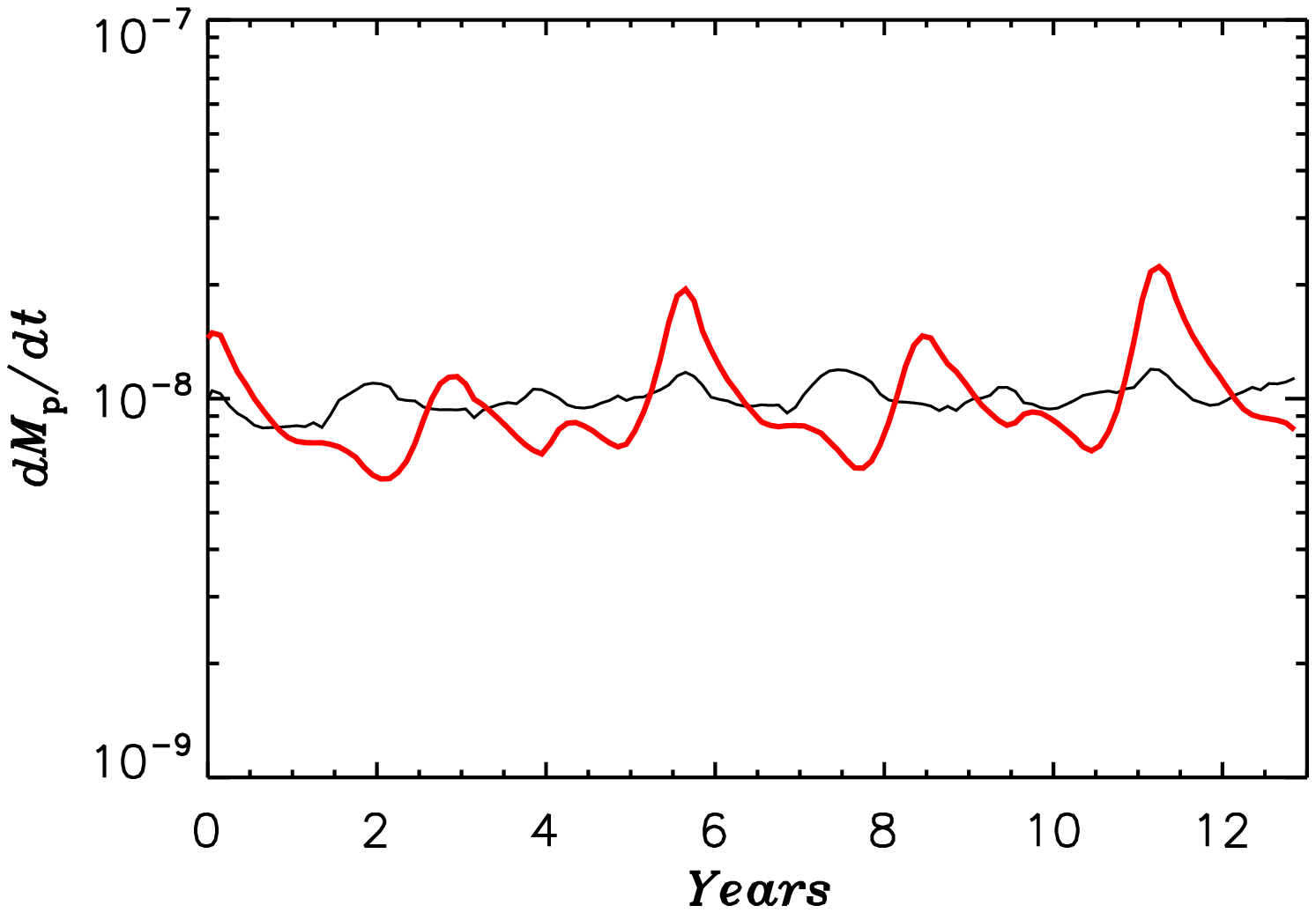}}
\caption{%
              Disk-limited gas accretion rates, in units of $\Ms\,\mathrm{yr}^{-1}$, of the
              interior (\textit{thin curve}) and exterior (\textit{thick curve}) planets.
              The pair is locked in the 3:2 mean motion resonance and
              the year count starts after about $6700$ years of evolution.
              The plot shows only a small time interval to highlight the accretion modulation.
              The numerical resolution is such that there are $\sim 27^3$ and $\sim 20^3$ 
              grid cells in the Hill sphere of the interior and exterior planets, respectively.
             }
\label{fig:acc}
\end{figure}
The evolution of one high-resolution 3D calculation was continued 
by allowing the two planets 
to accrete gas following the procedure outlined in \citetalias{gennaro2008},
modified to account for the different local dynamical times\footnote{%
If accretion proceeds through a disk around the planet, 
as suggested by the bottom-left panel of Figure~\ref{fig:zoom},
$\dot{M}_{p}$ is equal to the rate at which the nebula supplies
this disk, hence the connection with the orbital frequency.}
(i.e., the timescale for mass removal depends on the planet's
orbital frequency).
Disk conditions are similar to those applied to the reference model of 
Section~\ref{sec:OMR}. The disk-limited accretion rates of the two planets
are shown in Figure~\ref{fig:acc}, where the thicker curve refers
to the exterior planet. Both accretion rates are modulated 
(notice the logarithmic scale) over the
orbital period due to the eccentric orbits \citep[][]{gennaro2006}, although 
additional modulations may be present due to the resonant forcing.
For conditions simulated in Figure~\ref{fig:acc}, the integrated values
of $\dot{M}_{p}$ are very similar, differing by less than $10$\%.
These rates would 
yield a growth time for the exterior planet of a few times $10^{4}$ years, 
typically much shorter than the migration time (see Figure~\ref{fig:a1vst}). 
For comparison,
the growth time of the interior planet would be $\sim 10^{5}$ years.
The net effect of gas accretion could produce a mass ratio
$M_{2}/M_{1}$ close to $1$, or possibly larger (since gas starvation 
would likely occur for the interior planet first, see Figure~\ref{fig:Svst}).

The orbital evolution of the accreting planets, monitored over 
$\sim 1000$ years, does not show any significant deviation from the
evolution of the non-accreting planets (the planet masses are fixed in this
case, on account of the little variations expected over that timescale).
However, as explained in Section~\ref{sec:SPoA}, one or more accreting 
planets may change the steady-state structure of the inner disk, 
affecting the migration behavior of the planets.

\section{Gas Accretion and Effects on the Disk}
\label{sec:SPoA}

Resolving the problem of the rapid growth of the exterior planet,
in a still relatively massive disk,
would remove only one issue posed by gas accretion. In fact, even if
Saturn suddenly stopped accreting, gas accretion onto Jupiter would 
continue to pose a problem. The issue here is not related to the growth 
of the interior planet's mass, but rather to the modification of the mass
flux through the disk, across the planet's orbit.

As explained in Section~\ref{sec:DD}, in a stationary disk the accretion
rate is $\dot{M}=3\pi\nu\Sigma$ and nearly independent of the radius $r$.
Accretion on the interior (or exterior) planet would change $\dot{M}$. 
This phenomenon
was analyzed in detail by \citealt{lubow2006} (hereafter 
\citetalias{lubow2006}) for the case of a single planet.
The generalization to two planets can be performed by introducing an 
average accretion efficiency that quantifies the amount of gas accretion 
onto both planets, relative to an average local accretion rate through 
the disk\footnote{As pointed out by \citetalias{lubow2006}, for a given
disk, there is a planet mass ($M_{1}+M_{2}$, in this case) beyond 
which the couple exerted by the planet(s) will make the accretion 
disk evolve toward a decretion disk \citep{pringle1991}.}.
However, here we wish to consider the situation in which the interior
planet accretes gas, but the exterior planet does not. Therefore, 
the formalism of \citetalias{lubow2006} can be applied in a straightforward manner.
Let us indicate with $\dot{M}_{\mathrm{e}}=\dot{M}=3\pi\nu\Sigma$ the
accretion rate sufficiently far from the exterior planet's orbit (so that it is
basically unperturbed) and with $\dot{M}_{\mathrm{i}}$ the accretion
rate inside the orbit of the interior planet. If there was no sink in the disk,
then $\dot{M}_{\mathrm{i}}=\dot{M}_{\mathrm{e}}$. Yet, since some material
is removed by the planet, in general 
$\dot{M}_{\mathrm{i}}\le\dot{M}_{\mathrm{e}}$, which implies
a reduction of $\Sigma$ in the inner disk with respect to the same disk
without the planet. 
This reduction depends on the planet's accretion efficiency $\mathcal{E}$,
defined as the ratio of $\dot{M}_{p}$ to the accretion rate interpolated 
at the planet's orbital radius. Since 
$\dot{M}_{\mathrm{i}}=\dot{M}_{\mathrm{e}}-\dot{M}_{p}$,
one finds that $\dot{M}_{\mathrm{i}}=\dot{M}_{\mathrm{e}}/(\mathcal{E}+1)$.
This result formally applies if the disk's inner boundary, $r_{\mathrm{min}}$,
is much smaller than $a_{1}$. In general, one finds that 
\citepalias[see][]{lubow2006}
\begin{equation}
\dot{M}_{\mathrm{i}}=%
\frac{\dot{M}_{\mathrm{e}}}{1+(1-\sqrt{r_{\mathrm{min}}/a_{1}})\mathcal{E}}.
\label{eq:dotMi}
\end{equation}
According to the Equation~(\ref{eq:dotMi}),
the surface density in the inner disk is then expected to be reduced by
a factor of order $\mathcal{E}+1$ (assuming that $\nu$ remains unchanged),
relative to the situation in which $\dot{M}_{p}=0$. 
Accordingly, the (positive) Lindblad torque 
($\propto \Sigma$, see Equation~\ref{eq:Tos}) exerted by the inner disk on the
(inner) planet is also expected to decrease. 
Such effect may slow down outward migration and may even tip the torque 
balance in favor of the negative torque acting on the interior planet.

We estimate the efficiency of accretion, $\mathcal{E}$, allowing for accretion 
on the interior planet only by applying the steady-state solution 
given by Equation~(19) of \citetalias{lubow2006} as initial condition and 
using the iteration procedure outlined in \citetalias{lubow2006}.
The disk configuration is as that of the reference model in Section~\ref{sec:OMR}, 
except for a somewhat smaller disk's inner radius of $0.15\,a_{1}$ and for 
the initial location of the exterior planet ($a_{2}/a_{1}=1.315$, see Figrue~\ref{fig:oo}).
We find that $\mathcal{E}\sim 6$.
There are fluctuations over time of both $\dot{M}_{p}$ and $\dot{M}$ 
(see Figure~\ref{fig:acc}) and therefore
the accretion efficiency is taken as the ratio of averaged quantities. The estimated 
variation is $\Delta\mathcal{E}\sim 1$.
The accretion rate ratio is $\dot{M}_{\mathrm{i}}/\dot{M}_{\mathrm{e}}\sim 0.2$, 
whereas the corrected value, that is the ratio estimated for 
$r_{\mathrm{min}}/a_{1}\ll 1$, is $\sim 0.14$ (see Equation~\ref{eq:dotMi}). 

\begin{figure*}[t!]
\centering%
\resizebox{\figlen}{!}{%
\includegraphics{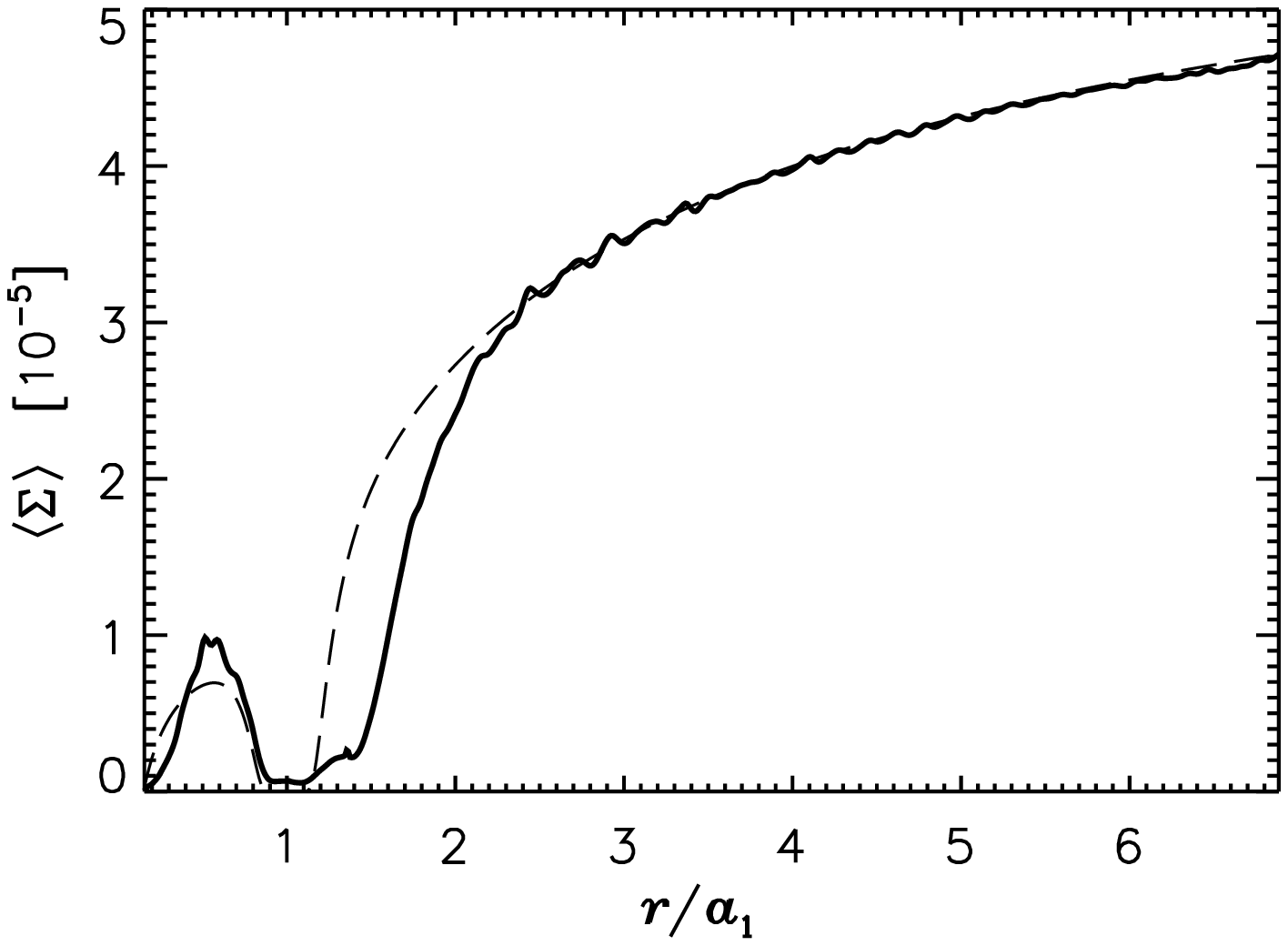}%
\includegraphics{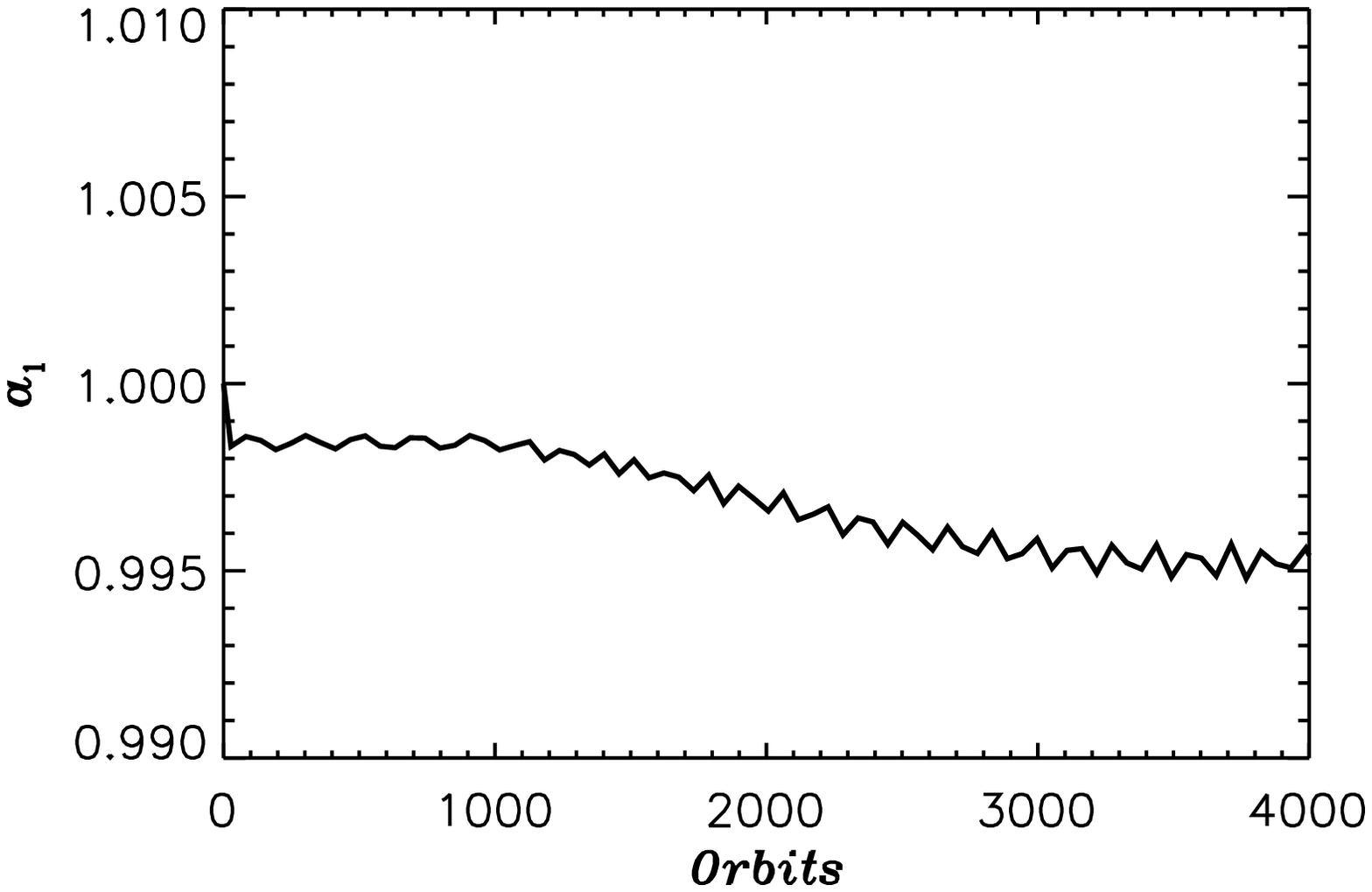}}
\caption{%
              \textit{Left:}
              azimuthally averaged surface density (\textit{solid line}), 
              in units of $\Ms\,a^{-2}_{1}$, of a disk with a pair of planets,
              initially placed in the 3:2 commensurability. 
              The exterior planet is non-accreting. The interior 
              planet accretes gas with an efficiency parameter 
              $\mathcal{E}\sim 6$ (see the text).
              The dashed
              line indicates the steady-state   solution of \citetalias{lubow2006}
              (with a single planet) for $\mathcal{E}=6$ and an inner disk radius of 
              $0.15\,a_{1}$.
              \textit{Right:} migration track of the interior planet. Data are averaged 
              over $\sim 50$ orbital periods. The planet begins migrating after $1000$
              (initial) orbits. The depletion
              of the inner disk (compare with Figure~\ref{fig:dtdm}, long-dashed
              line) is sufficient to deactivate the outward migration mechanism
              (compare with Figure~\ref{fig:aref}, \textit{top}, which has different
              units on the time axis).
             }
\label{fig:eff}
\end{figure*}
In Figure~\ref{fig:eff} (\textit{left panel}), the surface density after $1000$ orbits
of the interior planet (\textit{solid line}) is compared with the steady-state solution
(Equation~19) of \citetalias{lubow2006} for a disk with a single planet (\textit{dashed line}).
The reduced mass accretion past the planets, $\dot{M}_{\mathrm{i}}$,
results in a lower surface density (compare with the long-dashed line in the left
panel of Figure~\ref{fig:dtdm}).
The migration of the resonant-orbit pair (initially placed in the 3:2 mean motion
resonance) in the stationary surface density of Figure~\ref{fig:eff} 
(\textit{left panel}) is shown in the right panel. The reduced positive Lindblad 
torque by the inner disk cannot overcome the negative torque by the disk outside
the planet's orbit, resulting in a migration speed $\dot{a}_{1}\lesssim 0$ 
(note that the units of time in Figure~\ref{fig:eff} are initial orbits of the inner
planet, not years).

\section{The 2:1 Mean Motion Resonance}
\label{sec:21}

A pair of planets undergoing convergent migration will first cross
the 2:1 commensurability, before approaching the 3:2 mean motion 
resonance. While the latter resonant-orbit configuration may activate
the outward migration mechanism discussed here (if $M_{1}/M_{2}\approx 3$), 
the former may not (see Figures~\ref{fig:reso} and \ref{fig:mass_con}). 
But a pair of giant planets interacting with a gaseous disk can indeed be 
caught in this resonance, as shown by several studies
\citep[see, e.g.,][]{kley2003,kley2004,pierens2008,zhang2010}.
We therefore seek conditions such that the outer planet may, or may not, 
overcome the barrier represented by capture in the 1:2 mean motion 
resonance, while migrating toward the inner planet.

In the case of convergent migration, the condition for capture of the
exterior planet in a resonant orbit with the interior planet requires that 
the relative migration speed be such that
\begin{equation}
\left|\frac{da_{\mathrm{rel}}}{dt}\right|< \frac{\Delta a_{\mathrm{res}}}{T_{l}},
\label{eq:con_cap}
\end{equation}
where $\dot{a}_{\mathrm{rel}}=\dot{a}_{2}-\dot{a}_{1}$, $\Delta a_{\mathrm{res}}$ is the 
resonance amplitude in semimajor axis, and $T_{l}$ is the resonant libration 
period, i.e., the period of the critical angle \citep[e.g.,][]{michtchenko2008}
$\psi_{1}=2(\mathcal{M}_{2}+\varpi_{2})-(\mathcal{M}_{1}+\varpi_{1})-\varpi_{1}$,
where $\mathcal{M}$ indicates the mean anomaly and $\varpi$ the argument of periapsis.
Recall that, in the case of a more massive interior planet, $\psi_{1}$ is a real resonant
angle, as it displays dynamical libration.
For low-eccentricity orbits ($e\lesssim 0.1$), \citet{mustill2011} 
\citep[see also][]{quillen2006} approximate the critical relative velocity for capture 
of the outer planet in the 1:2 mean motion resonance as
\begin{equation}
\left|\frac{da_{\mathrm{rel}}}{dt}\right|\lesssim 1.2\left(\frac{M_{1}}{\Ms}\right)^{4/3}%
                                                                         a_{2}\Omega_{2},
\label{eq:con_cap_q}
\end{equation}
where it is assumed that $\Omega_{1}/\Omega_{2}=2$.
Capture appears to be probabilistic at higher eccentricities, with a critical relative
velocity that becomes somewhat larger for larger eccentricities.
Here we should stress, however, that the estimate of the critical velocity in 
Equation~(\ref{eq:con_cap_q}) assumes captures of ``particles'', i.e., 
$M_{1}\gg M_{2}$.
If the planet transits the 1:2 orbital resonance with the interior planet, 
capture in the next first-order resonance, the 2:3, requires that
$|\dot{a}_{\mathrm{rel}}|\lesssim 14 (M_{1}/\Ms)^{4/3}a_{2}\Omega_{2}$.

We shall assume that the inward migration speed of the interior planet
is negligible compared to that of the exterior planet, thus 
$\dot{a}_{\mathrm{rel}}\approx\dot{a}_{2}$. The exterior planet may in principle
undergo a mode of rapid migration dominated by corotation torques (\textit{type~III}).
\citet{gennaro2008} separated this mode of migration from type~I mode by
analyzing the torque density distributions and the fluid trajectories 
in the corotation region of the planet. They found that two conditions must 
be satisfied for the activation of type~III migration. 
The first is that the migration timescale 
across the coorbital region is shorter than the timescale required to clear 
a gap over that same region. The second is that the unperturbed surface 
density at the planet location is such that 
\begin{equation}
\left(\frac{a^{2}\Sigma}{\Ms}\right)\gtrsim \left(\frac{H}{a}\right)^{2},
\label{eq:con_typeIII}
\end{equation}
where disk quantities are sampled at $a=a_{2}$\footnote{%
The condition represented by Equation~(\ref{eq:con_typeIII}) is also
consistent with the case reported by \citet{masset2001}, which
show an exterior planet migration dominated by corotation torques.
In that case, 
$a_{1}^{2}\Sigma/\Ms=6\times 10^{-4}$, $H/r=0.04$, and $a_{2}=2$
(in their units).
Hence, evaluating at the initial position of the exterior planet, one has
$a_{2}^{2}\Sigma/\Ms=2.4\times 10^{-3}>(H/a_{2})^2$.}.
If the first condition is fulfilled, the second condition requires that
$\Sigma\gtrsim 10^{-3}\,\Ms\,a^{-2}$ for $H/a\gtrsim 0.03$, 
which corresponds to a density in excess of $10^{3}\,\mathrm{g\,cm}^{-2}$ 
in the disk region within $3\,\AU$ of the star. 
According to Figures~\ref{fig:S1} and \ref{fig:S1_hist},
this requirement is not met, thus we can assume that the outer planet migrates
at a rate in between type~I and type~II migration rates. Assuming a speed
of order type~I and using Equation~(\ref{eq:TI}), one finds
\begin{equation}
\left|\frac{da_{\mathrm{rel}}}{dt}\right|\sim   \left(\frac{a}{H}\right)^{2}\left(\frac{M_{2}}{\Ms}\right)%
\left(\frac{a^{2}\Sigma}{\Ms}\right)a_{2}\Omega_{2},
\label{eq:a2_typeI}
\end{equation}
all disk quantities being evaluated at $a=a_{2}$.
In the equation above, a numerical factor of order unity multiplying the
right-hand side is neglected. This is done to account for the fact that 
the density is partly depleted and the actual migration rate deviates 
somewhat from the type~I rate \citepalias[see][]{gennaro2008}.
Moreover, we find that Equation~(\ref{eq:a2_typeI}) gives 
a reasonable order-of-magnitude approximation to numerical 
results.

Therefore, capture of the exterior planet in the 1:2 mean motion 
resonance with the interior planet may occur if the unperturbed 
surface density is lower than a critical value, so that
\begin{equation}
\left(\frac{a^{2}\Sigma}{\Ms}\right)\lesssim \left(\frac{M_{1}}{\Ms}\right)^{4/3}%
\left(\frac{H}{a}\right)^{2}\left(\frac{\Ms}{M_{2}}\right).
\label{eq:S21}
\end{equation}
If $H/a\gtrsim 0.03$, then a surface density 
$\Sigma\lesssim 3\times 10^{-4}\,\Ms\,a^{-2}$, 
or about $650\,\mathrm{g\,cm}^{-2}$ at $\approx 2\,\AU$, 
may be sufficient to allow for capture in this resonance.
If the exterior planet crossed the 1:2 commensurability while
it had a much lower mass, say $\sim 20$ Earth masses,
this critical density would not decrease,
even accounting for a numerical factor of $4$--$5$ in 
Equation~(\ref{eq:a2_typeI}) and restoring a full type~I migration.

The inequality in Equation~(\ref{eq:S21}) suggests that if
$a_{2}^{2}\Sigma/\Ms\gtrsim 5\times 10^{-4}$ in a disk with
$H/r\sim 0.04$, the exterior planet may transit
the 1:2 mean motion resonance with the interior planet.
These conditions are realized, for example, in the model
of \citet{masset2001} (see their Figure~1) and in the models 
of \citet{pierens2008} (see their Figure~5), considering that
the density must be rescaled at the radius of the resonance 
crossing, where $a_{2}\approx 1.5$ \citep[in the units of][]{masset2001}
and $a_{2}\approx 1.3$ \citep[in the units of][]{pierens2008}.
The solar nebula models considered here, however, suggest that if 
the orbits approach the 1:2 commensurability in the inner disk,
after $\sim 1\,\mathrm{Myr}$, then capture is likely in a statistical sense
(see Figure~\ref{fig:S1_hist}).

\begin{figure}[t!]
\centering%
\resizebox{\linewidth}{!}{%
\includegraphics[bb=  0 15 360 252,clip]{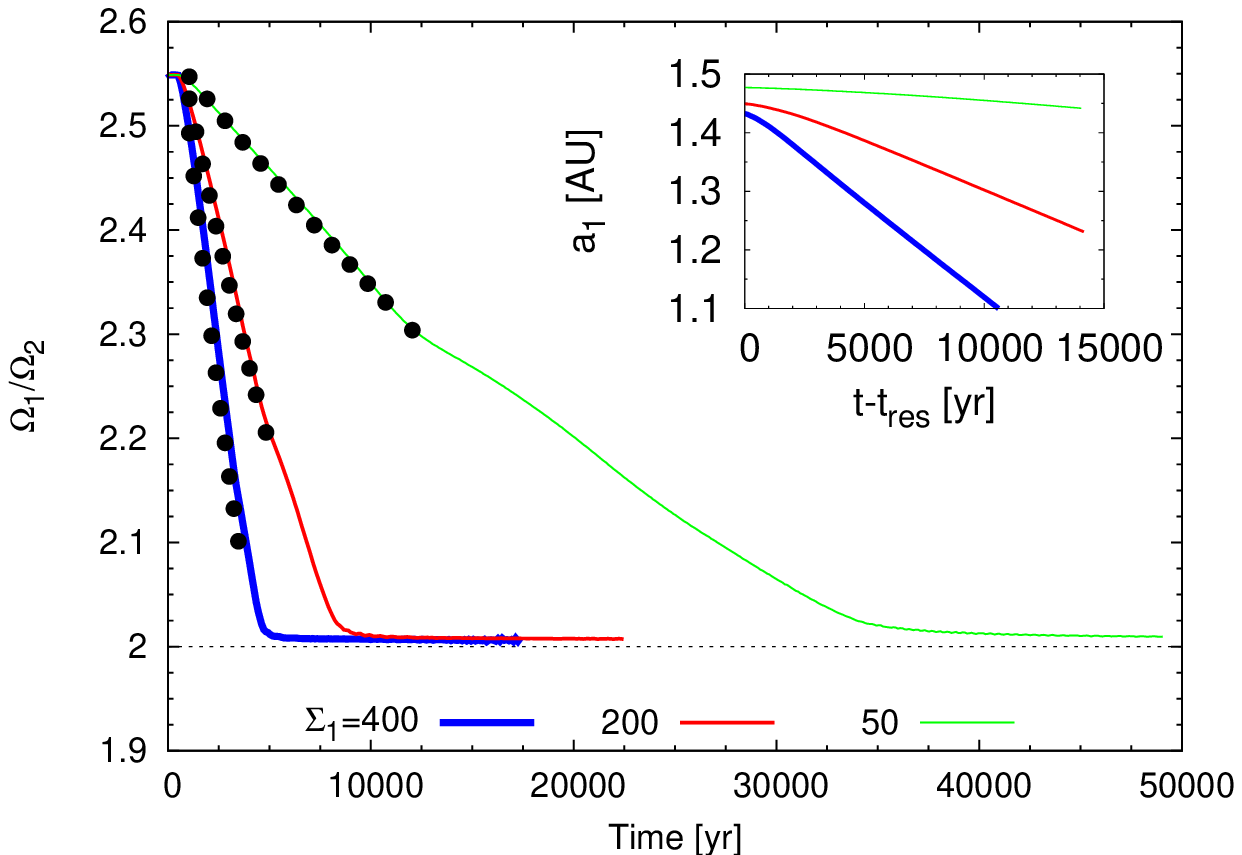}}
\resizebox{\linewidth}{!}{%
\includegraphics[bb= 50 49 410 295,clip]{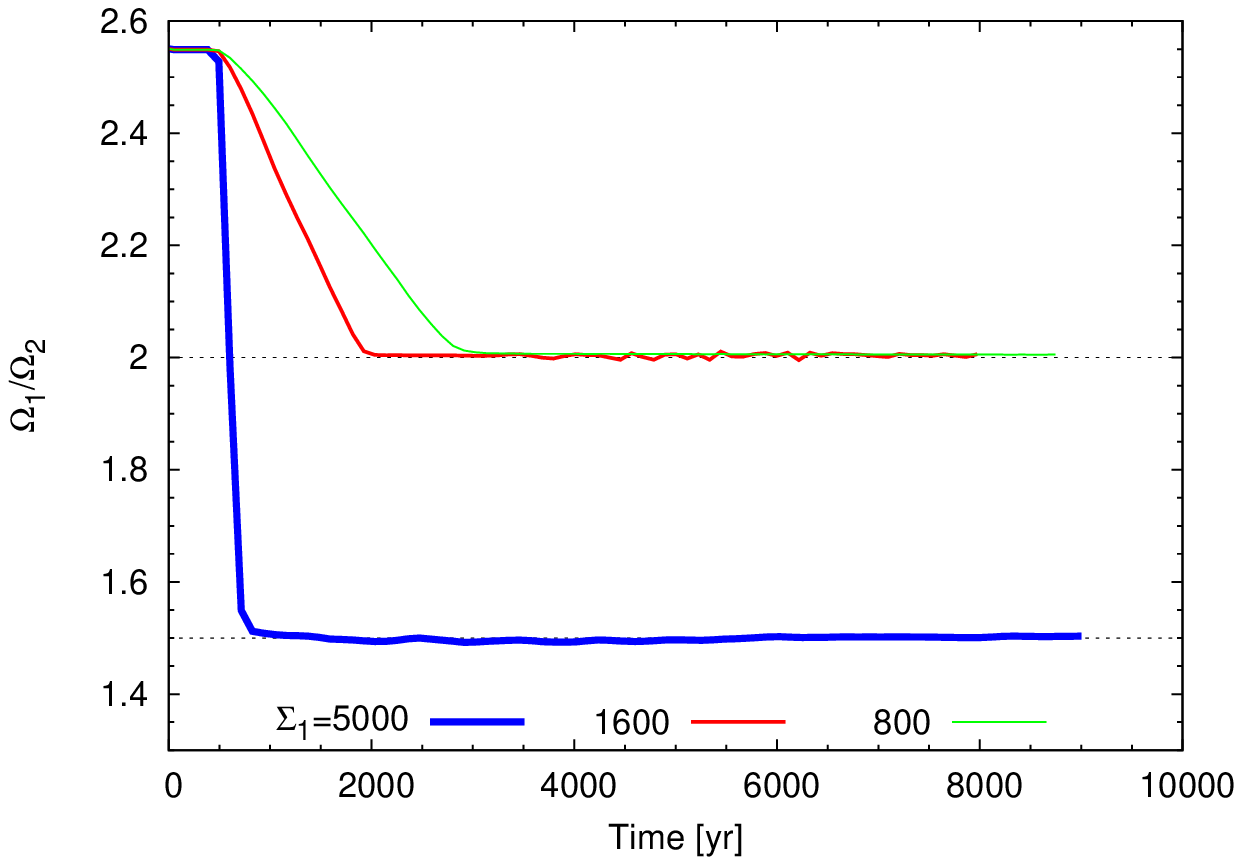}}
\caption{%
             Ratio of the mean motions vs.\ time for a pair of planets
             undergoing convergent migration. The value of the disk's
             surface density at $1\,\AU$ is indicated at the bottom of each
             panel, in units of $\mathrm{g\,cm}^{-2}$, for the curves of 
             different thickness. 
             The top panel illustrates
             evolutions for typical densities obtained from the disk models of
             Section~\ref{sec:MR}. Circles are predictions from 
             Equation~(\ref{eq:Omegadot}).
             The inset shows the migration tracks of the interior planet
             after resonance locking occurs. 
             At higher densities, capture of the exterior planet transitions 
             from the 1:2 to the 2:3 commensurability, as illustrated in 
             the bottom panel.
             }
\label{fig:12mmr}
\end{figure}
Equation~(\ref{eq:S21}) represents only an approximate condition
for resonance locking, because orbital eccentricity may
play some role and because Equation~(\ref{eq:con_cap_q}) was
not derived for the capture of similar mass bodies. 
In order to provide a further test on conditions that may lead to
locking of the exterior planet in the 1:2 commensurability, 
calculations along the 
lines of those presented in Section~\ref{sec:OMR} are performed 
for varying initial surface density at $1\,\AU$, $\Sigma_{1}$. The
disk conditions are those used for the reference model in Figure~\ref{fig:aref},
except that the exterior planet is placed initially on a circular orbit at a distance 
of $2.8\,\AU$ from the star, outside the resonance location with
the interior planet, and both begin migrating after $500$ years. 
Results from these calculations are illustrated in 
Figure~\ref{fig:12mmr}, which shows the ratio of the mean
motions. The top panel refers to values of $\Sigma_{1}$
compatible with those found in the disk evolution models at the time 
of planet formation (see Figure~\ref{fig:S1_hist}, \textit{left}). 
In these cases, the exterior planet becomes locked in the 1:2 orbital 
resonance.
Since $\dot{\Omega}=-(3/2)\Omega \dot{a}/a$, assuming that 
$\Omega_{1}$ is nearly constant, then
\begin{equation}
\frac{d}{dt}\!\left(\frac{\Omega_{1}}{\Omega_{2}}\right)\sim%
-\frac{3}{2} \left(\frac{\Omega_{1}}{\Omega_{2}}\right)\!%
\left(\frac{a}{H}\right)^{2}\!\left(\frac{M_{2}}{\Ms}\right)\!%
\left(\frac{a^{2}\Sigma}{\Ms}\right) \Omega_{2},
\label{eq:Omegadot}
\end{equation}
where quantities depending on $a$ are evaluated at $a=a_{2}$.
Predictions from Equation~(\ref{eq:Omegadot}) are superimposed
(circles) to $\Omega_{1}/\Omega_{2}$ curves in Figure~\ref{fig:12mmr},
indicating that the exterior planet does approach the interior
planet, at least initially, with a radial speed on the order of the relative
velocity given by Equation~(\ref{eq:a2_typeI}).

The bottom panel of Figure~\ref{fig:12mmr} 
shows cases with higher initial densities in which capture of the exterior
planet is in the 1:2 or the 2:3 orbital resonance.
There is overall agreement with Equation~(\ref{eq:S21}),
and transit across the 1:2 orbital resonance is obtained for 
$r_{1}^{2}\Sigma_{1}/\Ms=6\times 10^{-4}$.
We did not investigate the exact density value at which locking 
transitions from one to the other resonant configuration, but
it is likely that there exists an interval of values for which the result is stochastic.

The inset in the top panel of Figure~\ref{fig:12mmr} illustrates the migration 
of the interior planet after resonance locking. In the bottom panel, 
migration is outward for the case that shows locking of the exterior planet 
in the 2:3 mean motion resonance with the interior planet, 
inward in the other cases. 
These results confirm what was argued above and suggested by 
Figures~\ref{fig:reso} and \ref{fig:mass_con}: the 2:3 orbital resonance 
leads to outward migration, the 1:2 resonance does not. 

A small disk aspect ratio may help preventing capture of the exterior
planet in the 1:2 commensurability with the interior planet. 
However, the condition for gap formation 
(Equation~\ref{eq:gcon}) suggests that the migration of the exterior 
planet may transition to type~II, and $\dot{a}_{1}$ cannot be neglected.
In this case, since $\nu\propto a^{\beta}$ and 
$\dot{a}\sim\nu/a \propto a^{\beta-1}$, convergent migration
requires that $\beta>1$. By applying Equation~(\ref{eq:con_cap_q}),
one finds that if the kinematic viscosity at $1\,\AU$ is
$\nu_{1}\lesssim (M_{1}/\Ms)^{4/3}(1\,\AU/a_{2})^{(\beta-1/2)}$
then convergent type~II migration may still lead to orbital locking 
in the 1:2 mean motion resonance. 
However, this estimate is complicated by the fact that
if the planet mass becomes larger than the local disk mass,  
a likely situation at late evolutionary times,
$\dot{a}$ is also proportional to $a^{2}\Sigma/\Mp$ due 
to intervening inertia effects \citep[see, e.g.,][]{syer1995,ivanov1999}.
Hence, the interior planet would likely slow down before the exterior
planet would.

\subsection{Saturn Formation within the 1:2 Orbital Resonance with Jupiter}
\label{sec:formation12}

We shall consider here the possibility that Saturn forms within the 1:2, 
but outside the 2:3, commensurability with Jupiter, while both planets 
are beyond several $\AU$ from the Sun. 
If during the course of its evolution Saturn remains inside the 1:2 mean 
motion resonance with Jupiter, then capture in the 2:3 orbital resonance 
is still possible.

In order to maintain such a compact orbital configuration throughout 
the evolution of the two planets, the migration rates must be very similar 
over time. In fact, if the constraint $1.31 < a_{2}/a_{1} < 1.59$
has to be preserved, a change of this ratio of at most $20$\%
over the formation timescales basically implies that 
$a_{2}/a_{1}$ is roughly constant. Therefore, by taking the time derivative, 
one has 
\begin{equation}
\frac{\dot{a}_{2}}{a_{2}}\sim \frac{\dot{a}_{1}}{a_{1}}.
\label{eq:a2a1const}
\end{equation}
But since the interior planet has to grow faster than the exterior planet 
does, their migration rates are bound to differ, at some point in time 
at least. Thus, the condition in Equation~(\ref{eq:a2a1const}) is unlikely 
to be (always) satisfied and the difference $a_{2}-a_{1}$ will either
increase or decrease.

If Saturn forms within the 1:2 orbital resonance and Jupiter has still 
to acquire most of its mass, there will be a phase when the migration 
rate of Jupiter significantly exceeds that of Saturn 
\citepalias[presumably, around the time of runaway gas accretion; see][]{gennaro2008}. 
Hence, it is most likely that Saturn is left behind and becomes probably 
trapped  in the 1:2 commensurability. 
Instead, if Jupiter has already acquired the bulk of its mass, the opposite 
is likely to occur and most probably Saturn becomes locked in the 2:3 
orbital resonance while it is still growing.

\begin{figure*}[t!]
\centering%
\resizebox{\linewidth}{!}{%
\includegraphics[clip, bb=15 13 481 205]{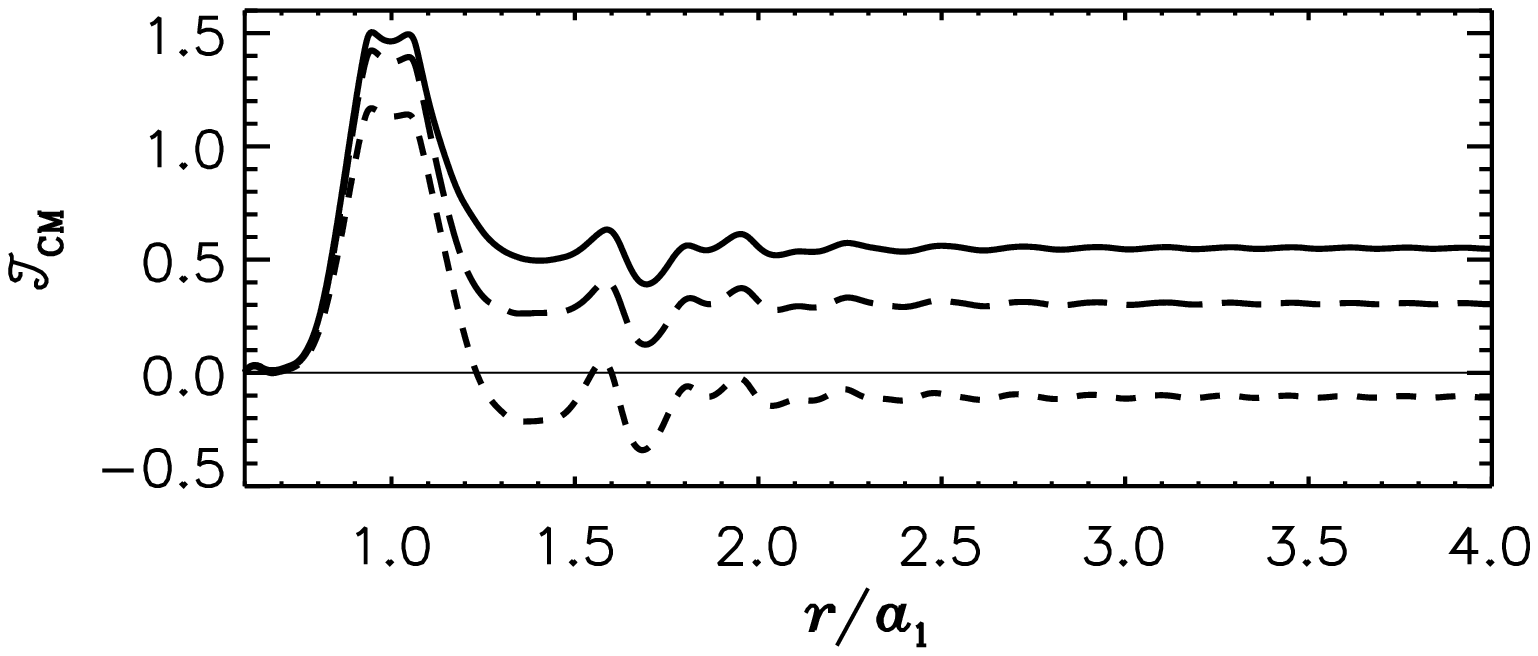}%
\includegraphics[clip, bb=15 13 481 205]{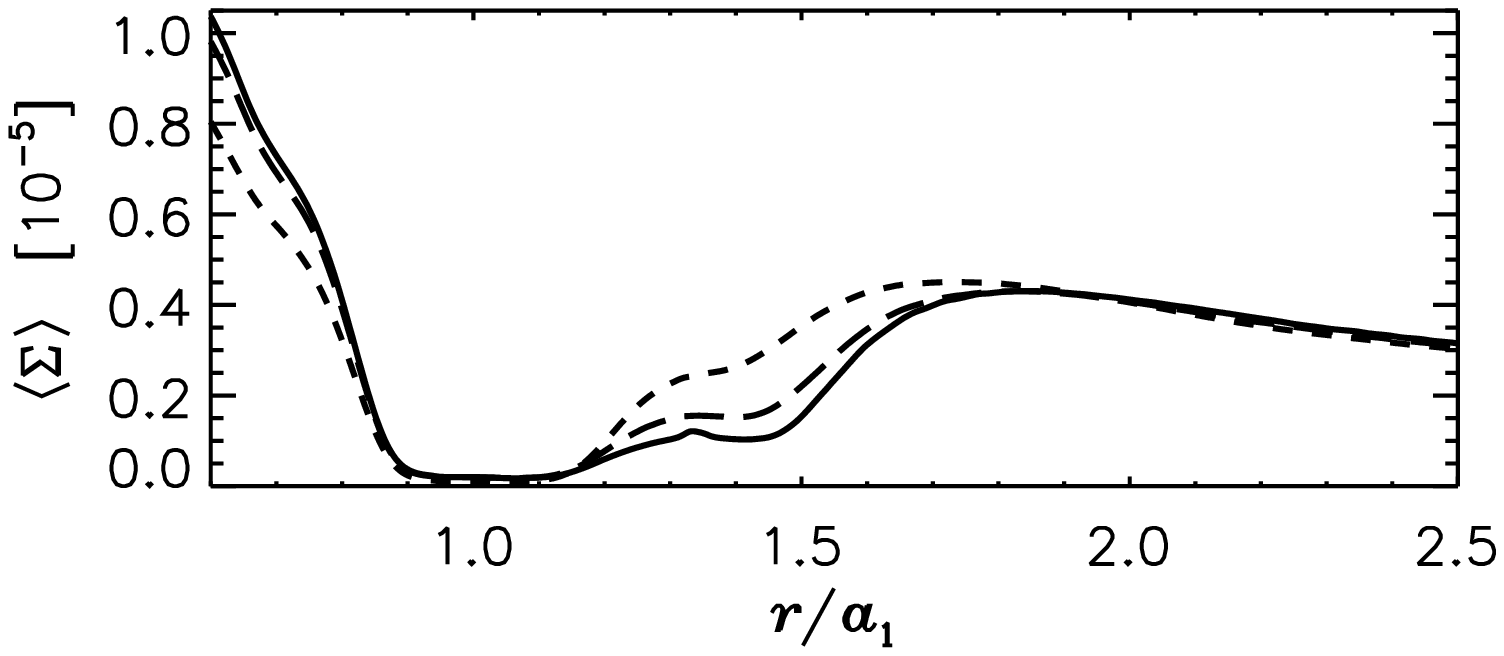}}
\resizebox{\linewidth}{!}{%
\includegraphics[clip,bb=-35 -15 501 216]{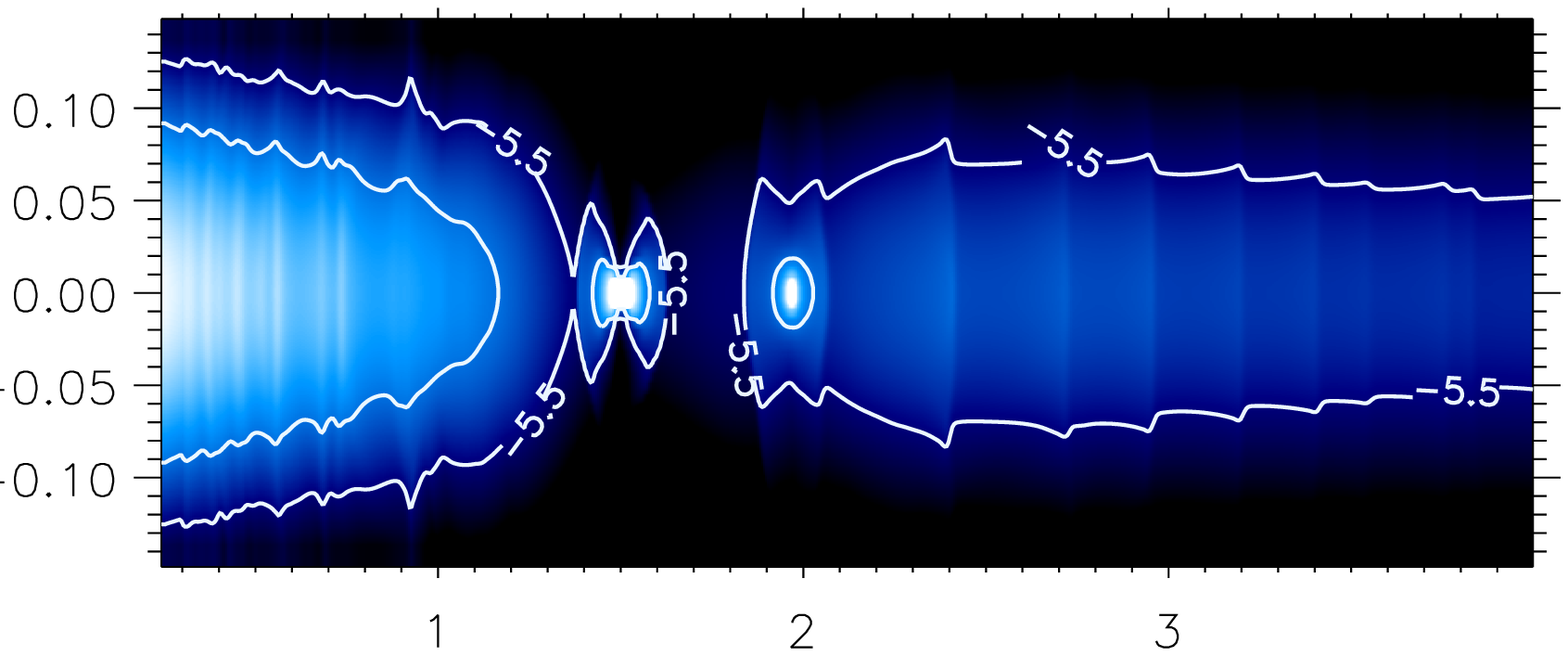}%
\includegraphics[clip,bb= 45 -15 561 216]{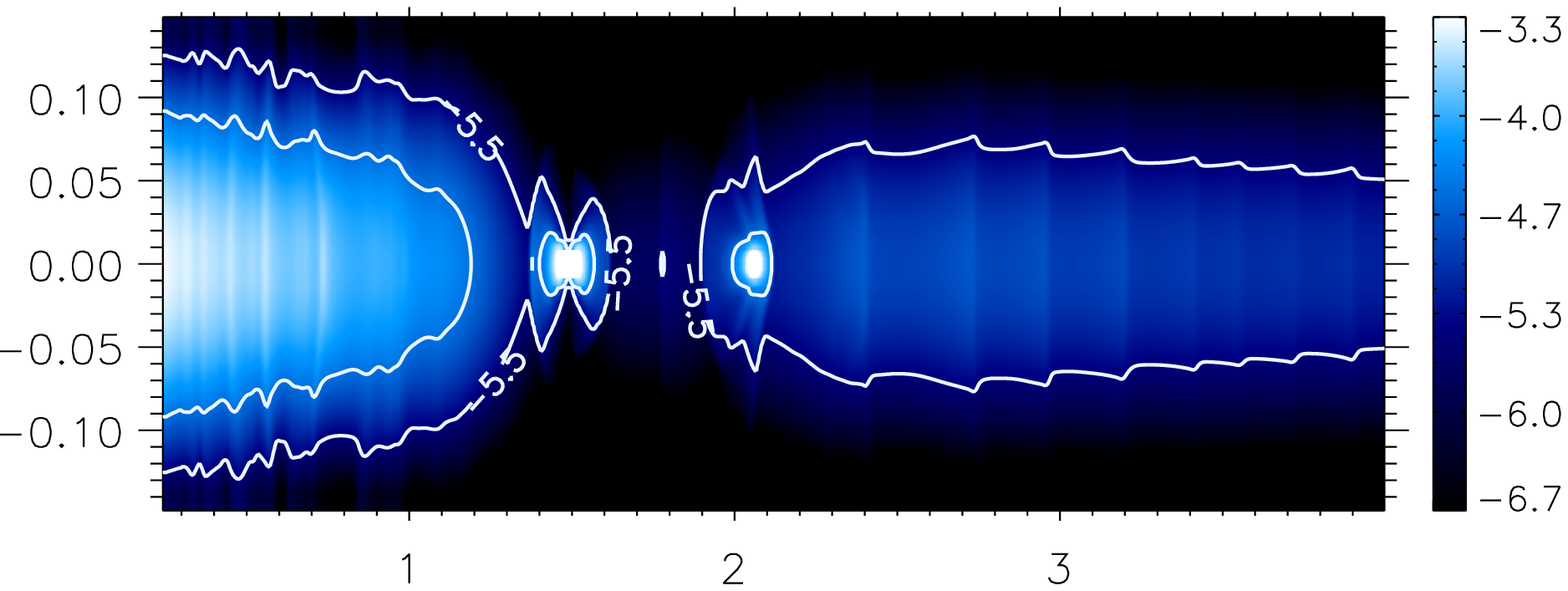}}
\caption{%
             \textit{Top:}
             cumulative torques (\textit{left}) exerted on the interior 
             planet and averaged surface density (\textit{right}), obtained from 
             3D calculations whose parameters are as in the reference model 
             of Section~\ref{sec:OMR}. Different lines types indicate
             different masses of the exterior planet ($M_{2}/\Ms$): 
             $2.9\times 10^{-4}$ (\textit{solid}), $2\times 10^{-4}$  
             (\textit{long-dashed}),  and $10^{-4}$ (\textit{short-dashed}).
             The torque is normalized to 
             $10^{-3}G M^{2}_{\mathrm{s}}(M_{1}/\Ms)^{2}/a_{1}$ and
             $\Sigma$ is in units of $\Ms\,r^{-2}_{1}$.
             \textit{Bottom:}
             vertical stratification of the mass density at disk azimuth 
             $\phi_{\mathrm{J}}=\phi_{\mathrm{S}}$, in units of $\Ms\,r^{-3}_{1}$,
             for $M_{2}=10^{-4}\,\Ms$ (\textit{left}) and
             $2\times 10^{-4}\,\Ms$ (\textit{right}).
             }
\label{fig:0102}
\end{figure*}
We consider this last situation in some details by performing 3D calculations similar 
to those of the reference model of Section~\ref{sec:OMR}, but applying 
a lower mass to the exterior planet: $q=M_{2}/\Ms=10^{-4}$ and 
$2\times 10^{-4}$. Equation~(\ref{eq:ggref}), appropriately modified 
for mass ratios $q\ne q_{\mathrm{ref}}$ (see discussion after 
Equation~\ref{eq:ggref}), suggests that if $q\approx 10^{-4}$ 
migration of the pair is inward, but if $q\approx 2\times 10^{-4}$ 
then migration is outward. Direct calculations agree with these predictions,
as indicated by the cumulative torques shown in the top-left panel of
Figure~\ref{fig:0102} for three different values of $M_{2}$ 
(see caption for details).
The density depletion due to the tidal torques of the exterior planet amounts 
to $\sim 35$\% for $M_{2}=10^{-4}\,\Ms$ and to $\sim 60$\% for 
$M_{2}=2\times 10^{-4}\,\Ms$ (\textit{top-right panel}), in accord
with the expectations of Equation~(\ref{eq:gcon}).
The bottom panels of Figure~\ref{fig:0102} show vertical distributions
of the mass density (see figure's caption).

Therefore, if Saturn grows while locked in the 2:3 orbital resonance with 
Jupiter, there is a mass smaller than Saturn's final mass for which 
migration stalls and then reverses, as suggested by 
$\mathcal{T}_{\mathrm{CM}}$ as function of $M_{2}$ in Figure~\ref{fig:0102}
(\textit{top-left panel}). 
This situation would prevent Jupiter from reaching the inner disk regions, 
unless Saturn achieved the mass for migration reversal when Jupiter is 
already there. 
For $H/r\approx 0.04$, such mass is between $10^{-4}\,\Ms$ 
and $2\times 10^{-4}\,\Ms$ (and presumably closer to $10^{-4}\,\Ms$ 
for $H/r\approx 0.03$), which implies that a substantial fraction of Saturn's 
envelope is acquired in the inner regions of the solar nebula. 

This circumstance, however, would be likely at odds with the elemental abundances 
of some species measured in Saturn's atmosphere \citep[see][]{hersant2008}. 
The abundances relative to hydrogen of elements such as C, N, S, As, and P 
are a few to several times as high as the solar abundances 
\citep[see][and references therein]{lodders2003,asplund2009}. 
In fact, the presence in large amount of these elements is believed to have
arisen from accretion of gas \citep[e.g.,][]{hueso2006} and/or solids 
\citep[e.g.,][]{hersant2008} in a cold disk environment.

\section{Summary and Discussion}
\label{sec:SaD}

This paper presents results of thermodynamical models of 
protoplanetary disks that are constructed by applying ranges of parameters
that may have characterized the early solar nebula (see Section~\ref{sec:DEM}).
Disk evolution is driven by viscous torques and photo-evaporation
originating from the central star (see Section~\ref{sec:DD}). 
Thermal balance in the disk is achieved by equating viscous and 
stellar irradiation heating with radiative cooling in the vertical direction 
(see Section~\ref{sec:DT}). 
Only models that predict disk lifetimes between $1$ and $20\,\mathrm{Myr}$ 
are considered viable representations of the solar nebula
(see Table~\ref{tbl:tD}), in line with observations 
and core-nucleated accretion calculations of gas giants.
Such models provide the physical conditions (see, e.g., Figure~\ref{fig:Svst}) 
at the time and after 
the planets acquired most of their mass (see Section~\ref{sec:MR})
and can be used to simulate their long-term orbital migration.

Two and three dimensional hydrodynamical calculations are used
to quantify the migration rates of a pair of planets with
mass ratios corresponding to Jupiter's and Saturn's,
$M_{1}/\Ms\approx 10^{-3}$ and $M_{2}/\Ms\approx 3\times 10^{-4}$
(see Section~\ref{sec:TI}).
The orbits of the planets are initially placed in proximity of the 3:2 
mean motion resonance (see, e.g., Figure~\ref{fig:zoom}). 
As described by \citet{masset2001}, a necessary condition to 
activate outward migration is the overlap of the tidal gap carved 
in the disk by a less massive, exterior planet with the gap of the 
more massive, interior planet (see Figure~\ref{fig:dtdm}). 
The relative depth and width
of the gaps provide the sufficient condition (see Section~\ref{sec:ToC}).
High temperatures and kinematic viscosities inhibit gap formation, 
either promoting inward migration (see Section~\ref{sec:OMR}) 
or stopping outward migration (see Section~\ref{sec:ROM})

If a near 3:2 commensurability is preserved, there are stalling radii
in the disk, toward which the pair will converge (see Figure~\ref{fig:adepg}). 
In general, to first approximation, these radii depend on a combination 
of the turbulence viscosity parameter, disk thickness, 
and mass of the outer planet (see Figures~\ref{fig:gg0} and \ref{fig:0102}).

For planets moving outward from the inner disk region
($r\lesssim 2\,\AU$) at a time between $1$ and $3\,\mathrm{Myr}$, 
the interior planet may reach beyond $\sim 5\,\AU$ only
if viscosity is low enough, the surface density is not too steep,
and the disk not too warm (see Section~\ref{sec:LTM}). 
However, the probability for this to happen appears low
(see Table~\ref{tbl:ainf_1}). 
Experiments performed on random samples of the parameter 
space suggest that in $98$\%
of the cases the interior planet stops within $4\,\AU$
(see Figure~\ref{fig:a1_tau}).

At least three requirements must be satisfied to establish and maintain
the 3:2 commensurability and hence promote outward migration: 
\textit{1)} the exterior planet must stop growing (see Section~\ref{sec:PPoA}),
\textit{2)} the interior planet must do the same (see Section~\ref{sec:SPoA}), 
and
\textit{3)} the relative migration speed prior to capture must be large, so
that the exterior planet can transit the 1:2 orbital resonance (see Section~\ref{sec:21}).
If requirement \textit{1)} is violated, the mass ratio $M_{2}/M_{1}$ may approach
$1$ on a timescale shorter than the migration timescale (see Figure~\ref{fig:acc}),
and the outward motion is interrupted.
If requirement \textit{2)} is violated, the accretion rate through the disk, 
past the interior planet, is reduced. The surface density inside the orbit 
of the inner planet drops and so does the positive Lindblad torque exerted
on the planet, inhibiting outward migration (see Figure~\ref{fig:eff}).
If requirement \textit{3)} is violated, migration is not reversed and both planets
continue moving toward the star (see Figure~\ref{fig:12mmr}). The 1:2 orbital resonance
may still induce outward migration, but at masses larger than Jupiter's and Saturn's
(see Figure~\ref{fig:mass_con}).

The outward migration mechanism is operable, as also argued in previous studies,
but the limitations can be severe. In particular,
it is difficult to reconcile the absence of accretion on both giant planets with the
presence of gas around the planets \citep[see, e.g.,][and references therein]{lissauer2009}.
Two processes capable of shutting down the accretion of gas must be invoked
although, in principle, they need not be different.
Additionally, since envelope collapse begins once the envelope mass exceeds 
the core mass, it is reasonable to assume that before growth stops both
planets were undergoing runaway gas accretion, i.e., digesting all the gas 
the nebula could provide.
Presumably, the sought processes are not ``internal'', i.e., related 
to the structure of the envelopes, because otherwise they would likely occur 
around similar envelope masses\footnote{%
The fast contraction phase that initiates runaway gas accretion is not much 
influenced by boundary conditions, i.e., by the thermodynamical state of the disk. 
Furthermore, given the compact orbital configuration of the planets, 
it is unlikely that disk conditions would be very different at the two locations.}, 
which is obviously not the case. 
But if the processes are of an ``external'' nature, i.e., related to
the supply of gas, then the interior planet would probably undergo through
said process before the exterior planet would, since disk gas removal within 
several \AU\ proceeds from the inside out. Therefore, it seems as 
though the problem of stopping gas accretion on both planets does not 
admit a trivial solution.

The transit of the exterior planet across the 1:2 commensurability 
with the interior planet, within a few $\AU$ of the star, requires surface
densities far in excess of those predicted by the disk evolution models 
constructed here. The 1:2 resonant-orbit configuration is then favored,
in a statistical sense, over the 2:3 one.
It is worth mentioning that core-nucleated accretion models necessitate
high enough surface densities of solid material, but in the form of 
planetesimals not of dust. Since it takes time to turn dust into planetesimals
and the gaseous disk evolves during that time, dust-to-gas mass ratios do
not provide useful information about gas densities at the time giant planets
acquired most of their gaseous contents. 
Disk-limited accretion rates depend linearly on the gas surface density,
but tend to be rather large. 
At $\sim 5\,\AU$, if $\Sigma$ was of order $10\,\mathrm{g\,cm}^{-2}$ 
around the time of the runaway gas accretion phase, it would take 
$\sim 10^{5}$ years to deliver over one half of the current mass of Jupiter
\citep[see, e.g.,][and references therein]{lissauer2009}.
At $\sim 9\,\AU$, a few $\mathrm{g\,cm}^{-2}$ would be sufficient to deliver 
about a Saturn mass worth of gas in $\sim 10^{5}$ years.
Hence, it is reasonable to assume that low gas densities in the 
solar nebula do not prevent the giant planets from reaching 
their final masses.

Even though it appears unlikely that Saturn can transit the 1:2 commensurability 
with Jupiter in an evolved nebula, there is the possibility that it forms 
within this orbital resonance. 
In such case, however, we argue (see Section~\ref{sec:formation12}) that 
it may be difficult for the pair to reach the $1$--$2\,\AU$ disk region. 
In fact, capture in the 2:3 mean motion resonance with Jupiter and ensuing 
migration reversal can occur before Saturn attains its full mass
(see Figure~\ref{fig:0102}), probably when it has between about 
$1/3$ and $2/3$ of the final mass.

\acknowledgments

We thank the referee for a prompt response and helpful suggestions.
G.D.\ thanks Los Alamos National Laboratory for its hospitality.
G.D.\ acknowledges support from NASA Outer Planets Research Program
grant 202844.02.02.01.75 and from NASA Origins of solar systems Program 
grants NNX11AD20G and NNX11AK54G.
Resources supporting this work were provided by the NASA High-End
Computing (HEC) Program through the NASA Advanced Supercomputing
(NAS) Division at Ames Research Center.





\end{document}